\documentclass[11pt,a4paper]{article}
\pdfoutput=1
\usepackage[utf8]{inputenc}
\usepackage{jheppub}
\hypersetup{unicode=true,bookmarksopen=true}

\usepackage{amsthm}
\usepackage{mathtools}
\usepackage{amsthm}
\usepackage{lmodern}
\usepackage[T1]{fontenc}
\usepackage{bbm}
\DeclareMathAlphabet{\mathbfi}{OML}{cmm}{b}{it}

\let\originalleft\left
\let\originalright\right
\renewcommand{\left}{\mathopen{}\mathclose\bgroup\originalleft}
\renewcommand{\right}{\aftergroup\egroup\originalright}
\makeatletter
\newcommand{\biggg}{\bBigg@\thr@@}
\newcommand{\Biggg}{\bBigg@{3.5}}
\makeatother

\makeatletter
\newenvironment{equations}[1][]{\subequations\ifx\relax#1\relax\else\label{#1}\fi\align\ignorespaces}{\endalign\ignorespacesafterend\endsubequations}
\def\@spliteq#1{\begin{equation}\begin{split}#1\end{split}\end{equation}}
\def\@spliteqstar#1{\begin{equation*}\begin{split}#1\end{split}\end{equation*}}
\def\splitequation{\collect@body\@spliteq}
\expandafter\def\csname splitequation*\endcsname{\collect@body\@spliteqstar}

\expandafter\def\csname endsplitequation*\endcsname{\ignorespacesafterend}
\makeatother

\makeatletter
\g@addto@macro\@floatboxreset\centering
\makeatother

\renewcommand{\vec}[1]{{\ifnum9<1#1\mathbf{#1}\else\ifcat\noexpand#1\relax\boldsymbol{#1}\else\mathbfi{#1}\fi\fi}}
\newcommand{\mathe}{\mathrm{e}}
\newcommand{\mathi}{\mathrm{i}}
\let\oldre\Re
\let\oldim\Im
\renewcommand{\Re}{\oldre\mathfrak{e}\,}
\renewcommand{\Im}{\oldim\mathfrak{m}\,}
\newcommand{\total}{\mathop{}\!\mathrm{d}}
\newcommand{\laplace}{\mathop{}\!\bigtriangleup}
\newcommand{\abs}[1]{{\left\lvert{#1}\right\rvert}}
\newcommand{\norm}[1]{{\left\lVert{#1}\right\rVert}}
\newcommand{\sgn}{\operatorname{sgn}}

\newcommand{\arctanh}{\operatorname{arctanh}}

\newcommand{\1}{\mathbbm{1}}
\newcommand{\tr}{\operatorname{tr}}
\newcommand{\eqend}[1]{\,#1}
\newcommand{\bigo}[1]{\mathcal{O}\left({#1}\right)}

\newcommand{\bessel}[3]{\mathop{}\!\mathrm{#1}_{#2}\left(#3\right)}
\newcommand{\hypergeom}[2]{\,{}_{#1}\mathrm{F}_{#2}}
\newcommand{\op}{\mathcal{O}}
\newcommand{\normord}[1]{\mathopen{:}{#1}\mathclose{:}}

\newcommand{\supp}{\operatorname{supp}}

\usepackage{tikz}
\usepackage{storebox}

\allowdisplaybreaks

\begin{document}

\makeatletter

\storebox{\n@pastcone}{\tikz[baseline=0pt]{\draw (0,0) arc(190:350:5pt and 2pt) -- (5pt,9pt) -- cycle; \draw[densely dotted] (0,0) arc(170:10:5pt and 2pt);}}
\storebox{\s@pastcone}{\tikz[baseline=0pt,very thin,scale=0.5]{\draw (0,0) arc(190:350:5pt and 2pt) -- (5pt,8pt) -- cycle; \draw[dash pattern=on \pgflinewidth off 0.5pt] (0,0) arc(170:10:5pt and 2pt);}}

\storebox{\n@futurecone}{\tikz[baseline=0pt]{\draw (0,8pt) arc(170:10:5pt and 2pt) -- (5pt,-1pt) -- cycle; \draw (0,8pt) arc(190:350:5pt and 2pt);}}
\storebox{\s@futurecone}{\tikz[baseline=0pt,very thin,scale=0.5]{\draw (0,7pt) arc(170:10:5pt and 2pt) -- (5pt,-1pt) -- cycle; \draw (0,7pt) arc(190:350:5pt and 2pt);}}

\storebox{\n@doublecone}{\tikz[baseline=0.5pt]{\draw (0,4pt) arc(190:350:5pt and 1.5pt); \draw (0,4pt) -- (5pt,9pt) -- (10pt,4pt) -- (5pt,-1pt) -- cycle; \draw[densely dotted] (0,4pt) arc(170:10:5pt and 1.5pt);}}
\storebox{\s@doublecone}{\tikz[baseline=0pt,very thin,scale=0.5]{\draw (0,4pt) arc(190:350:5pt and 1.5pt); \draw (0,4pt) -- (5pt,9pt) -- (10pt,4pt) -- (5pt,-1pt) -- cycle; \draw[dash pattern=on \pgflinewidth off 0.5pt] (0,4pt) arc(170:10:5pt and 1.5pt);}}

\protected\def\pastcone{\protect{\mathchoice{\usestorebox\n@pastcone}{\usestorebox\n@pastcone}{\usestorebox\s@pastcone}{\usestorebox\s@pastcone}}}
\protected\def\futurecone{\protect{\mathchoice{\usestorebox\n@futurecone}{\usestorebox\n@futurecone}{\usestorebox\s@futurecone}{\usestorebox\s@futurecone}}}
\protected\def\doublecone{\protect{\mathchoice{\usestorebox\n@doublecone}{\usestorebox\n@doublecone}{\usestorebox\s@doublecone}{\usestorebox\s@doublecone}}}

\makeatother


\title{Modular Hamiltonian for de~Sitter diamonds}

\author{Markus B. Fr\"ob}

\affiliation{Institut f{\"u}r Theoretische Physik, Universit{\"a}t Leipzig, Br{\"u}derstra{\ss}e 16, 04103 Leipzig, Germany}

\emailAdd{mfroeb@itp.uni-leipzig.de}

\abstract{We determine the Tomita--Takesaki modular data for CFTs in double cone and light cone regions in conformally flat spacetimes. This includes in particular the modular Hamiltonian for diamonds in the de Sitter spacetime. In the limit where the diamonds become large, we show that the modular automorphisms become time translations in the static patch. As preparation, we also provide a pedagogical rederivation of the known results for Minkowski spacetime. With our results and using the Araki formula, it becomes possible to compute relative entanglement entropies for CFTs in these regions.}


\maketitle

\section{Introduction}

Entropy as a fundamental thermodynamic concept is a measure of randomness or uncertainty, i.e., of the lack of detailed knowledge about a system. In quantum mechanics, a microscopic definition of entropy can be given as the von Neumann entropy
\begin{equation}
\mathcal{S}_\text{vN}(\rho) = - \tr\left( \rho \ln \rho \right)
\end{equation}
associated to a density matrix $\rho$ that describes the state of the system, generalizing the classical statistical definition in terms of probability measures. A particularly important case is the entanglement entropy $\mathcal{S}_\text{E}$, which measures the degree to which two subsystems that comprise the full system are independent, and which is obtained as the von Neumann entropy of the reduced density matrix associated to one of the subsystems. If the full system can be described by a product state, then clearly the reduced density matrices will be pure and hence the von Neumann entropy vanishes. Otherwise, the two subsystems are entangled, which is a key feature distinguishing quantum mechanics from classical statistical physics. 

Entanglement and entanglement entropy have been extensively studied in both quantum mechanical systems and quantum field theory models, see for example Refs.~\cite{holzheylarsenwilczek1994,calabresecardy2004,calabresecardy2009} for early work in two-dimensional conformal field theories, Ref.~\cite{casinihuerta2023} for a review, or Refs.~\cite{bergesfloerchingervenugopalan2018,katsinispastrastetradis2023} for recent computations. However, a straightforward generalization of the von Neumann entropy and hence the entanglement entropy results in ultraviolet divergences, and a typical result for the entanglement entropy computed for a $(d+1)$-dimensional quantum field theory with a UV cutoff $\epsilon \to 0$ has the form (see for example Ref.~\cite{marolfwall2016} and references therein)
\begin{equation}
\label{eq:intro_ee_expansion}
\mathcal{S}_\text{E}(V) = g_{d-1}[\partial V] \epsilon^{-(d-1)} + \cdots + g_1[\partial V] \epsilon^{-1} + g_0[\partial V] \ln \epsilon + \mathcal{S}_0(V) + \bigo{\epsilon} \eqend{.}
\end{equation}
Here, $V$ is the region to which the observer has access (such that the reduced density matrix is obtained by tracing over the degrees of freedom outside of $V$), the $g_i$ are local functions of its boundary $\partial V$ which are homogeneous of degree $i$, and $\mathcal{S}_0(V)$ is the finite part of the entanglement entropy; while $g_0$ is independent of the precise regulator, the other $g_i$ are not. Physically, the UV divergences stem from vacuum fluctuations of high-energy modes of the quantum field, and are therefore expected to be universal and state-independent.\footnote{See however Ref.~\cite{marolfwall2016} for situations where this does not hold.} Mathematically, the difference lies in the type of operator algebras that one considers: while in quantum mechanics one usually deals with a type I von Neumann algebra $\mathfrak{A}$ of operators, the algebras of quantum fields are typically of type III~\cite{buchholzfredenhagendantoni1987} which do not admit a trace. For details, we refer the reader to Refs.~\cite{witten2018,sorce2023a} and references therein, and we also note that intermediate type II algebras have recently attracted attention~\cite{witten2022,aliahmadjefferson2023,jensensorcesperanza2023,chandrasekaranpeningtonwitten2023,chandrasekaranlongopeningtonwitten2023,klingerleigh2023,witten2023} in connection with the proper mathematical description of quantum gravity.

Entanglement entropy also plays a role in the AdS/CFT correspondence, where the famous Ryu--Takayanagi proposal~\cite{ryutakayanagi2006a,ryutakayanagi2006b} (or a recent covariant generalization~\cite{hubenyrangamanitakayanagi2007}, see also Ref.~\cite{casinihuertamyers2011}) can be used to compute it via a holographic correspondence. Moreover, similar studies have been carried out in the context of de Sitter holography~\cite{gauryajnik2023,boutivaspastrastetradis2023,ageevetal2023,frankenpartoucherondeautoumbas2023,kawamotoruansuzukitakayanagi2023} or brane world scenarios~\cite{gengetal2023}. Entanglement and entanglement entropy have furthermore been used to derive gravitational laws~\cite{lashkarimcdermottvanraamsdonk2014,jacobson2016,buenominsperanzavisser2017,svesko2019,alonsoserranoliska2022,alonsoserranoliska2020,alonsoserranoliska2023}, based on the older ideas of emergent gravity of Jacobson~\cite{jacobson1995}, Padmanabhan~\cite{padmanabhan2010} and Verlinde~\cite{verlinde2011}. These ideas are based on the fact that for a thermal density matrix $\rho = Z^{-1} \mathe^{- \beta H}$ with the Hamiltonian $H$, the inverse temperature $\beta$ and the partition function $Z = \mathe^{- \beta F}$ with the free energy $F$, the von Neumann entropy $\mathcal{S}_\text{vN}(\rho) = \beta \tr( \rho H ) - \beta F$ agrees with the thermodynamic entropy. One therefore expects that relations analogous to the laws of thermodynamics also hold for the entanglement entropy (as a special case of the von Neumann entropy), even in situations more general than thermodynamic equilibrium. In fact, analogues of the first and second law involving entanglement entropy have been derived~\cite{blancocasinihungmyers2013,allahbakhshialishahihanaseh2013}, and from those relations one obtains gravitational laws by either imposing a condition of entanglement equilibrium~\cite{jacobson2016} or assuming the AdS/CFT correspondence~\cite{lashkarimcdermottvanraamsdonk2014}.

Even though one can define a renormalized entanglement entropy by subtracting divergences from the result~\eqref{eq:intro_ee_expansion}, the resulting quantity might not be unique, and in some cases the divergences can even be state-dependent~\cite{marolfwall2016}. It is therefore important to study quantities which are finite also in quantum field theory, such as the relative entropy
\begin{equation}
\mathcal{S}\left( \rho \Vert \sigma \right) = \tr\left( \rho \ln \rho - \rho \ln \sigma \right) \eqend{,}
\end{equation}
which can be interpreted as the expected gain of information when passing from the description of the subsystem via the density matrix $\sigma$ to $\rho$. While it is not clear from this definition that this quantity is well-defined and finite also for quantum field theories, this follows from a reformulation in terms of Tomita--Takesaki or modular theory as explained below. For some recent studies of relative entropy in quantum field theory, we refer the reader to Refs.~\cite{hollands2019,dangelo2021,ciollilongoranalloruzzi2022,galandamuchverch2023,garbarzpalau2023}. In fact, the relative entanglement entropy can be used instead of the von Neumann entropy to derive thermodynamic laws~\cite{floerchingerhaas2020,dowlingfloerchingerhaas2020}, and both the relative entropy and its rigorous formulation using modular theory can be employed in the holographic correspondence~\cite{jafferissuh2016,jafferislewkowyczmaldacenasuh2016,verlindezurek2020,boussoetal2020} instead of the divergent entanglement entropy.

Due to the connection with relative entropy, Tomita--Takesaki or modular theory~\cite{tomita1967,takesaki1970} has also received renewed attention. In its most general form, modular theory considers a von Neumann algebra $\mathfrak{A} \subset \mathcal{B}(\mathcal{H})$ of bounded operators on a Hilbert space $\mathcal{H}$ and a cyclic and separating vector $\Omega \in \mathcal{H}$, i.e., the set $\mathfrak{A} \Omega \equiv \{ a \Omega\colon a \in \mathfrak{A} \}$ is dense in $\mathcal{H}$ and if $a \Omega = 0$ for some $a \in \mathfrak{A}$ then $a = 0$. The antilinear operator $S$ defined as the closure of the map $S_0\colon \mathfrak{A} \Omega \to \mathfrak{A} \Omega$, $a \Omega \mapsto a^\dagger \Omega$ is called the Tomita operator, and its polar decomposition $S = J \Delta^\frac{1}{2}$ with $\Delta = S^\dagger S$ gives the modular operator $\Delta$ and the modular conjugation $J$; while $\Delta$ is a positive operator, $J$ is antilinear. One of the main results of Tomita--Takesaki theory is that the modular operator induces an automorphism of $\mathfrak{A}$, the modular automorphism or modular flow $\sigma_s$, via
\begin{equation}
\label{eq:intro_automorphism}
\mathfrak{A} \ni a \mapsto \sigma_s(a) \equiv \Delta^{\mathi s} \, a \, \Delta^{- \mathi s} \in \mathfrak{A} \eqend{,}
\end{equation}
whose generator $\ln \Delta$ is the modular Hamiltonian. Moreover, the vector state $\omega$ on $\mathfrak{A}$ defined by $\Omega$ is a KMS (thermal) state with respect to this automorphism: for any two $a,b \in \mathfrak{A}$, the function
\begin{equation}
F(s) \equiv \omega\left( \sigma_s(a) b \right) = \left( \Omega, \sigma_s(a) b \, \Omega \right)
\end{equation}
admits an analytic continuation to the strip $-1 \leq \Im s \leq 0$, and its boundary value at $\Im s = - 1$ is given by
\begin{equation}
\label{eq:intro_kms}
\omega\left( \sigma_{s-\mathi}(a) b \right) = F(s-\mathi) = \omega\left( b \, \sigma_s(a) \right) \eqend{.}
\end{equation}
That is, Tomita--Takesaki theory shows that any pair of von Neumann algebra $\mathfrak{A}$ and cyclic and separating state $\Omega$ admits a natural time evolution (with the time being the parameter $s$ of the modular automorphism), with respect to which $\Omega$ is a thermal state of inverse temperature $\beta = 1$. Note that while the modular automorphism $\sigma_s$~\eqref{eq:intro_automorphism} maps the algebra $\mathfrak{A}$ into itself, the unitaries $\Delta^{\mathi s}$ do not belong to $\mathfrak{A}$, i.e., the automorphism is outer. For more details and pedagogical derivations of these results, we refer the reader to Refs.~\cite{witten2018,sorce2023b} and references therein.

\begin{figure}
\includegraphics[]{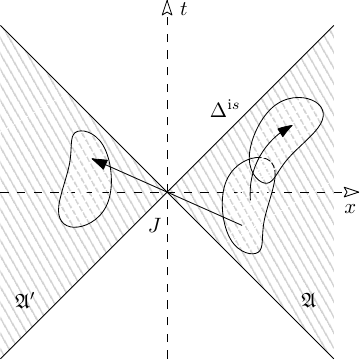}
\caption{Modular data for the Rindler wedge in Minkowski spacetime. Shown are the action of the modular automorphism induced by the modular operator $\Delta$, mapping the algebra $\mathfrak{A}$ of fields localized in the right wedge into itself, and the action of the modular conjugation $J$, mapping $\mathfrak{A}$ into its commutant $\mathfrak{A}'$.}
\label{fig:rindler}
\end{figure}

A prominent example where all appearing objects are explicitly known is the algebra of quantum fields in the Rindler wedge in Minkowski spacetime, see Fig.~\ref{fig:rindler}.\footnote{While the fields $\phi(f)$ are unbounded operators (even when smeared with a test function $f$), the Weyl operators $\exp\left[ \mathi \phi(f) \right]$ are bounded and form a von Neumann algebra. Via a limit argument, the conclusions of modular theory then also apply to the fields $\phi(f)$ themselves. See also~\cite{inoue1998} for a direct extension of modular theory to algebras of unbounded operators.} By the Reeh--Schlieder theorem~\cite{reehschlieder1961}, the Minkowski vacuum $\Omega$ is cyclic and separating, and it is well known that it appears as a thermal state with respect to evolution in Rindler time, at least for free fields~\cite{unruh1976}. Since evolution in Rindler time is generated by the Lorentz boosts, it is clear that the modular Hamiltonian $\ln \Delta$ is the boost generator $K$, properly rescaled to ensure an inverse temperature $\beta = 1$. The modular conjugation $J$ maps from the Rindler wedge to the opposite wedge, i.e., from the right one into the left one and vice versa. Because the left and right Rindler wedge are spacelike separated, operators localized in these two wedges commute with each other. Therefore, $J$ maps $\mathfrak{A}$ into its commutant $\mathfrak{A}' \equiv \{ a \in \mathcal{B}(\mathcal{H}) \colon [a,b] = 0 \text{ for all } b \in \mathfrak{A} \}$, verifying another result of modular theory in this case. By the Bisognano--Wichmann theorem~\cite{bisognanowichmann1975,bisognanowichmann1976}, these results hold for arbitrary Wightman quantum field theories, which may include interactions.\footnote{We refer the reader to Refs.~\cite{moretti1997,michel2016,blommaertmertensverscheldezakharov2018} and references therein for issues that arise for gauge fields.}

Given two states $\Phi$ and $\Psi$, one can define analogously the relative Tomita operator $S_{\Phi\vert\Psi}$ as the closure of the map $a \Phi \mapsto a^\dagger \Psi$ for $a \in \mathfrak{A}$. While $\Phi$ still must by cyclic and separating for $\mathfrak{A}$, $\Psi$ can in principle be arbitrary. The polar decomposition of $S_{\Phi\vert\Psi} = J_{\Phi\vert\Psi} \Delta_{\Phi\vert\Psi}^\frac{1}{2}$ with $\Delta_{\Phi\vert\Psi} = S_{\Phi\vert\Psi}^\dagger S_{\Phi\vert\Psi}$ then gives the relative modular operator $\Delta_{\Phi\vert\Psi}$ and the relative modular conjugation $J_{\Phi\vert\Psi}$; if $\Psi$ is not cyclic and separating, one has to be careful with the exact definition of the polar decomposition~\cite{witten2018}. Via the Araki formula~\cite{araki1975,araki1976}, the relative entropy $\mathcal{S}(\Phi\Vert\Psi)$ between the two states $\Phi$ and $\Psi$ is given by
\begin{equation}
\label{eq:intro_araki}
\mathcal{S}(\Phi\Vert\Psi) = - \left( \Phi, \ln \Delta_{\Phi\vert\Psi} \Phi \right) \eqend{,}
\end{equation}
that is the expectation value of the relative modular Hamiltonian in the state $\Phi$. This is the connection to relative entropy mentioned at the beginning; for further details, and in particular the relation to more known definitions of relative entropy we refer the reader to the reviews~\cite{hollandssanders2017,witten2018}. In practice, an important case is when $\Phi = u u' \Omega$ and $\Psi = v v' \Omega$ for unitary operators $u,v \in \mathfrak{A}$ and $u',v' \in \mathfrak{A}'$ commuting with $u$ and $v$. One easily verifies that the relative Tomita operator reads $S_{\Phi\vert\Psi} = v' u S_\Omega v^\dagger (u')^\dagger$, with $S_\Omega$ the Tomita operator associated to $\mathfrak{A}$ and $\Omega$. It follows that $\Delta_{\Phi\vert\Psi} = u' v \Delta_\Omega v^\dagger (u')^\dagger$ and
\begin{equation}
\label{eq:intro_araki2}
\mathcal{S}(\Phi\Vert\Psi) = - \left( v^\dagger u \Omega, \ln \Delta_\Omega v^\dagger u \Omega \right) \eqend{,}
\end{equation}
such that the relative entropy between two ``excited'' states relative to a ``vacuum'' state $\Omega$ can be computed using only the modular Hamiltonian $\ln \Delta_\Omega$ of the ``vacuum'' state. In particular, one can in this way compute the relative entropy between the Minkowski vacuum and a coherent state with $v = \exp\left[ \mathi \phi(f) \right]$~\cite{lashkariliurajagopal2021,longo2019,casinigrillopontello2019}.

Despite this important connection, the modular data is only known explicitly in some cases. Most of them are instances of so-called geometric modular action~\cite{buchholzsummers1993,buchholzdreyerflorigsummers2000}, where the modular Hamiltonian is the generator of a spacetime symmetry or a simple combination of such. Apart from the above-mentioned Rindler wedge, this includes the following cases:
\begin{itemize}
\item Massless free scalar fields in the Minkowski vacuum state, with $\mathfrak{A}$ being the algebra of fields inside the future light cone. The modular Hamiltonian $\ln \Delta$ is proportional to the generator of dilations, and the modular conjugation $J$ maps fields into the past light cone, see Fig.~\ref{fig:lightcone1}. This was shown by Buchholz~\cite{buchholz1977}, and we rederive it in section~\ref{sec:minkowski_modular_light_cone}.
\item Massless free scalar fields in the Minkowski vacuum state, with $\mathfrak{A}$ being the algebra of fields inside a double cone or diamond. The modular Hamiltonian $\ln \Delta$ is proportional to the generator of the flow along a conformal Killing vector field, and the modular conjugation $J$ maps fields into various parts of other cones, see Fig.~\ref{fig:diamond2}. This was shown by Hislop and Longo~\cite{hisloplongo1982,hislop1988}, and we rederive it in section~\ref{sec:minkowski_modular_double_cone}.
\item Scalar fields in the de~Sitter Euclidean (or Bunch--Davies) vacuum state~\cite{chernikovtagirov1968,schomblondspindel1976,bunchdavies1978}, with $\mathfrak{A}$ being the algebra of fields in a region obtained as the intersection of the de~Sitter hyperboloid with a Rindler wedge in the embedding Minkowski spacetime. Since the boost generator in the wedge is tangent to the de~Sitter hyperboloid, the modular data can be obtained as the restriction of the data of the Rindler wedge to the hyperboloid. This was shown by Bros, Epstein and Moschella~\cite{brosepsteinmoschella1998} and independently by Borchers, Buchholz, Dreyer, Florig and Summers~\cite{dreyer1996,florig1999,borchersbuchholz1999,buchholzdreyerflorigsummers2000}, and analogous results hold also for anti-de Sitter spacetime~\cite{buchholzflorigsummers2000}.
\item Scalar fields in an FLRW spacetime with closed spatial sections, embedded in de~Sitter. Both the quantum state, the algebra of fields and the modular data are obtained by restricting the previous results from the de~Sitter wedge regions to FLRW~\cite{buchholzmundsummers2001,buchholzmundsummers2002}. However, it is not clear whether this state or the algebra of fields correspond to any physically reasonable theory; only for massless, conformally coupled fields it is known that this construction gives the conformal vacuum.
\item Scalar fields in the de~Sitter Euclidean (or Bunch--Davies) vacuum state, with $\mathfrak{A}$ being the algebra of fields in the static patch. The modular Hamiltonian $\ln \Delta$ is proportional to the generator of time translations in the static patch, and the modular conjugation $J$ maps fields from the static patch embedded in the expanding Poincaré patch to the complementary static patch embedded in the contracting Poincaré patch. This follows from the fact that the Euclidean vacuum state in de~Sitter is a thermal state with respect to the translations in time $T$ of the static patch, which was shown by Figari, Höegh-Krohn and Nappi~\cite{figarihoeghkrohnnappi1975} in two and by Gibbons and Hawking~\cite{gibbonshawking1977} in four dimensions. The explicit connection to modular theory was spelled out in~\cite{chandrasekaranlongopeningtonwitten2023}, and we will obtain it as a limit of the construction for diamonds in the static patch in section~\ref{sec:desitter_modular_double_cone}.
\item Free scalar fields in the Hartle--Hawking state of the Kruskal extension of the Schwarz\-schild metric, with $\mathfrak{A}$ being the algebra of fields in either the right or left Kruskal wedge. The modular Hamiltonian $\ln \Delta$ is proportional to the generator of translations in Schwarzschild time $t$, and the modular conjugation $J$ maps fields between the two Kruskal wedges. This was shown explicitly by Kay~\cite{kay1985a,kay1985b}, based on an axiomatic approach of Sewell~\cite{sewell1982}; see also Kay and Wald~\cite{kaywald1991}.
\end{itemize}
More examples are known in one or two dimensions, both for free fields (scalars and fermions) and conformal field theories in various quantum states and for the algebras of fields restricted to complicated subregions (also including defects and boundaries), see for example Refs.~\cite{borchersyngvason1999,peschel2002,casinihuerta2008,casinihuerta2009a,casinihuerta2009b,longomartinettirehren2010,casinihuerta2011,rehrentedesco2013,cardytonni2016,ariasblancocasinihuerta2017,casinitestetorroba2017,tonnirodriguezlagunasierra2018,ariascasinihuertapontello2018,abterdmenger2018,eislertonnipeschel2019,hollands2021,blancopereznadal2019,friesreyes2019a,friesreyes2019b,digiuliotonni2020,eislerdigiuliotonnipeschel2020,erdmengerfriesreyessimon2020,mintchevtonni2021a,mintchevtonni2021b,gravakelstonni2021,javerzattonni2022,eislertonnipeschel2022,mintchevtonni2022,abateblancokoifmanpereznadal2023,rottolimurcianotonnicalabrese2023,digiulioerdmenger2023,huertavandervelde2023a,huertavandervelde2023b} and references therein.

The purpose of this work is to extend the above list of known examples, and derive the modular data for the algebra of conformal fields restricted to light cone and double cones in conformally flat spacetimes. In particular, this includes the case of diamonds (of arbitrary size) in de~Sitter spacetime, which is important as a model for both the primordial and present exponentially expanding phases of our universe.

The remainder of this article is structured as follows: in section~\ref{sec:minkowski} we first present a new derivation of the modular data for massless free scalar fields in the Minkowski vacuum state restricted to light cones and double cones. We then transplant the results to conformally flat spacetimes and with the fields in the conformal vacuum state in section~\ref{sec:conformal}, and specialize to de~Sitter spacetime in section~\ref{sec:desitter}. The conclusions and an outlook comprise section~\ref{sec:conclusions}, and we relegate various computations to the appendices~\ref{app:inversion}, \ref{app:modularfock} and~\ref{app:alt}. We use the $(-,+,\ldots,+)$ convention for the metric, such that $p^2 = \vec{p}^2 - (p^0)^2$ for the momentum and $x^2 = \vec{x}^2 - (x^0)^2$ for the coordinate interval, occasionally denote $x^0 = t$ and $\abs{\vec{x}} = r$, and work in $d$ dimensions. The Fourier transform is defined by $\tilde{f}(p) = \int f(x) \, \mathe^{- \mathi p x} \total^d x$ with inverse $f(x) = \int \tilde{f}(p) \, \mathe^{\mathi p x} \total^d p/(2\pi)^d$.

\section{Minkowski spacetime}
\label{sec:minkowski}

In this section we rederive and slightly generalize the results of Buchholz~\cite{buchholz1977} and Hislop and Longo~\cite{hisloplongo1982,hislop1988} for the modular data of light cones and double cones in Minkowski spacetime, for massless scalar fields in the Minkowski vacuum state. We first recall the quantization of the free real scalar field, the discrete CPT symmetries, the (improved) stress tensor and the implementation of Lorentz symmetries via unitaries constructed from the contraction of the stress tensor with the corresponding Killing vector field of Minkowski spacetime. Using these symmetries, we then construct candidates for the modular data according to the principle of geometric modular action, and show that they fulfill all the required conditions.

\subsection{Free real scalar field}
\label{sec:minkowski_scalarfield}

As usual, we will work on the bosonic Fock space $\mathcal{F}$ over the single-particle Hilbert space $\mathcal{H} = L^2(\mathbb{R}^{d-1})$, the space of complex square-integrable functions on $\mathbb{R}^{d-1}$ with scalar product $(f,g) \equiv \int f^*(\vec{x}) g(\vec{x}) \total^{d-1} \vec{x}$. We denote the vacuum vector in $\mathcal{F}$ by $\Omega$, and consider the usual (smeared) annihilation and creation operators $a(f)$, $a^\dagger(g)$ with $f,g \in \mathcal{H}$. Their commutation relations read
\begin{equation}
\label{eq:minkowski_aadagger}
[ a(f), a(g) ] = 0 = [ a^\dagger(f), a^\dagger(g) ] \eqend{,} \quad [ a(f), a^\dagger(g) ] = (f,g) \1 \eqend{,}
\end{equation}
and we note that $a(f)$ is antilinear in $f$, while $a^\dagger(f)$ is linear. Creation and annihilation operators with sharp momentum can be obtained in the limit where $f(\vec{x}) \to \mathe^{\mathi \vec{p} \vec{x}}$, but those are not square integrable and thus not elements of $\mathcal{H}$. To treat such operators rigorously one needs to consider Gel'fand triples, also known as rigged Hilbert spaces. In this context, one can then rigorously define
\begin{equations}
a(f) &= \int a(\vec{p}) \int f^*(\vec{x}) \, \mathe^{\mathi \vec{p} \vec{x}} \total^{d-1} \vec{x} \frac{\total^{d-1} \vec{p}}{(2\pi)^{d-1}} \eqend{,} \\
a^\dagger(f) &= \int a^\dagger(\vec{p}) \int f(\vec{x}) \, \mathe^{- \mathi \vec{p} \vec{x}} \total^{d-1} \vec{x} \frac{\total^{d-1} \vec{p}}{(2\pi)^{d-1}} \eqend{,}
\end{equations}
while in a strict Hilbert space context the smeared operators are fundamental. In terms of these, for any square integrable function $f \in \mathcal{H}$ we define the (smeared) field
\begin{equation}
\label{eq:minkowski_phi_def}
\phi(f) \equiv a(F_-) + a^\dagger(F_+) \eqend{,} \quad \widetilde{F_+}(\vec{p}) \equiv \left. \frac{\tilde{f}(p)}{\sqrt{2 p^0}} \right\rvert_{p^0 = \abs{\vec{p}}} \eqend{,} \quad \widetilde{F_-}(\vec{p}) \equiv \left. \frac{\left[ \tilde{f}(-p) \right]^*}{\sqrt{2 p^0}} \right\rvert_{p^0 = \abs{\vec{p}}} \eqend{,}
\end{equation}
where $\tilde{f}(p)$ is the Fourier transform of the smearing function $f(x)$, and we have imposed the on-shell condition $p^0 = \abs{\vec{p}}$. In the following, we will always work on-shell, such that $p^0$ is always just a shortcut for $\abs{\vec{p}}$.

On the Fock space $\mathcal{F}$, we can implement discrete symmetries: parity, time reversal and charge conjugation. Parity is unitary and acts according to
\begin{equation}
\label{eq:minkowski_parity_fock}
P a(f) P^{-1} = a\left( f^P \right) \eqend{,} \quad P a^\dagger(f) P^{-1} = a^\dagger\left( f^P \right) \quad\text{with}\quad f^P(\vec{x}) \equiv f(-\vec{x}) \eqend{,}
\end{equation}
and hence when acting on the field
\begin{equation}
\label{eq:minkowski_parity}
P \phi(f) P^{-1} = a\left( F_-^P \right) + a^\dagger\left( F_+^P \right) = \phi(f_P) \quad\text{with}\quad f_P(t,\vec{x}) \equiv f(t,-\vec{x}) \eqend{.}
\end{equation}
Time reversal is antiunitary and acts according to
\begin{equation}
\label{eq:minkowski_timereversal_fock}
T a(f) T^{-1} = a\left( f^T \right) \eqend{,} \quad T a^\dagger(f) T^{-1} = a^\dagger\left( f^T \right) \quad\text{with}\quad f^T(\vec{x}) \equiv f^*(\vec{x}) \eqend{,}
\end{equation}
and hence when acting on the field
\begin{equation}
\label{eq:minkowski_timereversal}
T \phi(f) T^{-1} = a\left( F_-^T \right) + a^\dagger\left( F_+^T \right) = \phi(f_T) \quad\text{with}\quad f_T(t,\vec{x}) \equiv f^*(-t,\vec{x}) \eqend{,}
\end{equation}
since we have to switch the sign of the time integration in the Fourier transform to compensate for the complex conjugation. For a real scalar field, charge conjugation (a unitary operator) does not do anything, so
\begin{equation}
\label{eq:minkowski_chargeconjugation}
C a(f) C^{-1} = a(f) \eqend{,} \quad C a^\dagger(f) C^{-1} = a^\dagger(f) \eqend{,} \quad C \phi(f) C^{-1} = \phi(f) \eqend{.}
\end{equation}
They all keep the vacuum invariant: $P \Omega = T \Omega = C \Omega = \Omega$.

Continuous symmetries are determined via the Noether method from conserved currents, which in turn are obtained by contracting the stress tensor with Killing vector fields or conformal Killing vector fields. The canonical stress tensor for free scalar fields reads
\begin{equation}
\Theta_{\mu\nu} \equiv \partial_\mu \phi \partial_\nu \phi - \frac{1}{2} \eta_{\mu\nu} \partial_\rho \phi \partial^\rho \phi \eqend{,}
\end{equation}
but for conformal symmetries we consider conformally coupled fields and take the improved stress tensor
\begin{splitequation}
\label{eq:minkowski_stress_improved}
T_{\mu\nu} &\equiv \Theta_{\mu\nu} - \frac{d-2}{4 (d-1)} \left( \partial_\mu \partial_\nu - \eta_{\mu\nu} \partial^2 \right) \phi^2 \\
&\approx \frac{d}{2 (d-1)} \partial_\mu \phi \partial_\nu \phi - \frac{d-2}{2 (d-1)} \phi \partial_\mu \partial_\nu \phi - \frac{1}{2 (d-1)} \eta_{\mu\nu} \partial_\rho \phi \partial^\rho \phi \eqend{,}
\end{splitequation}
where we dropped a term proportional to the field equation $\partial^2 \phi$ in the second expression; we denote on-shell equality by an $\approx$ sign. The improved stress tensor is both conserved and traceless on-shell:
\begin{equations}
\partial^\mu T_{\mu\nu} &= \partial^2 \phi \partial_\nu \phi \approx 0 \eqend{,} \\
\eta^{\mu\nu} T_{\mu\nu} &= \frac{d-2}{2} \phi \partial^2 \phi \approx 0 \eqend{,}
\end{equations}
and both the improved and the canonical stress tensor are manifestly symmetric.

Continuous symmetries of the theory are determined via the Noether method by computing conserved currents. Given any Killing vector $\xi^\mu$, which satisfies the Killing equation $\partial_\mu \xi_\nu + \partial_\nu \xi_\mu = 0$, the current
\begin{equation}
\label{eq:minkowski_jxi_def}
J_\xi^\mu \equiv T^{\mu\nu} \xi_\nu
\end{equation}
is conserved on-shell:
\begin{equation}
\partial_\mu J_\xi^\mu = \partial_\mu T^{\mu\nu} \xi_\nu + \frac{1}{2} T^{\mu\nu} \left( \partial_\mu \xi_\nu + \partial_\nu \xi_\mu \right) \approx 0 \eqend{.}
\end{equation}
The classical Noether charge is then obtained by integrating the time component of the current over the hypersurface $t = 0$. In this way, the Killing vector of time translations $\xi^\mu \partial_\mu = \partial_t$ results in the Hamiltonian $H$, the Killing vectors of space translations $\xi^\mu \partial_\mu = \partial_i$ result in the total momentum $P^i$, and the Killing vectors corresponding to rotations and boosts result in the angular momentum $M^{ij}$ and the boost charge $M^{0i}$, whose conservation gives the movement of the relativistic center of mass.

For a conformally coupled field, also conformal Killing vectors $\xi^\mu$ satisfying the conformal Killing equation $\partial_\mu \xi_\nu + \partial_\nu \xi_\mu = \frac{2}{d} \eta_{\mu\nu} \partial_\rho \xi^\rho$ result in conserved charges. The current is defined in the same way as before~\eqref{eq:minkowski_jxi_def}, and since the improved stress tensor is traceless the current is also conserved on-shell:
\begin{equation}
\partial_\mu J_\xi^\mu = \partial_\mu T^{\mu\nu} \xi_\nu + \frac{1}{2} T^{\mu\nu} \left( \partial_\mu \xi_\nu + \partial_\nu \xi_\mu \right) \approx \frac{1}{d} \eta_{\mu\nu} T^{\mu\nu} \partial_\rho \xi^\rho \approx 0 \eqend{.}
\end{equation}
In this way, we obtain the dilation charge $D$ from the conformal Killing vector $\xi^\mu \partial_\mu = x^\mu \partial_\mu$, and the conserved charges of special conformal transformations $K^\mu$ from the special conformal Killing vectors $\xi^\mu = 2 c_\rho x^\rho x^\mu - x^2 c^\mu = c^\rho ( 2 x_\rho x^\mu - x^2 \delta^\mu_\rho )$ for constant $c^\mu$.

In the quantum theory, composite operators are defined by normal-ordering, or more generally by point-splitting and subtraction of the Hadamard parametrix~\cite{hollandswald2002,hollandswald2015}. A formal computation results for the relevant charges $H$, $D$ and $K = K^\mu$ with $c^\mu = \delta^\mu_0$ in
\begin{equations}[eq:minkowski_hdk_def]
H &= \int p^0 \, a^\dagger(\vec{p}) a(\vec{p}) \frac{\total^{d-1} \vec{p}}{(2\pi)^{d-1}} \eqend{,} \\
D &= - \frac{\mathi}{2} \int \left[ a^\dagger(\vec{p}) \vec{p}^i \partial_{\vec{p}^i} a(\vec{p}) - \left( \vec{p}^i \partial_{\vec{p}^i} a^\dagger(\vec{p}) \right) a(\vec{p}) \right] \frac{\total^{d-1} \vec{p}}{(2\pi)^{d-1}} \eqend{,} \\
K &= \frac{1}{2} \int \left[ p^0 \left( \laplace_{\vec{p}} a^\dagger(\vec{p}) \right) a(\vec{p}) + p^0 a^\dagger(\vec{p}) \laplace_{\vec{p}} a(\vec{p}) - \frac{1}{2 p^0} a^\dagger(\vec{p}) a(\vec{p}) \right] \frac{\total^{d-1} \vec{p}}{(2\pi)^{d-1}} \eqend{,}
\end{equations}
see for example~\cite{siboldeden2010}. To properly define these as operators on Fock space $\mathcal{F}$, we first compute the commutators
\begin{equations}[eq:minkowski_hdk_commutator]
[ H, a^\dagger(f) ] &= a^\dagger(f_H) \quad\text{with}\quad \widetilde{f_H}(\vec{p}) \equiv p^0 \tilde{f}(\vec{p}) \eqend{,} \\
[ D, a^\dagger(f) ] &= a^\dagger(f_D) \quad\text{with}\quad \widetilde{f_D}(\vec{p}) \equiv - \mathi \left( \vec{p}^i \partial_{\vec{p}^i} + \frac{d-1}{2} \right) \tilde{f}(\vec{p}) \eqend{,} \\
[ K, a^\dagger(f) ] &= a^\dagger(f_K) \quad\text{with}\quad \widetilde{f_K}(\vec{p}) \equiv p^0 \laplace_{\vec{p}} \tilde{f}(\vec{p}) + \frac{\vec{p}^i}{p^0} \partial_{\vec{p}^i} \tilde{f}(\vec{p}) + \frac{2d-5}{4 p^0} \tilde{f}(\vec{p}) \eqend{.}
\end{equations}
Together with the fact that they all annihilate the vacuum, $H \Omega = D \Omega = K \Omega = 0$, this defines $H$, $D$ and $K$ on the dense subset $\mathcal{D} \subset \mathcal{F}$ comprised of smoothly smeared states of finite particle number, i.e., finite linear combinations of vectors $a^\dagger(f_1) \cdots a^\dagger(f_k) \Omega \in \mathcal{F}$ with finite $k$ and Schwartz functions $f_i \in \mathcal{S}(\mathbb{R}^{d-1})$ for all $i$. From their formal definition~\eqref{eq:minkowski_hdk_def}, $H$, $D$ and $K$ are also clearly symmetric, which results in $[ H, a(f) ] = - a(f_H)$, $[ D, a(f) ] = - a(f_D)$ and $[ K, a(f) ] = - a(f_K)$ with the same functions $f_H$, $f_D$ and $f_K$.

Computing the inverse Fourier transforms in the commutators~\eqref{eq:minkowski_hdk_commutator} explicitly, we also obtain the commutators with the field $\phi$~\eqref{eq:minkowski_phi_def}:
\begin{equations}[eq:minkowski_hdk_phi_commutator]
[ H, \phi(f) ] &= \phi(f_H) \quad\text{with}\quad f_H(x) \equiv \mathi \partial_t f(x) \eqend{,} \\
[ D, \phi(f) ] &= \phi(f_D) \quad\text{with}\quad f_D(x) \equiv \mathi \left( x^\mu \partial_\mu + d-\Delta \right) f(x) \eqend{,} \\
[ K, \phi(f) ] &= \phi(f_K) \quad\text{with}\quad f_K(x) \equiv - 2 t f_D(x) - x^2 f_H(x) \eqend{,}
\end{equations}
where $\Delta = (d-2)/2$ is the scaling dimension of the field, assuming that $f \in \mathcal{S}(\mathbb{R}^d)$. This is nothing else but the adjoint of the transformation of the field operator, and agrees with the well-known transformation laws of scalar primary operators~\cite{francescomathieusenechal}. We can also compute the commutator of any pair of the commutators~\eqref{eq:minkowski_hdk_phi_commutator} [or~\eqref{eq:minkowski_hdk_commutator}], which results in the subalgebra
\begin{equation}
\label{eq:minkowski_hdk_algebra}
[ H, D ] = - \mathi H \eqend{,} \quad [ H, K ] = 2 \mathi D \eqend{,} \quad [ D, K ] = - \mathi K
\end{equation}
of the conformal algebra when acting on vectors in the dense subset $\mathcal{D}$.

On $\mathcal{D}$, we can now define a one-parameter unitary group $U(\lambda)$ whose generator is a mixture of $H$, $D$ and $K$:
\begin{equation}
\label{eq:minkowski_q_def}
Q = Q_{abc} \equiv a H + 2 b D + c K \eqend{,} \quad a,b,c \in \mathbb{R} \eqend{,} \quad U(\lambda) \equiv \mathe^{\mathi \lambda Q} \eqend{.}
\end{equation}
Its adjoint action
\begin{equation}
\label{eq:minkowski_ulambda_phi}
U(\lambda) \phi(f) U(-\lambda) = \phi(f_\lambda)
\end{equation}
can be determined from the commutators~\eqref{eq:minkowski_hdk_phi_commutator}. Taking a derivative of Eq.~\eqref{eq:minkowski_ulambda_phi} with respect to $\lambda$, we obtain the first-order differential equation
\begin{splitequation}
\partial_\lambda \phi(f_\lambda) &= \mathi U(\lambda) [ Q, \phi(f) ] U(-\lambda) \\
&= \mathi a U(\lambda) \phi(f_H) U(-\lambda) + 2 \mathi b U(\lambda) \phi(f_D) U(-\lambda) + \mathi c U(\lambda) \phi(f_K) U(-\lambda) \\
&= \phi[ \mathi a (f_H)_\lambda + 2 \mathi b (f_D)_\lambda + \mathi c (f_K)_\lambda ] \eqend{,}
\end{splitequation}
which can now be solved for $f_\lambda$. Since we consider a conformal theory, we are only guaranteed an action up to conformal transformations, i.e., rescalings for the scalar $\phi$, and hence make the ansatz
\begin{equation}
\label{eq:minkowski_flambda_def}
f_\lambda(x) = A(x_\lambda) f(x_\lambda) \eqend{.}
\end{equation}
This results in the differential equations
\begin{equations}
\partial_\lambda t_\lambda &= - a - 2 b t_\lambda + c t_\lambda^2 + c r_\lambda^2 \eqend{,} \\
\partial_\lambda x^i_\lambda &= 2 ( - b + c t_\lambda ) x^i_\lambda \eqend{,} \\
\partial_\lambda A(x_\lambda) &= 2 ( d-\Delta ) ( - b + c t_\lambda ) A(x_\lambda)
\end{equations}
for the coordinates $x_\lambda$ (with $r_\lambda^2 = \delta_{ij} x^i_\lambda x^j_\lambda$) and the rescaling $A$, which are easily solved. In particular, in spherical coordinates the angles are unchanged, and only $r$ depends on $\lambda$. From the second equation, we obtain $t_\lambda$ in terms of $r_\lambda$, and plugging this into the first equation it can be solved for $r_\lambda$. Finally, we insert the result in the last equation to determine $A$, and the final solution normalized to $x^\mu_0 = x^\mu$ and $A(x_0) = 1$ reads
\begin{equations}[eq:minkowski_flambda_sol]
r_\lambda &= r \omega \left[ \omega + \frac{(b - c t)^2 + \omega^2 - c^2 r^2}{2 \omega} \left[ \cosh\left( 2 \omega \lambda \right) - 1 \right] + (b - c t) \sinh\left( 2 \omega \lambda \right) \right]^{-1} \eqend{,} \\
t_\lambda &= \frac{b}{c} - \frac{r_\lambda}{c r} \left[ (b - c t) \cosh\left( 2 \omega \lambda \right) + \frac{(b - c t)^2 + \omega^2 - c^2 r^2}{2 \omega} \sinh\left( 2 \omega \lambda \right) \right] \eqend{,} \\
A(x_\lambda) &= \left( \frac{r_\lambda}{r} \right)^{d-\Delta} \eqend{,}
\end{equations}
where $\omega \equiv \sqrt{b^2 + a c}$. It is a somewhat lengthy computation to check that $(f_\lambda)_\sigma = f_{\lambda+\sigma}$, i.e., that transforming a function $f$ twice is equal to a single transformation with the sum of the two parameters. We have thus shown that the operators $U(\lambda)$ defined by the adjoint action~\eqref{eq:minkowski_ulambda_phi} fulfill $U(\lambda+\sigma) = U(\lambda) U(\sigma)$, and consequently $U(\lambda)^{-1} = U(-\lambda) = U(\lambda)^\dagger$ when acting on vectors in $\mathcal{D}$. Moreover, since $U(\lambda) \Omega = \Omega$ because $Q$ annihilates the vacuum, and since the adjoint action~\eqref{eq:minkowski_ulambda_phi}, with the solution for $f_\lambda$ given by Eqs.~\eqref{eq:minkowski_flambda_def} and \eqref{eq:minkowski_flambda_sol}, is clearly continuous as $\lambda \to 0$, the one-parameter unitary group $U(\lambda)$ is strongly continuous when acting on $\mathcal{D}$. However, $U(\lambda)$ can now be extended to the full Fock space $\mathcal{F}$ by continuity, and by the Stone--von Neumann theorem~\cite{stone1932,vonneumann1932}, its generator is a self-adjoint operator. Choosing $a$, $b$ and $c$ appropriately, this shows that there exist distinguished self-adjoint extensions of $H$, $D$ and $K$, which we will continue to denote by the same letter, and their combinations. They are defined on all vectors $\psi \in \mathcal{F}$ where the limit $\lim_{\lambda \to 0} (\mathi \lambda)^{-1} \norm{ \left[ U(\lambda) - \1 \right] \psi }$ is finite, and in fact their action is defined by this limit. On $\mathcal{D}$ clearly their action is the same as before, defined by the commutators~\eqref{eq:minkowski_hdk_phi_commutator} [or~\eqref{eq:minkowski_hdk_commutator}] and the fact that they annihilate the vacuum.

Let us summarize: we consider the bosonic Fock space $\mathcal{F}$ of a free massless real scalar field $\phi$ in the Minkowski vacuum state $\Omega$. On this Hilbert space, we have an action of a one-parameter group of unitaries $U(\lambda)$, which is given by Eq.~\eqref{eq:minkowski_ulambda_phi} on the dense subspace $\mathcal{D}$ and extended to $\mathcal{F}$ by continuity. The generator of this action is (a self-adjoint extension of) $Q$~\eqref{eq:minkowski_q_def}, a linear combination of the generator of time translations $H$, the generator of dilations $D$ and a generator of special conformal transformations $K$, acting on $\mathcal{D}$ via the commutators~\eqref{eq:minkowski_hdk_phi_commutator}. The adjoint action~\eqref{eq:minkowski_ulambda_phi}, together with the explicit solution for $f_\lambda$ given by Eqs.~\eqref{eq:minkowski_flambda_def} and~\eqref{eq:minkowski_flambda_sol}, is thus a unitary implementation of a combination of time translations, dilations and special conformal transformations on the Fock space $\mathcal{F}$. In turn, $H$, $D$ and $K$ are obtained via Noether's method from the conserved currents $J_\xi$~\eqref{eq:minkowski_jxi_def}, which are given by the contraction of the (improved) stress tensor with a Killing vector or conformal Killing vector. In a similar way, one can derive one-parameter groups of unitaries that implement other Poincaré or conformal transformations.\footnote{It is known that in this way one can in general only obtain a representation of (part of) the conformal algebra~\cite{macksalam1969,ruehl1973}, but not the full conformal group, which can only be represented on fields living on an $\infty$-dimensional cover of Minkowski spacetime~\cite{hortacsuseilerschroer1972,schroerswieca1974,lueschermack1975}. However, the issue doesn't arise for free fields~\cite{swiecavoelkel1973}, in Euclidean space, or when one considers only certain subgroups of the conformal group~\cite{ciccariellosartori1974}, as we do here.}

In the following, we will only need the adjoint action, and may forget how it is obtained. To stress this, let us write $\op$ (a generic scalar primary operator) instead of $\phi$ and keep $\Delta$ generic as well. We then have the adjoint action
\begin{equation}
\label{eq:minkowski_ulambda_op}
U(\lambda) \op(f) U(\lambda)^\dagger = \op(f_\lambda)
\end{equation}
with $f_\lambda$ given by Eqs.~\eqref{eq:minkowski_flambda_def} and~\eqref{eq:minkowski_flambda_sol} and $\Delta$ generic. The above construction for $\op = \phi$ provides an explicit example where the action~\eqref{eq:minkowski_ulambda_op} is realized; another example would be given (on the same Hilbert space) by $\op = \normord{ \phi^2 }$ with $\Delta = d-2$. Analogous constructions can be made for fields of other spin, but then the commutators~\eqref{eq:minkowski_hdk_phi_commutator} would be different. Moreover, if $\op$ is not real, the action of charge conjugation~\eqref{eq:minkowski_chargeconjugation} changes, and we have
\begin{equation}
\label{eq:minkowski_chargeconjugation_op}
C \op(f) C^{-1} = \op^\dagger(f) \eqend{.}
\end{equation}
The only restriction on the scaling dimension is $\Delta \geq (d-2)/2$, which results from requiring that the Wightman two-point function is positive definite, $( \Omega, \op(f^*) \op(f) \Omega ) > 0$, as is well known~\cite{ruehl1973,ferraragattogrillo1974,mack1977a}. Namely, for fields transforming in an irreducible representation of the conformal algebra the two-point function in the vacuum state is completely fixed up to an overall constant and reads
\begin{equation}
\label{eq:minkowski_wightman}
( \Omega, \op(f) \op(g) \Omega ) = C_\op \lim_{\epsilon \to 0^+} \iint \frac{f(x) g(y)}{[ - (x^0-y^0 - \mathi \epsilon)^2 + (\vec{x}-\vec{y})^2 ]^\Delta} \total^d x \total^d y
\end{equation}
with $C_\op > 0$, and where the complex power $z^{-\Delta}$ is defined by $\abs{z}^{-\Delta} \mathe^{- \mathi \Delta \arg z}$. For $\Delta > \frac{d-2}{2}$, we can use the Fourier transform~\cite[Eq.~(8.730) converted to our conventions]{brychkovprudnikov}
\begin{equation}
\label{eq:minkowski_wightman_fourier}
\lim_{\epsilon \to 0^+} \int \frac{\mathe^{- \mathi p x}}{[ - (t-\mathi \epsilon)^2 + \vec{x}^2 ]^\Delta} \total^d x = \frac{2^{-2\Delta+d+1} \pi^{\frac{d}{2}+1}}{\Gamma\left( \Delta - \frac{d-2}{2} \right) \Gamma(\Delta)} \Theta(p^0) \Theta(-p^2) (-p^2)^{\Delta - \frac{d}{2}} \eqend{.}
\end{equation}
Since this is a positive function in Fourier space, we have
\begin{equation}
( \Omega, \op(f^*) \op(f) \Omega ) = C_\op \frac{2^{-2\Delta+d+1} \pi^{\frac{d}{2}+1}}{\Gamma\left( \Delta - \frac{d-2}{2} \right) \Gamma(\Delta)} \int \abs{ \tilde{f}(p) }^2 \Theta(p^0) \Theta(-p^2) (-p^2)^{\Delta - \frac{d}{2}} \frac{\total^d p}{(2\pi)^d} > 0
\end{equation}
as required from the positivity of the scalar product on Hilbert space. As $\Delta \to (d-2)/2$, we use that~\cite{mack1977b}
\begin{equation}
\label{eq:minkowski_wightman_fourier_limit}
\lim_{\Delta \to \frac{d-2}{2}} \frac{1}{\Gamma\left( \Delta - \frac{d-2}{2} \right)} \Theta(-p^2) \Theta(p^0) (-p^2)^{\Delta - \frac{d}{2}} = \Theta(p^0) \delta(p^2) \eqend{,}
\end{equation}
and recover the Fourier transform of the free massless scalar two-point function. However, we cannot continue $\Delta$ below that bound.

\subsection{Modular data for the future and past light cones}
\label{sec:minkowski_modular_light_cone}

Buchholz~\cite{buchholz1977} has shown that for the future light cone with tip at the origin the modular Hamiltonian is proportional to the generator of dilations, such that the principle of geometric modular action is fulfilled. However, we also want to consider cones with tip at $t = \tau \neq 0$, which are the regions
\begin{equations}[eq:minkowski_lightcones]
\futurecone &= \left\{ (t,\vec{x})\colon t > \tau, t-\tau > r = \abs{\vec{x}} \right\} \eqend{,} \\
\pastcone &= \left\{ (t,\vec{x})\colon t < \tau, t-\tau > r \right\} \eqend{.}
\end{equations}
The algebras $\mathfrak{A}_\futurecone$ and $\mathfrak{A}_\pastcone$ that we consider are generated by the identity $\1$ and smeared operators $\op(f)$ with $\supp f \subset \futurecone$ or $\supp f \subset \pastcone$, respectively, and the cyclic and separating vector $\Omega$ is the Minkowski vacuum.

In accordance with the principle of geometric modular action, for the modular Hamiltonian $\ln \Delta$ we thus make the ansatz
\begin{equation}
\ln \Delta_\futurecone = Q_{a_\futurecone b_\futurecone 0} = a_\futurecone H + 2 b_\futurecone D \eqend{,} \quad \ln \Delta_\pastcone = Q_{a_\pastcone b_\pastcone 0} = a_\pastcone H + 2 b_\pastcone D
\end{equation}
with $Q$ defined in Eq.~\eqref{eq:minkowski_q_def}, a linear combination of the dilation generator $D$ and the generator of time translations $H$. The modular automorphism is then given by
\begin{equation}
\label{eq:minkowski_modular_light_cone_automorphism_pre}
\Delta_\futurecone^{\mathi s} \op(f) \Delta_\futurecone^{- \mathi s} = \op(f_{s,\futurecone}) \eqend{,} \quad \Delta_\pastcone^{\mathi s} \op(f) \Delta_\pastcone^{- \mathi s} = \op(f_{s,\pastcone})
\end{equation}
with
\begin{equations}[eq:minkowski_modular_light_cone_fs]
f_{s,\futurecone}(x) &= \mathe^{- 2 (d-\Delta) b_\futurecone s} f(x_{s,\futurecone}) \eqend{,} \\
\vec{x}_{s,\futurecone} &= \mathe^{- 2 b_\futurecone s} \vec{x} \eqend{,} \\
t_{s,\futurecone} &= - \frac{a_\futurecone}{2 b_\futurecone} + \mathe^{- 2 b_\futurecone s} \left( t + \frac{a_\futurecone}{2 b_\futurecone} \right)
\end{equations}
and the analogous result for the past light cone $\pastcone$, where we used the adjoint action~\eqref{eq:minkowski_ulambda_op} and the solution for $f_\lambda$ given by Eqs.~\eqref{eq:minkowski_flambda_def} and~\eqref{eq:minkowski_flambda_sol}, taking into account that the angles do not change under the flow. To determine the constants $a_\futurecone$ and $b_\futurecone$, we first impose that as $s \to -\infty$ the modular flow reaches the tip of the light cone at $(t,\vec{x}) = (\tau,\vec{0})$. This gives $a_\futurecone = - 2 b_\futurecone \tau$ and $b_\futurecone > 0$, since then we have
\begin{equation}
\lim_{s \to -\infty} f_{s,\futurecone}(x) = \lim_{s \to -\infty} \mathe^{- 2 (d-\Delta) b_\futurecone s} f\left[ \tau + \mathe^{- 2 b_\futurecone s} (t-\tau), \mathe^{- 2 b_\futurecone s} \vec{x} \right] = 0 \quad\text{for}\quad (t,\vec{x}) \neq (\tau,\vec{0}) \eqend{.}
\end{equation}
This becomes clearer if we take the formal limit where $f(x) \to \delta^d(x-x_0)$ for some $x_0 \in \futurecone$. In this limit, we have $\op(f) \to \op(x_0)$ and
\begin{equation}
\Delta_\futurecone^{\mathi s} \op(f) \Delta_\futurecone^{- \mathi s} \to \mathe^{2 \Delta b_\futurecone s} \op\left( \tau + \mathe^{2 b_\futurecone s} (t_0-\tau), \mathe^{2 b_\futurecone s} \vec{x}_0 \right) \eqend{.}
\end{equation}
So in the limit $s \to -\infty$, if $b_\futurecone > 0$ the operator gets pushed to the tip of the cone $(\tau,\vec{0})$ regardless of its initial position $(t_0,\vec{x}_0)$.

To fully determine $b_\futurecone$, we then impose that the modular automorphism~\eqref{eq:minkowski_modular_light_cone_automorphism_pre} can be analytically continued to $s = - \mathi/2$, and after this continuation maps $\op(f)$ into an element of the commutant $\mathfrak{A}'$. The analytic continuation is easily obtained from Eqs.~\eqref{eq:minkowski_modular_light_cone_fs}, and reads
\begin{equation}
\label{eq:minkowski_modular_light_cone_deltahalf_pre}
\Delta_\futurecone^\frac{1}{2} \op(f) \Delta_\futurecone^{- \frac{1}{2}} = \mathe^{\mathi (d-\Delta) b_\futurecone} \op\left( \underline{f} \right) \eqend{,} \quad \underline{f}(x) = f\left( x_{- \frac{\mathi}{2},\futurecone} \right)
\end{equation}
with
\begin{equation}
\label{eq:minkowski_modular_light_cone_deltahalf_coords}
\vec{x}_{- \frac{\mathi}{2},\futurecone} = \mathe^{\mathi b_\futurecone} \vec{x} \eqend{,} \quad t_{- \frac{\mathi}{2},\futurecone} = \tau + \mathe^{\mathi b_\futurecone} (t-\tau) \eqend{.}
\end{equation}
If Huygen's principle is valid, a massless field propagates on the light cone. In this case, all elements in the algebra $\mathfrak{A}_\futurecone$ commute with all elements in $\mathfrak{A}_\pastcone$, such that $\mathfrak{A}_\futurecone' \supset \mathfrak{A}_\pastcone$ and vice versa. We thus obtain the required mapping into the commutant by choosing $b_\futurecone$ such that $\mathe^{\mathi b_\futurecone} = -1$, since then $f(x_{- \frac{\mathi}{2},\futurecone})$ has support in the past light cone if $f$ is supported in the future light cone. The smallest $b_\futurecone$ for which this holds is $b_\futurecone = \pi$. We thus have the candidate
\begin{equation}
\label{eq:minkowski_modular_light_cone_lndelta}
\ln \Delta_\futurecone = 2 \pi ( D - \tau H )
\end{equation}
for the modular Hamiltonian of the future light cone with tip at $(t,\vec{x}) = (\tau,\vec{0})$, assuming Huygen's principle, and the modular automorphism
\begin{equation}
\label{eq:minkowski_modular_light_cone_automorphism}
\Delta_\futurecone^{\mathi s} \op(f) \Delta_\futurecone^{- \mathi s} = \op(f_{s,\futurecone}) \quad\text{with}\quad f_{s,\futurecone}(t,\vec{x}) = \mathe^{- 2 \pi s (d-\Delta)} f\left( \tau + \mathe^{- 2 \pi s} ( t - \tau ), \mathe^{- 2 \pi s} \vec{x} \right) \eqend{.}
\end{equation}

To verify Huygen's principle, we compute the commutator of two operators and check that it is non-vanishing only for lightlike separations. From the two-point function~\eqref{eq:minkowski_wightman}, we compute the expectation value of the commutator, and simplify it by interchanging the integration variables $x \leftrightarrow y$ in one term and bringing both terms onto a common denominator. Shifting then $x \to x+y$, we obtain
\begin{equation}
( \Omega, [ \op(f), \op(g) ] \Omega ) = C_\op \lim_{\epsilon \to 0^+} \iint f(x+y) \frac{\mathe^{\mathi \Delta \arg[ r^2 - (t + \mathi \epsilon)^2 ]} - \mathe^{\mathi \Delta \arg[ r^2 - (t - \mathi \epsilon)^2 ]}}{\left[ (r-t)^2 + \epsilon^2 \right]^\frac{\Delta}{2} \left[ (r+t)^2 + \epsilon^2 \right]^\frac{\Delta}{2}} g(y) \total^d x \total^d y \eqend{.}
\end{equation}
If the supports of $f$ and $g$ are spacelike related, the integral over $x$ has support only for $r^2 > t^2$. We can then take the limit $\epsilon \to 0$ inside the integral, and obtain $\lim_{\epsilon \to 0} \arg[ r^2 - (t \pm \mathi \epsilon)^2 ] = 0$, such that the commutator vanishes for spacelike separations as it must for a relativistic field theory. On the other hand, if the supports of $f$ and $g$ are timelike related, the integral over $x$ has support only for $r^2 < t^2$. We can again take the limit $\epsilon \to 0$ inside the integral, but now obtain $\lim_{\epsilon \to 0^+} \arg[ r^2 - (t \pm \mathi \epsilon)^2 ] = \mp \sgn(t) \pi$ and
\begin{equation}
\lim_{\epsilon \to 0^+} \left[ \mathe^{\mathi \Delta \arg[ r^2 - (t + \mathi \epsilon)^2 ]} - \mathe^{\mathi \Delta \arg[ r^2 - (t - \mathi \epsilon)^2 ]} \right] = - 2 \mathi \sgn(t) \sin(\pi \Delta) \eqend{.}
\end{equation}
For generic $f$ and $g$, this is non-vanishing only if $\Delta \in \mathbb{N}$, which for example holds for the free scalar field in even dimensions $d = 2k$ such that $\Delta = k-1$.

For a free theory, the commutator of two fields $\phi$ is proportional to the identity operator, and thus equal to its expectation value. In this case, the only condition that we need to impose to ensure Huygen's principle is thus that we are in even dimensions. In a more general theory, we cannot exclude that more restrictions need to be made. However, for other reasonable examples such as $\op = \normord{ \phi^2 }$ in the free massless scalar theory, one easily checks that the commutator $[ \op(f), \op(g) ]$ vanishes for $f$ and $g$ with timelike related support, since its computation ultimately leads to powers of the free-field commutator $[ \phi(f), \phi(g) ] = 0$. In the following, we thus assume that only the condition $\Delta \in \mathbb{N}$ is needed.

\begin{figure}
\includegraphics[]{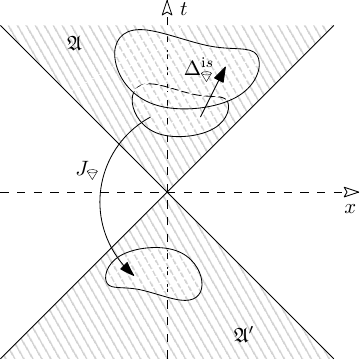}
\caption{Modular data for the future light cone in Minkowski spacetime. Shown are the action of the modular automorphism induced by the modular operator $\Delta_\futurecone$, mapping the algebra $\mathfrak{A}$ of fields localized in the future light cone into itself, and the action of the modular conjugation $J_\futurecone$, mapping $\mathfrak{A}$ into its commutant $\mathfrak{A}'$.}
\label{fig:lightcone1}
\end{figure}

Therefore, for scaling dimension $\Delta \in \mathbb{N}$ the modular Hamiltonian of the future light cone with tip at $(t,\vec{x}) = (\tau,\vec{0})$ is given by Eq.~\eqref{eq:minkowski_modular_light_cone_lndelta}, which is $\ln \Delta_\futurecone = 2 \pi ( D - \tau H )$. The analytic continuation $s \to - \mathi/2$ of the modular automorphism is given in Eqs.~\eqref{eq:minkowski_modular_light_cone_deltahalf_pre} and~\eqref{eq:minkowski_modular_light_cone_deltahalf_coords} with $b_\futurecone = \pi$, and reads
\begin{equation}
\label{eq:minkowski_modular_light_cone_deltahalf}
\Delta_\futurecone^\frac{1}{2} \op(f) \Delta_\futurecone^{- \frac{1}{2}} = \op\left( \underline{f} \right) \eqend{,} \quad \underline{f}(t,\vec{x}) = (-1)^{d-\Delta} f(2\tau-t,-\vec{x}) \eqend{,}
\end{equation}
mapping a test function $f$ with support in the future light cone into the past light cone with the same tip. It only remains to determine the modular conjugation $J_\futurecone$. Because $J_\futurecone$ is an antilinear operator, and the only antilinear operator that we have at our disposal is time reversal $T$~\eqref{eq:minkowski_timereversal}, it will certainly contain $T$. However, since $T$ changes the sign of time but we need the reflection $t \to 2 \tau - t$, it must also contain a time translation through the adjoint action~\eqref{eq:minkowski_ulambda_op} with the generator $Q$~\eqref{eq:minkowski_q_def} proportional to $H$, taking $a = 1$ and $b = c = 0$. The reflection of the spatial coordinates is done using the parity operator $P$~\eqref{eq:minkowski_parity}, and we also need to compensate for the phase factor $(-1)^{d-\Delta}$. For this, we define an operator $W$ that acts on the dense subspace $\mathcal{D}$ of Fock space via
\begin{equation}
\label{eq:minkowski_modular_w_def}
W \op(f) W^{-1} = (-1)^{\Delta-d} \op(f) \eqend{,} \quad W \Omega = \Omega \eqend{,}
\end{equation}
is extended to the full Fock space by continuity, and clearly satisfies $W^2 = \1$ and $W^\dagger = W = W^{-1}$. It is then easy to verify that the definition
\begin{equation}
\label{eq:minkowski_modular_light_cone_j_def}
J_\futurecone \equiv U(\tau) C P T W U(-\tau)
\end{equation}
results in
\begin{equation}
\label{eq:minkowski_modular_light_cone_j}
J_\futurecone \op(f) J_\futurecone^{-1} = \op^\dagger(f_J) \eqend{,} \quad f_J(t,\vec{x}) \equiv (-1)^{\Delta-d} f^*(2\tau-t, -\vec{x}) \eqend{,}
\end{equation}
and we also included charge conjugation $C$~\eqref{eq:minkowski_chargeconjugation_op} in the general case. From the explicit expression~\eqref{eq:minkowski_modular_light_cone_j} for $f_J$, it is clear that $J_\futurecone^2 = \1$ as required, and that $J_\futurecone$ maps $\mathfrak{A}_\futurecone$ into its commutant $\mathfrak{A}_\futurecone' \supset \mathfrak{A}_\pastcone$. The action of the modular automorphism and the modular conjugation are depicted in Fig.~\ref{fig:lightcone1}.


From $\Delta_\futurecone$ and $J_\futurecone$, we can now assemble the Tomita operator $S_\futurecone = J_\futurecone \Delta_\futurecone^\frac{1}{2}$ with action
\begin{equation}
S_\futurecone \op(f) S_\futurecone^{-1} = \op^\dagger(f^*)
\end{equation}
following from the actions~\eqref{eq:minkowski_modular_light_cone_deltahalf} and~\eqref{eq:minkowski_modular_light_cone_j}. Since both the modular Hamiltonian and the modular conjugation leave the vacuum invariant, $\Delta_\futurecone^{-\frac{1}{2}} \Omega = \Omega$ and $J_\futurecone \Omega = \Omega$, it follows that $S_\futurecone a \Omega = a^\dagger \Omega$ for all $a \Omega$ in the dense set $\mathcal{D}$ whose support lies in the future light cone, i.e. for all $a \in \mathfrak{A}_\futurecone$ as required. For the past light cone with tip at $(t,\vec{x}) = (\tau,\vec{0})$, the analogous computation establishes that $a_\pastcone = 2 \pi \tau$ and $b_\pastcone = - \pi$, and hence
\begin{equation}
\label{eq:minkowski_modular_light_cone_past_lndelta_j}
\ln \Delta_\pastcone = 2 \pi ( - D + \tau H ) \eqend{,} \quad J_\pastcone = U(\tau) C P T W U(-\tau)
\end{equation}
and the modular automorphism
\begin{equation}
\label{eq:minkowski_modular_light_cone_past_automorphism}
\Delta_\pastcone^{\mathi s} \op(f) \Delta_\pastcone^{- \mathi s} = \op(f_{s,\pastcone}) \quad\text{with}\quad f_{s,\pastcone}(t,\vec{x}) = \mathe^{2 \pi s (d-\Delta)} f\left( \tau + \mathe^{2 \pi s} ( t - \tau ), \mathe^{2 \pi s} \vec{x} \right) \eqend{.}
\end{equation}

Lastly, we show that the vacuum state $\Omega$ is a KMS state with respect to the modular flow, i.e., we verify the KMS condition~\eqref{eq:intro_kms}. Using the two-point function~\eqref{eq:minkowski_wightman} and the modular automorphism~\eqref{eq:minkowski_modular_light_cone_automorphism}, we compute
\begin{splitequation}
\label{eq:minkowski_kms_pre}
&\left( \Omega, \sigma^\futurecone_s\left[ \op(f) \right] \op(g) \Omega \right) = \left( \Omega, \Delta_\futurecone^{\mathi s} \op(f) \Delta_\futurecone^{- \mathi s} \op(g) \Omega \right) = \left( \Omega, \op(f_{s,\futurecone}) \op(g) \Omega \right) \\
&\quad= C_\op \lim_{\epsilon \to 0^+} \iint \frac{f_{s,\futurecone}(x) g(y)}{[ - (x^0-y^0 - \mathi \epsilon)^2 + (\vec{x}-\vec{y})^2 ]^\Delta} \total^d x \total^d y \\
&\quad= C_\op \, \mathe^{- 2 \pi s (d-\Delta)} \lim_{\epsilon \to 0^+} \iint \frac{f\left( \tau + \mathe^{- 2 \pi s} ( x^0 - \tau ), \mathe^{- 2 \pi s} \vec{x} \right) g(y)}{[ - (x^0-y^0 - \mathi \epsilon)^2 + (\vec{x}-\vec{y})^2 ]^\Delta} \total^d x \total^d y \\
&\quad= C_\op \, \mathe^{2 \pi s \Delta} \lim_{\epsilon \to 0^+} \iint \frac{f(x^0+\tau,\vec{x}) g(y^0+\tau,\vec{y})}{\left[ - \left( \mathe^{2 \pi s} x^0 - y^0 - \mathi \epsilon \right)^2 + \left( \mathe^{2 \pi s} \vec{x} - \vec{y} \right)^2 \right]^\Delta} \total^d x \total^d y \\
&\quad= C_\op \lim_{\epsilon \to 0^+} \iint \frac{f(x^0+\tau,\vec{x}) g(y^0+\tau,\vec{y})}{\left[ - \left( \mathe^{\pi s} x^0 - \mathe^{- \pi s} y^0 - \mathi \epsilon \right)^2 + \left( \mathe^{\pi s} \vec{x} - \mathe^{- \pi s} \vec{y} \right)^2 \right]^\Delta} \total^d x \total^d y \eqend{,}
\end{splitequation}
where we changed variables $x^0 \to \mathe^{2 \pi s} x^0 + \tau$, $\vec{x} \to \mathe^{2 \pi s} \vec{x}$, $y^0 \to y^0 + \tau$ to arrive at the fourth line, and all manipulations were valid for real $s$. However, in the last form we can now easily perform the analytic continuation $s \to s - \mathi$, whose effect is to change the sign of $\epsilon$. Note that this continuation is valid because the denominator $- \left( \mathe^{\pi s} x^0 - \mathe^{- \pi s} y^0 - \mathi \epsilon \right)^2 + \left( \mathe^{\pi s} \vec{x} - \mathe^{- \pi s} \vec{y} \right)^2 \neq 0$ for all $\Im s \in [-1,0]$, such that no singularities are encountered. To see this explicitly, we use the Fourier transform~\eqref{eq:minkowski_wightman_fourier}, which gives (dropping an overall constant)
\begin{splitequation}
\label{eq:minkowski_kms_finite}
&\lim_{\epsilon \to 0^+} \frac{1}{\left[ - \left( \mathe^{\pi s} x^0 - \mathe^{- \pi s} y^0 - \mathi \epsilon \right)^2 + \left( \mathe^{\pi s} \vec{x} - \mathe^{- \pi s} \vec{y} \right)^2 \right]^\Delta} \\
&\quad\sim \int \Theta(p^0) \Theta(-p^2) (-p^2)^{\Delta - \frac{d}{2}} \mathe^{- \mathi p^0 \left( \mathe^{\pi s} x^0 - \mathe^{- \pi s} y^0 \right) + \mathi \vec{p} \left( \mathe^{\pi s} \vec{x} - \mathe^{- \pi s} \vec{y} \right)} \frac{\total^{d-1} \vec{p}}{(2\pi)^{d-1}} \eqend{.}
\end{splitequation}
The exponential can be rewritten in the form
\begin{equation}
\label{eq:minkowski_kms_finite2}
\mathe^{\mathi \cos(\pi \, \Im s) \left( \mathe^{\pi \Re s} p x - \mathe^{- \pi \Re s} p y \right)} \, \mathe^{- \sin(\pi \, \Im s) \left( \mathe^{\pi \Re s} p x + \mathe^{- \pi \Re s} p y \right)} \eqend{,}
\end{equation}
and while the first term is bounded, the second one can potentially blow up and lead to a divergence. However, since both $p$ and $x$ lie in the closed forward light cone such that $p^2 \leq 0$, $p^0 \geq 0$ (from the explicit Heaviside $\Theta$ functions) and $x^2 \leq 0$, $x^0 \geq 0$ (since the support of $f$ is restricted to this region), we have the estimate
\begin{equation}
p x = - p^0 x^0 + \vec{p} \vec{x} \leq - p^0 x^0 + \abs{\vec{p}} \abs{\vec{x}} \leq - p^0 x^0 + \abs{p^0} \abs{x^0} = 0 \eqend{,}
\end{equation}
and the analogous one $p y \leq 0$ for $y$. It follows that the exponential~\eqref{eq:minkowski_kms_finite2} and hence the integrand of the Fourier transform~\eqref{eq:minkowski_kms_finite} is bounded if $\sin(\pi \, \Im s) \leq 0$, which gives exactly the condition $\Im s \in [-1,0]$. We thus obtain
\begin{splitequation}
\label{eq:minkowski_kms_final}
&\left( \Omega, \sigma^\futurecone_{s - \mathi}\left[ \op(f) \right] \op(g) \Omega \right) \\
&\quad= C_\op \lim_{\epsilon \to 0^+} \iint \frac{f(x^0+\tau,\vec{x}) g(y^0+\tau,\vec{y})}{\left[ - \left( \mathe^{\pi s} x^0 - \mathe^{- \pi s} y^0 + \mathi \epsilon \right)^2 + \left( \mathe^{\pi s} \vec{x} - \mathe^{- \pi s} \vec{y} \right)^2 \right]^\Delta} \total^d x \total^d y \\
&\quad= C_\op \, \mathe^{2 \pi s \Delta} \lim_{\epsilon \to 0^+} \iint \frac{f(y^0+\tau,\vec{y}) g(x^0+\tau,\vec{x})}{\left[ - \left( \mathe^{2 \pi s} y^0 - x^0 + \mathi \epsilon \right)^2 + \left( \mathe^{2 \pi s} \vec{y} - \vec{x} \right)^2 \right]^\Delta} \total^d x \total^d y \\
&\quad= C_\op \, \mathe^{2 \pi s (\Delta-d)} \lim_{\epsilon \to 0^+} \iint \frac{f\left( \mathe^{- 2 \pi s} (y^0 - \tau) + \tau, \mathe^{- 2 \pi s} \vec{y} \right) g(x)}{\left[ - \left( y^0 - x^0 + \mathi \epsilon \right)^2 + \left( \vec{y} - \vec{x} \right)^2 \right]^\Delta} \total^d x \total^d y \\
&\quad= C_\op \, \lim_{\epsilon \to 0^+} \iint \frac{f_{s,\futurecone}(y) g(x)}{\left[ - \left( x^0 - y^0 - \mathi \epsilon \right)^2 + \left( \vec{x} - \vec{y} \right)^2 \right]^\Delta} \total^d x \total^d y \\
&\quad= \left( \Omega, \op(g) \op\left( f_{s,\futurecone} \right) \Omega \right) = \left( \Omega, \op(g) \Delta_\futurecone^{\mathi s} \op(f) \Delta_\futurecone^{- \mathi s} \Omega \right) = \left( \Omega, \op(g) \sigma^\futurecone_s\left[ \op(f) \right] \Omega \right) \eqend{,}
\end{splitequation}
where we exchanged the integration variables $x \leftrightarrow y$ in the second equality and changed variables $y^0 \to \mathe^{- 2 \pi s} (y^0 - \tau)$, $\vec{y} \to \mathe^{- 2 \pi s} \vec{y}$, $x^0 \to x^0 - \tau$ in the third equality. This is exactly the KMS condition~\eqref{eq:intro_kms}, and its verification for the past light cone $\pastcone$ with the modular automorphism~\eqref{eq:minkowski_modular_light_cone_past_automorphism} proceeds analogously.

\subsection{Modular data for double cones}
\label{sec:minkowski_modular_double_cone}

For a double cone or diamond with center at the origin, Hislop and Longo~\cite{hisloplongo1982,hislop1988} have determined the modular automorphism and the modular conjugation, using a conformal mapping from the future light cone to the double cone. Martinetti and Rovelli~\cite{martinettirovelli2003} already remark that the modular flow in the double cone is a conformal transformation, and Casini and Huerta~\cite{casini2008,casinihuerta2011} assert that the modular Hamiltonian is proportional to the generator of conformal transformations leaving the diamond invariant. Casini, Huerta and Myers~\cite{casinihuertamyers2011} and later on Longo and Morsella~\cite{longomorsella2020} have then shown that the generator, the modular Hamiltonian, can be written as an integral over the (improved) stress tensor (which is implicit in~\cite{casinihuerta2011}), and de~Boer, Haehl, Heller and Myers~\cite{deboerhaehlhellermyers2016} have given an explicit link between the modular Hamiltonian and the generator of conformal transformations along the conformal Killing vector that leaves the diamond invariant, together with an explicit expression for this Killing vector.

\begin{figure}
\begin{minipage}[t]{0.48\textwidth}
\centering
\includegraphics[]{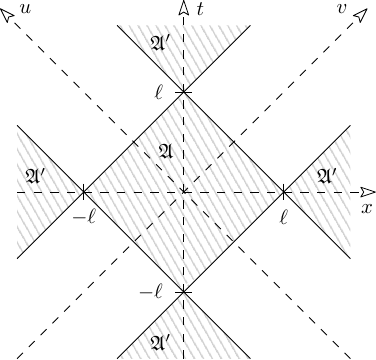}
\caption{The double cone of size $\ell$ in Minkowski spacetime. Shown is a cross section in the $t$-$x$ plane, the algebra $\mathfrak{A}$ of fields localized in the double cone and its commutant $\mathfrak{A}'$. In contrast to the Rindler wedge and the light cones, for the double cone $\mathfrak{A}'$ is localized in three disconnected regions, two timelike and one spacelike separated one (since the whole picture is rotationally symmetric around the time axis.)}
\label{fig:diamond1}
\end{minipage}
\hfill
\begin{minipage}[t]{0.48\textwidth}
\centering
\includegraphics[]{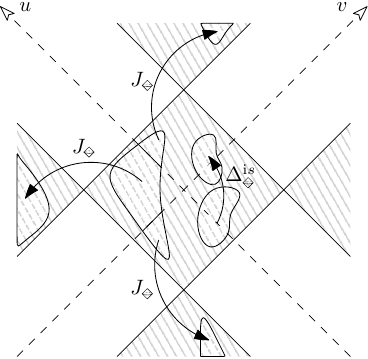}
\caption{Modular data for the double cone in Minkowski spacetime. Shown are the action of the modular automorphism induced by the modular operator $\Delta_\doublecone$, mapping the algebra $\mathfrak{A}$ of fields localized in the double cone into itself, and the action of the modular conjugation $J_\doublecone$, mapping $\mathfrak{A}$ into its commutant $\mathfrak{A}'$.}
\label{fig:diamond2}
\end{minipage}
\end{figure}

Here, we want to stress this connection, which is another instance of the principle of geometric modular action. We further generalize these results by considering double cones of size $\ell$ with center at $t = \tau \neq 0$, which are the regions
\begin{equation}
\label{eq:minkowski_doublecone}
\doublecone = \left\{ (t,\vec{x})\colon r = \abs{\vec{x}} \in [0, \ell), t \in (\tau - \ell + r, \tau + \ell - r) \right\} \eqend{.}
\end{equation}
The algebra $\mathfrak{A}_\doublecone$ that we consider is then generated by the identity $\1$ and smeared operators $\op(f)$ with $\supp f \subset \doublecone$, and the cyclic and separating vector $\Omega$ is the Minkowski vacuum. The double cone and the algebra $\mathfrak{A}_\doublecone$ are depicted in Fig.~\ref{fig:diamond1}.

In accordance with the principle of geometric modular action, for the modular Hamiltonian $\ln \Delta$ we thus make the ansatz
\begin{equation}
\ln \Delta_\doublecone = Q_{a_\doublecone b_\doublecone c_\doublecone} = a_\doublecone H + 2 b_\doublecone D + c_\doublecone K
\end{equation}
with $Q$ defined in Eq.~\eqref{eq:minkowski_q_def}, a linear combination of all three generators: dilations $D$, time translations $H$ and special conformal transformations $K$. The modular automorphism reads
\begin{equation}
\label{eq:minkowski_modular_double_cone_automorphism}
\Delta_\doublecone^{\mathi s} \op(f) \Delta_\doublecone^{- \mathi s} = \op(f_{s,\doublecone})
\end{equation}
with the solution for $f_\lambda$ given by Eqs.~\eqref{eq:minkowski_flambda_def} and~\eqref{eq:minkowski_flambda_sol}. To make this case tractable, and since the angles do not change under the flow, we may rotate the coordinate system such that the flow is contained in the $t$-$x$ plane. We then introduce null coordinates $u = t-\tau-x$ and $v = t-\tau+x$, such that the intersection of the $t$-$x$ plane with the double cone $\doublecone$~\eqref{eq:minkowski_doublecone} is the region $\left\{ (u,v) \colon \abs{u} < \ell, \abs{v} < \ell \right\}$, as can be seen in Fig.~\ref{fig:diamond1}. In terms of $u$ and $v$, the function $f_{s,\doublecone}$ appearing in the modular automorphism~\eqref{eq:minkowski_modular_double_cone_automorphism} reads
\begin{equation}
\label{eq:minkowski_modular_double_cone_fs}
f_{s,\doublecone}(u,v,\vec{x}^\perp) = \left( \frac{v_s-u_s}{v-u} \right)^{d-\Delta} f(u_s,v_s,\vec{x}^\perp) \eqend{,}
\end{equation}
where $\vec{x}^\perp$ denotes the remaining spatial coordinates, and where
\begin{equations}[eq:minkowski_modular_double_cone_fs_coords]
u_s &= \frac{b_\doublecone - c_\doublecone \tau}{c_\doublecone} + \frac{\omega}{c_\doublecone} \frac{2 c_\doublecone \omega (u-v) - k_- \, \mathe^{2 \omega s} + k_+ \, \mathe^{- 2 \omega s}}{2 \omega^2 - 2 k_0 + k_- \, \mathe^{2 \omega s} + k_+ \, \mathe^{- 2 \omega s}} \eqend{,} \\
v_s &= \frac{b_\doublecone - c_\doublecone \tau}{c_\doublecone} + \frac{\omega}{c_\doublecone} \frac{2 c_\doublecone \omega (v-u) - k_- \, \mathe^{2 \omega s} + k_+ \, \mathe^{- 2 \omega s}}{2 \omega^2 - 2 k_0 + k_- \, \mathe^{2 \omega s} + k_+ \, \mathe^{- 2 \omega s}} \eqend{,} \\
k_\alpha &\equiv (-b_\doublecone + c_\doublecone \tau + c_\doublecone u + \alpha \omega) (-b_\doublecone + c_\doublecone \tau + c_\doublecone v + \alpha \omega)
\end{equations}
with $\omega = \sqrt{b_\doublecone^2 + a_\doublecone c_\doublecone}$.

To determine the constants $b_\doublecone$ and $c_\doublecone$, we impose that the flow at early and late flow times reaches the lower and upper tips of the diamond, respectively. Since the only dependence on $s$ is through the exponentials, as $s \to \pm \infty$ we obtain $u_s \to u_{\pm \infty} = (b_\doublecone - c_\doublecone \tau)/c_\doublecone \mp \omega/c_\doublecone$ and $v_s \to v_{\pm \infty} = u_{\pm \infty}$. At early (late) flow times, the transformed function $f_{s,\doublecone}$~\eqref{eq:minkowski_modular_double_cone_fs} should vanish everywhere except at the lower (upper) tip, that is\footnote{To avoid confusion, we note that $u_s-v_s$ is proportional to $u-v$, such that the prefactor is finite also in the limit $u \to v$.}
\begin{equation}
\lim_{s \to \pm\infty} f_{s,\doublecone}(u,v,\vec{x}^\perp) = \lim_{s \to \pm\infty} \left( \frac{v_s-u_s}{v-u} \right)^{d-\Delta} f(u_s,v_s,\vec{x}^\perp) = 0 \quad\text{for}\quad (u,v) \neq (\pm\ell, \pm\ell) \eqend{,}
\end{equation}
which leads to $b_\doublecone = c_\doublecone \tau$ and $c_\doublecone = \omega/\ell$. This becomes again clearer if we take the formal limit where $f(x) \to \delta^d(x-x_0) = 2 \delta(u-u_0) \delta(v-v_0) \delta^{d-2}(\vec{x}^\perp-\vec{x}_0^\perp)$ for some $x_0 \in \doublecone$. In this limit, we have $\op(f) \to \op(x_0)$ and obtain
\begin{equation}
\Delta_\doublecone^{\mathi s} \op(f) \Delta_\doublecone^{- \mathi s} \to B_s(u_0,v_0) \, \op\left( u_{-s}(u_0), v_{-s}(v_0), \vec{x}_0^\perp \right)
\end{equation}
with some function $B_s(u_0,v_0)$ that could be determined in a lengthy but straightforward computation. So in the limit $s \to -\infty$ the operator gets pushed to the point $( u_\infty, v_\infty ) = (-\ell, -\ell)$ (the lower tip of the double cone) regardless of its initial position $(u_0,v_0)$, and analogously in the limit $s \to \infty$ the operator gets pushed to the upper tip.

To also determine $a_\doublecone$, we impose that the modular automorphism~\eqref{eq:minkowski_modular_double_cone_automorphism} can be analytically continued to $s = - \mathi/2$, and after this continuation maps $\op(f)$ into an element of the commutant $\mathfrak{A}'$. The analytic continuation is easily obtained from Eqs.~\eqref{eq:minkowski_modular_double_cone_fs} and~\eqref{eq:minkowski_modular_double_cone_fs_coords}, and reads
\begin{splitequation}
\label{eq:minkowski_modular_double_cone_deltahalf_pre}
\Delta_\doublecone^\frac{1}{2} \op(f) \Delta_\doublecone^{- \frac{1}{2}} &= \op\left( \underline{f} \right) \eqend{,} \\
\underline{f}(u,v,\vec{x}^\perp) &= \left( \frac{v_{- \frac{\mathi}{2}} - u_{- \frac{\mathi}{2}}}{v-u} \right)^{d-\Delta} f\left( u_{- \frac{\mathi}{2}},v_{- \frac{\mathi}{2}},\vec{x}^\perp \right)
\end{splitequation}
with
\begin{equations}
u_{- \frac{\mathi}{2}} &= \ell \frac{(\ell+u) \mathe^{\mathi \omega} - (\ell-u)}{(\ell+u) \mathe^{\mathi \omega} + (\ell-u)} \eqend{,} \\
v_{- \frac{\mathi}{2}} &= \ell \frac{(\ell+v) \mathe^{\mathi \omega} - (\ell-v)}{(\ell+v) \mathe^{\mathi \omega} + (\ell-v)} \eqend{,} \\
\omega &= \frac{a_\doublecone \ell}{\ell^2 - \tau^2} \eqend{.}
\end{equations}
Certainly both $u_{- \frac{\mathi}{2}}$ and $v_{- \frac{\mathi}{2}}$ must be real if the analytically continued modular automorphism maps $\op(f)$ into an element of the commutant. Imposing that their imaginary part vanishes for any $(u,v) \in \doublecone$, we obtain the condition $\sin \omega = 0$ whose smallest non-zero solution is $\omega = \pi$, and it follows that $a_\doublecone = \pi ( \ell - \tau^2/\ell )$. With this choice, the denominator of the transformed coordinates~\eqref{eq:minkowski_modular_double_cone_fs_coords} vanishes for
\begin{equation}
2 \omega s = \ln\left( \frac{\ell + u}{\ell - u} \right) + \mathi \pi (2k+1) \eqend{,} \quad 2 \omega s = \ln\left( \frac{\ell + v}{\ell - v} \right) + \mathi \pi (2k+1) \eqend{,} \quad k \in \mathbb{Z} \eqend{.}
\end{equation}
Since inside the double cone $\ell \pm u > 0$ and $\ell \pm v > 0$, the argument of the logarithm is always positive, and we can analytically continue from $s = 0$ to $s = - \mathi \pi/(2\omega)$ without encountering a singularity. For $\omega = \pi$, this gives exactly the continuation to $s = - \mathi/2$ that is needed.

Where does $\underline{f}$ have support? With $\omega = \pi$, we have $u_{- \mathi/2} = \ell^2/u$ and $v_{- \mathi/2} = \ell^2/v$, an inversion of the null coordinates in the plane of the flow. Since inside the double cone we have $\abs{u}, \abs{v} < \ell$, the map leads outside the diamond, but a point is mapped into three disconnected regions depending on where it lies inside the diamond. For $u > 0$ and $v > 0$, the upper quadrant of the double cone $\doublecone$, the point $x^\mu_{- \mathi/2}$ lies inside the future light cone $\futurecone$ with tip at $(t,\vec{x}) = (\ell,\vec{0})$, the upper tip of the double cone, and is thus timelike separated from $x^\mu$. Similarly, for $u < 0$ and $v < 0$ (the lower quadrant of $\doublecone$), the point $x_{- \mathi/2}$ lies inside the past light cone $\pastcone$ with tip at $(t,\vec{x}) = (-\ell,\vec{0})$, the lower tip of the double cone, and is again timelike separated from $x^\mu$. Note that both regions $u,v > 0$ and $u,v < 0$ are themselves double cones. For $u v < 0$ (the remainder of the couble cone, topologically a torus), the point $x^\mu_{- \mathi/2}$ lies instead in the region spacelike separated from the double cone. Assuming again Huygen's principle and thus the condition $\Delta \in \mathbb{N}$, we thus obtain the required mapping into the commutant algebra $\mathfrak{A}'$. The action of the modular automorphism and the mapping into the commutant are shown in Fig.~\ref{fig:diamond2}.

Summarizing, we have the candidate
\begin{equation}
\label{eq:minkowski_modular_double_cone_lndelta}
\ln \Delta_\doublecone = \frac{\pi}{\ell} \left[ ( \ell^2 - \tau^2 ) H + 2 \tau D + K \right]
\end{equation}
for the modular Hamiltonian of the double cone with center at $(t,\vec{x}) = (\tau,\vec{0})$, and the modular automorphism~\eqref{eq:minkowski_modular_double_cone_automorphism}, \eqref{eq:minkowski_modular_double_cone_fs}
\begin{equation}
\label{eq:minkowski_modular_double_cone_automorphism2}
\Delta_\doublecone^{\mathi s} \op(f) \Delta_\doublecone^{- \mathi s} = \op(f_{s,\doublecone}) \eqend{,} \quad f_{s,\doublecone}(u,v,\vec{x}^\perp) = \left( \frac{v_s-u_s}{v-u} \right)^{d-\Delta} f(u_s,v_s,\vec{x}^\perp)
\end{equation}
with the transformed null coordinates~\eqref{eq:minkowski_modular_double_cone_fs_coords}
\begin{equation}
\label{eq:minkowski_modular_double_cone_fs_coords1}
u_s = \ell \frac{(\ell+u) \, \mathe^{- \pi s} - (\ell-u) \, \mathe^{\pi s}}{(\ell+u) \, \mathe^{- \pi s} + (\ell-u) \, \mathe^{\pi s}} \eqend{,} \quad v_s = \ell \frac{\ell + v - (\ell-v) \, \mathe^{2 \pi s}}{\ell + v + (\ell-v) \, \mathe^{2 \pi s}} \eqend{.}
\end{equation}
Going back to Cartesian coordinates $t$ and $\vec{x}$, this reads
\begin{equation}
\label{eq:minkowski_modular_double_cone_fs2}
f_{s,\doublecone}(t,\vec{x}) = \left( \frac{\vec{x}_s^2}{\vec{x}^2} \right)^\frac{d-\Delta}{2} f(t_s,\vec{x}_s)
\end{equation}
with
\begin{equations}[eq:minkowski_modular_double_cone_fs_coords2]
t_s &= \tau + \ell \frac{2 \ell (t-\tau) \cosh(2 \pi s) - \left( \ell^2 + (t-\tau)^2 - \vec{x}^2 \right) \sinh(2 \pi s)}{2 \ell^2 + \left( \ell^2 + (t-\tau)^2 - \vec{x}^2 \right) [ \cosh(2 \pi s) - 1 ] - 2 \ell (t-\tau) \sinh(2 \pi s)} \eqend{,} \\
\vec{x}_s &= \frac{2 \ell^2 \vec{x}}{2 \ell^2 + \left( \ell^2 + (t-\tau)^2 - \vec{x}^2 \right) [ \cosh(2 \pi s) - 1 ] - 2 \ell (t-\tau) \sinh(2 \pi s)} \eqend{.}
\end{equations}
The analytic continuation $s \to - \mathi/2$ is well-defined and results in~\eqref{eq:minkowski_modular_double_cone_deltahalf_pre}
\begin{splitequation}
\label{eq:minkowski_modular_double_cone_deltahalf}
\Delta_\doublecone^\frac{1}{2} \op(f) \Delta_\doublecone^{- \frac{1}{2}} &= \op\left( \underline{f} \right) \eqend{,} \quad \underline{f}(u,v,\vec{x}^\perp) = \left( - \frac{\ell^2}{u v} \right)^{d-\Delta} f\left( \frac{\ell^2}{u}, \frac{\ell^2}{v}, \vec{x}^\perp \right) \eqend{,} \\
\underline{f}(t,\vec{x}) &= \left( \frac{\ell^2}{\vec{x}^2 - (t-\tau)^2} \right)^{d-\Delta} f\left( \tau - \frac{\ell^2 (t-\tau)}{\vec{x}^2 - (t-\tau)^2}, \frac{\ell^2 \vec{x}}{\vec{x}^2 - (t-\tau)^2} \right) \eqend{,}
\end{splitequation}
where we gave expressions both for null and Cartesian coordinates.

It only remains to determine the modular conjugation $J_\doublecone$. Since we need to compensate for the inversion of the coordinates~\eqref{eq:minkowski_modular_double_cone_deltahalf}, apart from time reversal $T$~\eqref{eq:minkowski_timereversal} which again is needed to obtain an antilinear operator, we also need a unitary operator $V$ that implements inversions. Such an operator was first constructed by Swieca and V{\"o}lkel~\cite{swiecavoelkel1973} on the Fock space $\mathcal{F}$ for the free massless scalar field, and reads
\begin{equation}
\label{eq:minkowski_modular_double_cone_inversion}
V a(f) V^{-1} = a(f_V) \eqend{,} \quad V a^\dagger(f) V^{-1} = a^\dagger(f_V) \eqend{,} \quad V \Omega = \Omega
\end{equation}
with
\begin{equation}
\label{eq:minkowski_modular_double_cone_inversion_fv}
\widetilde{f_V}(\vec{p}) = \int \sqrt{ \frac{q^0}{p^0} } \, \tilde{f}(\vec{q}) \int \left( \frac{\ell^2}{\vec{x}^2} \right)^\frac{d-2}{2} \mathe^{\mathi \vec{q} \vec{x} - \mathi \ell^2 (\vec{p} \vec{x})/\vec{x}^2} \total^{d-1} \vec{x} \frac{\total^{d-1} \vec{q}}{(2\pi)^{d-1}} \eqend{.}
\end{equation}
It is somewhat involved to show that the action of $V$ on the scalar field $\phi(f)$ defined by Eq.~\eqref{eq:minkowski_phi_def} from the creation and annihilation operators indeed results in a Lorentzian inversion according to
\begin{equation}
\label{eq:minkowski_modular_double_cone_inversion_phi}
V \phi(f) V^{-1} = \phi(f_V) \quad\text{with}\quad f_V(x) \equiv \left( \frac{\ell^2}{x^2} \right)^{d-\Delta} f\left( \frac{\ell^2 t}{x^2}, \frac{\ell^2 \vec{x}}{x^2} \right) \eqend{,}
\end{equation}
and we leave the demonstration to App.~\ref{app:inversion}, where we also show that $V^2 = \1$.\footnote{In the general case, we do not know whether such an operator exists, since conformal inversions are not continuously connected to the identity. However, inversions exist where the sign of time and one spatial coordinate $x^k$ are also flipped, since they can be obtained from the analytic continuation of the adjoint action whose generator is a linear combination of $P_k$ and $K_k$. Nevertheless, these are of no use for our purposes since time reversal already flips the sign of time.}

We also need again a time translation through the adjoint action~\eqref{eq:minkowski_ulambda_op} with the generator $Q$~\eqref{eq:minkowski_q_def} proportional to $H$, taking $a = 1$ and $b = c = 0$. One then verifies that
\begin{equation}
\label{eq:minkowski_modular_double_cone_j_def}
J_\doublecone \equiv U(\tau) VCT U(-\tau)
\end{equation}
results in
\begin{splitequation}
\label{eq:minkowski_modular_double_cone_j}
J_\doublecone \op(f) J_\doublecone^{-1} &= \op^\dagger(f_J) \eqend{,} \\
f_J(t,\vec{x}) &\equiv \left( \frac{\ell^2}{\vec{x}^2 - (t-\tau)^2} \right)^{d-\Delta} f^*\left( \tau - \frac{\ell^2 (t-\tau)}{\vec{x}^2 - (t-\tau)^2}, \frac{\ell^2 \vec{x}}{\vec{x}^2 - (t-\tau)^2} \right) \eqend{,}
\end{splitequation}
where we have again included charge conjugation $C$~\eqref{eq:minkowski_chargeconjugation_op} in the general case. Because $V^2 = \1$, we have $J_\futurecone^2 = \1$ as required, and since $f_J$~\eqref{eq:minkowski_modular_double_cone_j} involves the same Lorentzian inversion as $\underline{f}$ given in Eq.~\eqref{eq:minkowski_modular_double_cone_deltahalf}, we also obtain that $J_\doublecone$ maps the algebra $\mathfrak{A}_\doublecone$ into its commutant $\mathfrak{A}_\doublecone'$. The action of the modular automorphism and the modular conjugation are depicted in Fig.~\ref{fig:diamond2}.

From $\Delta_\doublecone$ and $J_\doublecone$, we again assemble the Tomita operator $S_\doublecone = J_\doublecone \Delta_\doublecone^\frac{1}{2}$ with action
\begin{equation}
S_\doublecone \op(f) S_\doublecone^{-1} = \op^\dagger(f^*)
\end{equation}
following from the actions~\eqref{eq:minkowski_modular_double_cone_deltahalf} and~\eqref{eq:minkowski_modular_double_cone_j}. Since both the modular Hamiltonian and the modular conjugation leave the vacuum invariant, $\Delta_\doublecone^{-\frac{1}{2}} \Omega = \Omega$ and $J_\doublecone \Omega = \Omega$, it follows that $S_\doublecone a \Omega = a^\dagger \Omega$ for all $a \Omega$ in the dense set $\mathcal{D}$ whose support lies in the double cone, i.e. for all $a \in \mathfrak{A}_\doublecone$ as required. Lastly, we also verify the KMS condition~\eqref{eq:intro_kms} for double cones, and compute using the two-point function~\eqref{eq:minkowski_wightman} and the modular automorphism~\eqref{eq:minkowski_modular_double_cone_automorphism} with the transformed function $f_{s,\doublecone}$~\eqref{eq:minkowski_modular_double_cone_fs2} that
\begin{splitequation}
&\left( \Omega, \sigma^\doublecone_s\left[ \op(f) \right] \op(g) \Omega \right) = \left( \Omega, \op\left( f_{s,\doublecone} \right) \op(g) \Omega \right) \\
&\quad= C_\op \lim_{\epsilon \to 0^+} \iint \frac{\left( \frac{\vec{x}_s^2}{\vec{x}^2} \right)^\frac{d-\Delta}{2} f(x_s) g(y)}{[ - (x^0-y^0 - \mathi \epsilon)^2 + (\vec{x}-\vec{y})^2 ]^\Delta} \total^d x \total^d y \\
&\quad= C_\op \lim_{\epsilon \to 0^+} \iint \frac{f(z^0+\tau, \vec{z}) g(y^0+\tau, \vec{y})}{\left[ - \left[ \frac{\ell h'_s(z)}{2 \pi \sqrt{h_s(z)}} - \sqrt{h_s(z)} \, y^0 - \mathi \epsilon \right]^2 + \left[ \frac{\vec{z}}{\sqrt{h_s(z)}} - \sqrt{h_s(z)} \, \vec{y} \right]^2 \right]^\Delta} \total^d z \total^d y
\end{splitequation}
with
\begin{equation}
h_s(x) = 1 + \frac{\ell^2 - x^2}{2 \ell^2} [ \cosh(2 \pi s) - 1 ] + \frac{t}{\ell} \sinh(2 \pi s) \eqend{.}
\end{equation}
To arrive at this expression, we changed variables $x^0 \to x^0_s + \tau$, $\vec{x} \to \vec{x}_s$, $y^0 \to y^0 + \tau$ with the transformed coordinates $x_s$ given by Eq.~\eqref{eq:minkowski_modular_double_cone_fs_coords2} and then renamed $x_s$ to $z$; the Jacobian of this coordinate change is $\abs{ \det\left( \frac{\partial x^\mu}{\partial x_s^\nu} \right)_{\mu,\nu=1}^d }_{x_s = z} = \left[ h_s(z) \right]^{-d}$. In this form, we can perform the analytic continuation $s \to s - \mathi$, under which we have $h_s(x) \to h_{s-\mathi}(x) = h_s(x)$ as well as $h'_s(x) \to h'_{s-\mathi}(x) = h'_s(x)$, but $\sqrt{h_s(x)} \to \sqrt{h_{s-\mathi}(x)} = - \sqrt{h_s(x)}$ since the phase of the square root only changes half as quickly. We see that the effect of the analytic continuation is again to change the sign of $\epsilon$, and performing the above steps in reverse we arrive at
\begin{splitequation}
\left( \Omega, \sigma^\doublecone_{s-\mathi}\left[ \op(f) \right] \op(g) \Omega \right) &= C_\op \lim_{\epsilon \to 0^+} \iint \frac{\left( \frac{\vec{x}_s^2}{\vec{x}^2} \right)^\frac{d-\Delta}{2} f(x_s) g(y)}{[ - (x^0-y^0 + \mathi \epsilon)^2 + (\vec{x}-\vec{y})^2 ]^\Delta} \total^d x \total^d y \\
&= \left( \Omega, \op(g) \sigma^\doublecone_s\left[ \op(f) \right] \Omega \right) \eqend{,}
\end{splitequation}
such that the KMS condition~\eqref{eq:intro_kms} holds.

\section{Conformally flat spacetimes}
\label{sec:conformal}

In this section we generalize the flat-space results for the modular data of light cones and double cones to a general conformally flat spacetime. We first determine the conformal transformations of the scalar field, stress tensor and (conformal) Killing vectors from Minkowski space, then quantize the scalar field in the conformal vacuum state and determine how the discrete and continuous symmetries are implemented in the quantum theory. Using these results, we show that the modular data is essentially determined by a suitable rescaling of the Minkowski modular data.

\subsection{Conformal rescaling}

We consider $\mathbb{R}^d$ with metric
\begin{equation}
\label{eq:conformal_metric}
g_{\mu\nu} = \mathe^{2 \omega} \eta_{\mu\nu} \eqend{,} \quad g^{\mu\nu} = \mathe^{- 2 \omega} \eta^{\mu\nu} \eqend{,} \quad \sqrt{-g} = \mathe^{d \omega} \eqend{,}
\end{equation}
where $\omega = \omega(x)$ is a real smooth function. The Levi-Civita connection (expressed using the Christoffel symbols) on this spacetime reads
\begin{equation}
\Gamma^\rho_{\mu\nu} = \delta_\mu^\rho \partial_\nu \omega + \delta_\nu^\rho \partial_\mu \omega - \eta_{\mu\nu} \partial^\rho \omega \eqend{,}
\end{equation}
and the curvature tensors are
\begin{equations}
\begin{split}
R^\rho{}_{\sigma\mu\nu} &= - 2 \delta_{[\mu}^\rho \partial_{\nu]} \partial_\sigma \omega + 2 \eta_{\sigma[\mu} \partial_{\nu]} \partial^\rho \omega + 2 \delta_{[\mu}^\rho \partial_{\nu]} \omega \partial_\sigma \omega - 2 \eta_{\sigma[\mu} \partial_{\nu]} \omega \partial^\rho \omega \\
&\quad- \left( \delta_\mu^\rho \eta_{\sigma\nu} - \delta_\nu^\rho \eta_{\sigma\mu} \right) \partial_\lambda \omega \partial^\lambda \omega \eqend{,}
\end{split} \\
R_{\mu\nu} &= - (d-2) \partial_\mu \partial_\nu \omega - \eta_{\mu\nu} \partial^2 \omega + (d-2) \partial_\mu \omega \partial_\nu \omega - (d-2) \eta_{\mu\nu} \partial_\lambda \omega \partial^\lambda \omega \eqend{,} \\
\mathe^{2 \omega} R &= - (d-1) \left[ 2 \partial^2 \omega + (d-2) \partial_\lambda \omega \partial^\lambda \omega \right] \eqend{.}
\end{equations}

We furthermore compute
\begin{equation}
\left( \nabla^2 - \xi R \right) \Phi = \mathe^{- \frac{d+2}{2} \omega} \partial^2 \phi + \mathe^{- 2 \omega} \left[ (d-1) \xi - \frac{d-2}{4} \right] \left[ 2 \partial^2 \omega + (d-2) \partial_\rho \omega \partial^\rho \omega \right] \Phi
\end{equation}
with the rescaled scalar field
\begin{equation}
\label{eq:conformal_phi}
\phi = \mathe^{\Delta \omega} \Phi = \mathe^{\frac{d-2}{2} \omega} \Phi \eqend{,}
\end{equation}
such that for conformal coupling $\xi = (d-2)/[4(d-1)]$ we have a simple conformal rescaling of the flat-space scalar equation of motion $\partial^2 \phi = 0$. This is of course the reason for the specific coefficient in the improved stress tensor~\eqref{eq:minkowski_stress_improved}, and we take this conformal coupling in the following. In curved spacetime, the improved stress tensor reads
\begin{splitequation}
\mathcal{T}_{\mu\nu} &\equiv \nabla_\mu \Phi \nabla_\nu \Phi - \frac{1}{2} g_{\mu\nu} \nabla_\rho \Phi \nabla^\rho \Phi - \xi \left( \nabla_\mu \nabla_\nu - g_{\mu\nu} \nabla^2 - R_{\mu\nu} + \frac{1}{2} R g_{\mu\nu} \right) \Phi^2 \\
&\approx \frac{d}{2 (d-1)} \nabla_\mu \Phi \nabla_\nu \Phi - \frac{d-2}{2 (d-1)} \Phi \nabla_\mu \nabla_\nu \Phi - \frac{1}{2 (d-1)} g_{\mu\nu} \nabla_\rho \Phi \nabla^\rho \Phi \\
&\quad+ \frac{d-2}{4 (d-1)} \left( R_{\mu\nu} - \frac{1}{2 (d-1)} R g_{\mu\nu} \right) \Phi^2
\end{splitequation}
where we dropped a term proportional to the field equation $( \nabla^2 - \xi R ) \Phi$ in the second equality. The improved stress tensor is both conserved and traceless on-shell:
\begin{equations}
\nabla^\mu \mathcal{T}_{\mu\nu} &= \frac{d}{2 (d-1)} \left( \nabla^2 - \xi R \right) \Phi \nabla_\nu \Phi - \frac{d-2}{2 (d-1)} \Phi \nabla_\nu \left( \nabla^2 - \xi R \right) \Phi \approx 0 \eqend{,} \\
g^{\mu\nu} \mathcal{T}_{\mu\nu} &= - \frac{d-2}{2 (d-1)} \Phi \left( \nabla^2 - \xi R \right) \Phi \approx 0 \eqend{,}
\end{equations}
and manifestly symmetric. In terms of the rescaled scalar $\phi$~\eqref{eq:conformal_phi}, we obtain
\begin{equation}
\label{eq:conformal_stresstensor}
\mathe^{(d-2) \omega} \mathcal{T}_{\mu\nu} = \frac{d}{2 (d-1)} \partial_\mu \phi \partial_\nu \phi - \frac{d-2}{2 (d-1)} \phi \partial_\mu \partial_\nu \phi - \frac{1}{2 (d-1)} \eta_{\mu\nu} \partial_\rho \phi \partial^\rho \phi = T_{\mu\nu} \eqend{,}
\end{equation}
so the classical improved stress tensor also differs just by a global rescaling from the flat-space one~\eqref{eq:minkowski_stress_improved}.

We next have a look at conformal Killing vectors. If $\Xi^\mu$ is a conformal Killing vector of $g$, such that
\begin{equation}
\label{eq:conformal_killing}
\nabla_\mu \Xi_\nu + \nabla_\nu \Xi_\mu = \frac{2}{d} g_{\mu\nu} \nabla^\rho \Xi_\rho \eqend{,}
\end{equation}
the rescaled vector $\xi^\mu = \Xi^\mu$ ($\xi_\mu = \mathe^{- 2 \omega} \Xi_\mu$) is a conformal Killing vector in flat space:
\begin{equation}
\partial_\mu \xi_\nu + \partial_\nu \xi_\mu = \frac{2}{d} \eta_{\mu\nu} \partial^\rho \xi_\rho \eqend{.}
\end{equation}
On the other hand, a Killing vector is generally not mapped into a Killing vector, but also into a conformal Killing vector: from $\nabla_\mu \Xi_\nu + \nabla_\nu \Xi_\mu = 0$ we obtain with the same rescaling
\begin{equation}
\partial_\mu \xi_\nu + \partial_\nu \xi_\mu = \frac{2}{d} \eta_{\mu\nu} \partial_\rho \xi^\rho \eqend{,} \quad \partial_\rho \xi^\rho = - d \xi^\rho \partial_\rho \omega \eqend{,}
\end{equation}
and vice versa from $\partial_\mu \xi_\nu + \partial_\nu \xi_\mu = 0$ we obtain
\begin{equation}
\nabla_\mu \Xi_\nu + \nabla_\nu \Xi_\mu = \frac{2}{d} g_{\mu\nu} \nabla_\rho \Xi^\rho \eqend{,} \quad \nabla_\rho \Xi^\rho = d \Xi^\rho \nabla_\rho \omega \eqend{.}
\end{equation}
The conserved currents are again determined via the Noether method. Defining $\mathcal{J}^\mu_\Xi \equiv \mathcal{T}^{\mu\nu} \Xi_\nu$, we compute as before
\begin{equation}
\label{eq:conformal_current_conserved}
\nabla_\mu \mathcal{J}^\mu_\Xi = \nabla_\mu \mathcal{T}^{\mu\nu} \Xi_\nu + \frac{1}{2} \mathcal{T}^{\mu\nu} \left( \nabla_\mu \Xi_\nu + \nabla_\nu \Xi_\mu \right) \approx \frac{1}{d} \mathcal{T}^{\mu\nu} g_{\mu\nu} \nabla^\rho \Xi_\rho \approx 0 \eqend{,}
\end{equation}
which holds at least classically.

\subsection{Free real scalar field}

Since the scalar field $\Phi$ can be obtained by the simple rescaling~\eqref{eq:conformal_phi} from $\phi$, it can be quantized by rescaling the quantized field $\phi$ in Minkowski spacetime. The corresponding state is known as the conformal vacuum, which we will continue to denote by $\Omega$. For this rescaling to work, it is clear that it must be possible to find a Cauchy surface in the curved spacetime with metric~\eqref{eq:conformal_metric} which can be identified with the corresponding one in Minkowski spacetime. Only in this case, the conformal rescaling results in an invertible mapping of the corresponding single-particle Hilbert spaces $\mathcal{H}$ and correspondingly the bosonic Fock space $\mathcal{F}$. For a clear discussion of this point we refer to Candelas and Dowker~\cite{candelasdowker1979}, who show that for example the flat and closed Robertson--Walker universes as well as the de~Sitter spacetime share a common Cauchy surface with Minkowski spacetime, while the open Robertson--Walker universe or the static patch of de~Sitter spacetime share a common Cauchy surface only with the Rindler wedge. In the following, we assume that such an identification has been made. We can thus define
\begin{equation}
\label{eq:conformal_phi_def}
\Phi(f) \equiv \phi(f_\omega) \quad\text{with}\quad f_\omega(x) \equiv \mathe^{(d-\Delta) \omega(x)} f(x) \eqend{,}
\end{equation}
such that we have (either formally or rigorously when using Gel'fand triples)
\begin{equation}
\label{eq:conformal_phi_rescaling}
\Phi(f) = \int \Phi(x) f(x) \sqrt{-g} \total^d x = \int \phi(x) f(x) \, \mathe^{(d-\Delta) \omega(x)} \total^d x = \int \phi(x) f_\omega(x) \total^d x = \phi(f_\omega)
\end{equation}
using the rescaling~\eqref{eq:conformal_phi}. It follows that the discrete symmetries of parity, time reserval and charge conjugation can also be simply related to the flat-space ones. Namely, from the parity transformation~\eqref{eq:minkowski_parity} together with the definition~\eqref{eq:conformal_phi_def} of $\Phi$ we obtain
\begin{equation}
\label{eq:conformal_parity}
P \Phi(f) P^{-1} = \Phi(f_P) \quad\text{with}\quad f_P(x) \equiv \mathe^{- (d-\Delta) [ \omega(x) - \omega(t,-\vec{x}) ]} f(t,-\vec{x}) \eqend{,}
\end{equation}
from the time reversal~\eqref{eq:minkowski_timereversal} it follows that
\begin{equation}
\label{eq:conformal_timereversal}
T \Phi(f) T^{-1} = \Phi(f_T) \quad\text{with}\quad f_T(x) \equiv \mathe^{- (d-\Delta) [ \omega(x) - \omega(-t,\vec{x}) ]} f^*(-t,\vec{x}) \eqend{,}
\end{equation}
and as in Minkowski spacetime charge conjugation does not do anything for a real scalar field:
\begin{equation}
\label{eq:conformal_chargeconjugation}
C \Phi(f) C^{-1} = \Phi(f) \eqend{.}
\end{equation}
The other discrete symmetry that we need, the Lorentzian inversion $V$~\eqref{eq:minkowski_modular_double_cone_inversion_phi}, can also be generalized in the same way, and we obtain
\begin{equation}
\label{eq:conformal_inversion}
V \Phi(f) V^{-1} = \Phi(f_V) \quad\text{with}\quad f_V(x) \equiv \mathe^{- (d-\Delta) \left[ \omega(x) - \omega\left( \frac{\ell^2 t}{x^2}, \frac{\ell^2 \vec{x}}{x^2} \right) \right]} \left( \frac{\ell^2}{x^2} \right)^{d-\Delta} f\left( \frac{\ell^2 t}{x^2}, \frac{\ell^2 \vec{x}}{x^2} \right) \eqend{.}
\end{equation}
All discrete symmetries thus are implemented by multiplying the flat-space results by appropriate powers of the conformal factor, and one checks that this works in such a way that all are self-inverse: $P^2 = T^2 = C^2 = V^2 = \1$.

The same approach for the continuous symmetries transforms the commutators~\eqref{eq:minkowski_hdk_phi_commutator} of the generators of time translations, dilations and special conformal transformations with the scalar field into
\begin{equations}[eq:conformal_hdk_phi_commutator]
[ H, \Phi(f) ] &= \Phi(f_H) \quad\text{with}\quad f_H(x) \equiv \mathi \left[ \partial_t + (d-\Delta) \partial_t \omega \right] f(x) \eqend{,} \\
[ D, \Phi(f) ] &= \Phi(f_D) \quad\text{with}\quad f_D(x) \equiv \mathi \left[ x^\mu \partial_\mu + (d-\Delta) \left( 1 + x^\mu \partial_\mu \omega \right) \right] f(x) \eqend{,} \\
[ K, \Phi(f) ] &= \Phi(f_K) \quad\text{with}\quad f_K(x) \equiv - 2 t f_D(x) - x^2 f_H(x) \eqend{.}
\end{equations}
However, we could also directly compute these using the Noether method in the curved spacetime. One issue that arises there is that while the stress tensor can always be made divergence-free in the quantum theory~\cite{hollandswald2005}, its trace will be generically non-vanishing even for conformal theories, which is the well-known conformal (or trace) anomaly~\cite{duff1994}. Therefore, the classically conserved currents~\eqref{eq:conformal_current_conserved} are not conserved anymore after quantization, and we have instead
\begin{equation}
\label{eq:conformal_anomaly}
\nabla_\mu \mathcal{J}^\mu_\Xi \approx \mathcal{A} \nabla^\rho \Xi_\rho
\end{equation}
with the conformal anomaly $\mathcal{A} \equiv \frac{1}{d} \mathcal{T}^{\mu\nu} g_{\mu\nu}$. Nevertheless, the anomaly is state-independent and thus proportional to the identity $\1$, such that commutators are unaffected. That is, we can use the Noether method to derive the commutator of charges with operators implementing the corresponding conformal transformation, even though the charges themselves are time-dependent. Integrating the current $\mathcal{J}^\mu_\Xi$ over a spacelike Cauchy hypersurface $\Sigma$, we obtain the charge
\begin{equation}
\label{eq:conformal_charge}
\mathcal{Q}_{\Xi,\Sigma} \equiv \int_\Sigma \mathcal{J}^\mu_\Xi n_\mu \sqrt{\gamma} \total^{d-1} \vec{x} \eqend{,}
\end{equation}
where $n_\mu$ is the future-directed normal to the hypersurface $\Sigma$ and $\gamma$ the induced metric on the hypersurface. Considering two such hypersurfaces, and using Gauss' divergence theorem, we have
\begin{equation}
\int_{V_{12}} \nabla_\mu \mathcal{J}^\mu_\Xi \sqrt{-g} \total^d x = \mathcal{Q}_{\Xi,\Sigma_2} - \mathcal{Q}_{\Xi,\Sigma_1} \eqend{,}
\end{equation}
where $V_{12}$ is the region whose boundary is composed of the Cauchy surfaces $\Sigma_1$ and $\Sigma_2$ (with $\Sigma_2$ lying to the future of $\Sigma_1$) and spatial infinity, where we assume everything to fall off sufficiently fast such that no boundary terms arise. Because of the conformal anomaly, the divergence~\eqref{eq:conformal_anomaly} of the current is non-vanishing, and thus $\mathcal{Q}_{\Xi,\Sigma_2} \neq \mathcal{Q}_{\Xi,\Sigma_1}$. Since $\mathcal{A} \sim \1$ as stated before, the commutators are nevertheless unaffected, and we obtain
\begin{equation}
\left[ \mathcal{Q}_{\Xi,\Sigma_2}, \Phi(f) \right] = \left[ \mathcal{Q}_{\Xi,\Sigma_1}, \Phi(f) \right] \eqend{.}
\end{equation}
So we are justified in taking any Cauchy surface $\Sigma$ to compute the charge $\mathcal{Q}_\Xi$. In other words, once we exponentiate to obtain the unitary that generates the required finite transformation, the conformal anomaly only results in a phase which cancels out in the adjoint action.

We can thus perform a conformal rescaling also in the quantum theory, and ignore the anomaly for the purposes of computing commutators. From the classical rescaling~\eqref{eq:conformal_stresstensor} of the stress tensor, the rescaling~\eqref{eq:conformal_metric} of the metric and the rescaling of the conformal Killing vector~\eqref{eq:conformal_killing}, we derive that $\mathcal{J}^\mu_\Xi = \mathe^{- d \omega} J^\mu_\xi$ with the flat-space current $J^\mu_\xi$~\eqref{eq:minkowski_jxi_def}. Using hypersurfaces of constant $t$ such that $n_\mu = \mathe^\omega \delta_\mu^0$ (which is properly normalized to $n^\mu n_\mu = -1$) and $\gamma_{ij} = \mathe^{2 \omega} \delta_{ij}$, $\sqrt{\gamma} = \mathe^{(d-1) \omega}$, it follows that classically the charge~\eqref{eq:conformal_charge} does not change at all under conformal rescalings. The commutators~\eqref{eq:conformal_hdk_phi_commutator} can thus equivalently be obtained from the Noether method in the curved spacetime, and the terms involving derivatives of $\omega$ arise solely from the different definition~\eqref{eq:conformal_phi_rescaling} of smeared operators (which includes the metric determinant $\sqrt{-g} = \mathe^{d \omega}$) and the conformal rescaling~\eqref{eq:conformal_phi} of the field $\Phi$, i.e., the definition~\eqref{eq:conformal_phi_def}. As in flat space, the above results hold more generally than just for the free scalar field, namely whenever a suitable action of the required generators of the conformal algebra is available. Consequently, in the following we consider a generic scalar primary operator $\hat{\op}$ related to the flat-space one $\op$ by the rescaling
\begin{equation}
\label{eq:conformal_op_def}
\hat{\op}(f) \equiv \op(f_\omega) \quad\text{with}\quad f_\omega(x) \equiv \mathe^{(d-\Delta) \omega(x)} f(x) \eqend{,}
\end{equation}
where $\Delta$ is the scaling dimension of $\op$. Contrary to all other discrete symmetries, the counting operator $W$~\eqref{eq:minkowski_modular_w_def} is the only one that is not rescaled:
\begin{equation}
\label{eq:conformal_modular_w_def}
W \hat{\op}(f) W^{-1} = (-1)^{\Delta-d} \hat{\op}(f) \eqend{,} \quad W \Omega = \Omega \eqend{.}
\end{equation}

While the changed commutators~\eqref{eq:conformal_hdk_phi_commutator} also result in a changed adjoint action of $U(\lambda)$~\eqref{eq:minkowski_q_def}, it is easy to see that this action is obtained by an appropriate rescaling similar to the discrete symmetries. Namely, from Eq.~\eqref{eq:minkowski_ulambda_op} we obtain
\begin{equation}
\label{eq:conformal_ulambda_op}
U(\lambda) \hat{\op}(f) U(\lambda)^\dagger = \hat{\op}(f_\lambda) \quad\text{with}\quad f_\lambda(x) \equiv \left[ \mathe^{\omega(x_\lambda) - \omega(x)} \frac{r_\lambda}{r} \right]^{d-\Delta} f(x_\lambda)
\end{equation}
and the solution~\eqref{eq:minkowski_flambda_sol} for the transformed coordinates $x_\lambda$. Just as in flat space, it follows that $(f_\lambda)_\sigma = f_{\lambda+\sigma}$, i.e., that transforming a function $f$ twice is equal to a single transformation with the sum of the two parameters, and thus $U(\lambda)$ indeed are unitaries whose definition can be extended to the full Hilbert space.

Finally, the two-point function of $\hat{\op}$ can also be obtained from the flat-space one~\eqref{eq:minkowski_wightman}. Using the relation~\eqref{eq:conformal_op_def} between the flat-space operator $\op$ and the curved-space one $\hat{\op}$, it follows that
\begin{equation}
\label{eq:conformal_wightman}
( \Omega, \hat{\op}(f) \hat{\op}(g) \Omega ) = C_\op \lim_{\epsilon \to 0^+} \iint \frac{f_\omega(x) g_\omega(y)}{[ - (x^0-y^0 - \mathi \epsilon)^2 + (\vec{x}-\vec{y})^2 ]^\Delta} \total^d x \total^d y \eqend{,}
\end{equation}
where the metric determinant has been absorbed in the rescaled test functions $f_\omega$ and $g_\omega$.

\subsection{Modular data for the future and past light cones}

Because the transformed coordinates $x_\lambda$ are the same as in flat space, and only the overall rescaling of the transformed function $f_\lambda$ changes in the adjoint action~\eqref{eq:conformal_ulambda_op}, the modular Hamiltonian is the same as in flat space. That is, for the future light cone $\futurecone$ with tip at $(t,\vec{x}) = (\tau,\vec{0})$ we have $\ln \Delta_\futurecone = 2 \pi ( D - \tau H )$~\eqref{eq:minkowski_modular_light_cone_lndelta}, and for the past light cone $\pastcone$ with tip at $(t,\vec{x}) = (\tau,\vec{0})$ we have $\ln \Delta_\pastcone = 2 \pi ( - D + \tau H )$~\eqref{eq:minkowski_modular_light_cone_past_lndelta_j}. Of course, the generators of time translations $H$ and of dilations $D$ now act according to Eq.~\eqref{eq:conformal_hdk_phi_commutator}, which depends on the conformal rescaling involving $\omega$. Therefore, also the modular automorphism $\sigma^\futurecone_s\left( \hat{\op}(f) \right) = \Delta_\futurecone^{\mathi s} \hat{\op}(f) \Delta_\futurecone^{- \mathi s}$ now depends on $\omega$ according to the adjoint action~\eqref{eq:conformal_ulambda_op}, and we obtain
\begin{splitequation}
\label{eq:conformal_modular_light_cone_automorphism}
&\Delta_\futurecone^{\mathi s} \hat{\op}(f) \Delta_\futurecone^{- \mathi s} = \hat{\op}(f_{s,\futurecone}) \eqend{,} \quad f_{s,\futurecone}(t,\vec{x}) = \mathe^{- 2 \pi s (d-\Delta)} \mathe^{- (d-\Delta) [ \omega(x) - \omega(x_s) ]} f(x_s) \eqend{,} \\
&\text{with}\quad t_s = \tau + \mathe^{- 2 \pi s} ( t - \tau ) \eqend{,} \quad \vec{x}_s = \mathe^{- 2 \pi s} \vec{x}
\end{splitequation}
for the future light cone and
\begin{equation}
\Delta_\pastcone^{\mathi s} \hat{\op}(f) \Delta_\pastcone^{- \mathi s} = \hat{\op}(f_{s,\pastcone}) \quad\text{with}\quad f_{s,\pastcone}(x) = f_{-s,\futurecone}(x)
\end{equation}
for the past light cone.

It follows that just as in flat space we can also perform the analytic continuation $s \to - \mathi/2$ of the modular automorphism without encountering singularities, and obtain
\begin{equation}
\label{eq:conformal_modular_light_cone_deltahalf}
\Delta_\futurecone^\frac{1}{2} \hat{\op}(f) \Delta_\futurecone^{-\frac{1}{2}} = \hat{\op}\left( \underline{f} \right) \eqend{,} \quad \underline{f}(t,\vec{x}) = (-1)^{d-\Delta} \mathe^{- (d-\Delta) [ \omega(t,\vec{x}) - \omega(2\tau-t,-\vec{x}) ]} f(2\tau-t,-\vec{x}) \eqend{.}
\end{equation}
Again this maps a test function $f$ with support in the future light cone into the past light cone with the same tip as required, i.e., $\mathfrak{A}_\futurecone$ gets mapped its commutant $\mathfrak{A}_\futurecone' \supset \mathfrak{A}_\pastcone$, and the analogous result holds for the past light cone. Moreover, using the explicit form of the discrete symmetries $C$~\eqref{eq:conformal_chargeconjugation}, $P$~\eqref{eq:conformal_parity}, and $T$~\eqref{eq:conformal_timereversal}, as well as the counting operator $W$~\eqref{eq:conformal_modular_w_def}, it is easy to see that also the same modular conjugation as in flat space~\eqref{eq:minkowski_modular_light_cone_j_def}, namely
\begin{equation}
\label{eq:conformal_modular_light_cone_j_def}
J_\futurecone = J_\pastcone \equiv U(\tau) C P T W U(-\tau)
\end{equation}
results in the correct transformation
\begin{equation}
\label{eq:conformal_modular_light_cone_j}
J_\futurecone \hat{\op}(f) J_\futurecone^{-1} = \hat{\op}^\dagger(f_J) \eqend{,} \quad f_J(t,\vec{x}) \equiv (-1)^{\Delta-d} \mathe^{- (d-\Delta) [ \omega(t,\vec{x}) - \omega(2\tau-t,-\vec{x}) ]} f^*(2\tau-t, -\vec{x}) \eqend{.}
\end{equation}
As in flat space, we obtain that $J_\futurecone^2 = \1 = J_\pastcone^2$, that $J_\futurecone$ maps $\mathfrak{A}_\futurecone$ into its commutant $\mathfrak{A}_\futurecone' \supset \mathfrak{A}_\pastcone$, and that $J_\pastcone$ maps $\mathfrak{A}_\pastcone$ into its commutant $\mathfrak{A}_\pastcone' \supset \mathfrak{A}_\futurecone$.

By inspection of the results~\eqref{eq:conformal_modular_light_cone_deltahalf} and~\eqref{eq:conformal_modular_light_cone_j} and the analogous results for the past light cone, we see that the Tomita operators $S_\futurecone = J_\futurecone \Delta_\futurecone^\frac{1}{2}$ and $S_\pastcone = J_\pastcone \Delta_\pastcone^\frac{1}{2}$ have the action
\begin{equation}
S_\futurecone \op(f) S_\futurecone^{-1} = \op^\dagger(f^*) \eqend{,} \quad S_\pastcone \op(f) S_\pastcone^{-1} = \op^\dagger(f^*)
\end{equation}
as required, and as in flat spacetime they also fulfill all further necessary properties.

Lastly, we also verify the KMS condition~\eqref{eq:intro_kms} for the two-point function~\eqref{eq:conformal_wightman}. Analogously to the flat-space computation~\eqref{eq:minkowski_kms_pre}, we determine first for real $s$
\begin{splitequation}
\label{eq:conformal_light_cone_kms_pre}
&\left( \Omega, \sigma^\futurecone_s\left[ \hat{\op}(f) \right] \hat{\op}(g) \Omega \right) = \left( \Omega, \hat{\op}(f_{s,\futurecone}) \hat{\op}(g) \Omega \right) \\
&\quad= C_\op \lim_{\epsilon \to 0^+} \iint \frac{(f_{s,\futurecone})_\omega(x) g_\omega(y)}{[ - (x^0-y^0 - \mathi \epsilon)^2 + (\vec{x}-\vec{y})^2 ]^\Delta} \total^d x \total^d y \\
&\quad= C_\op \, \mathe^{- 2 \pi s (d-\Delta)} \lim_{\epsilon \to 0^+} \iint \frac{\mathe^{(d-\Delta) [ \omega(x_s) + \omega(y) ]} f(x_s) g(y)}{[ - (x^0-y^0 - \mathi \epsilon)^2 + (\vec{x}-\vec{y})^2 ]^\Delta} \total^d x \total^d y \\
&\quad= C_\op \lim_{\epsilon \to 0^+} \iint \frac{\mathe^{(d-\Delta) \omega(x^0+\tau,\vec{x})} f(x^0+\tau,\vec{x}) \mathe^{(d-\Delta) \omega(y^0+\tau,\vec{y})} g(y^0+\tau,\vec{y})}{\left[ - \left( \mathe^{\pi s} x^0 - \mathe^{- \pi s} y^0 - \mathi \epsilon \right)^2 + \left( \mathe^{\pi s} \vec{x} - \mathe^{- \pi s} \vec{y} \right)^2 \right]^\Delta} \total^d x \total^d y
\end{splitequation}
using the definition of $f_\omega$~\eqref{eq:conformal_op_def}, the modular automorphism~\eqref{eq:conformal_modular_light_cone_automorphism}, and the same coordinate transformations as in~\eqref{eq:minkowski_kms_pre}. Performing then the analytic continuation $s \to s - \mathi$ again effectively changes the sign of $\epsilon$, and performing the same coordinate transformations as in~\eqref{eq:minkowski_kms_final} we obtain
\begin{splitequation}
\label{eq:conformal_light_cone_kms_final}
&\left( \Omega, \sigma^\futurecone_{s-\mathi}\left[ \hat{\op}(f) \right] \hat{\op}(g) \Omega \right) \\
&\quad= C_\op \, \mathe^{- 2 \pi s (d-\Delta)} \lim_{\epsilon \to 0^+} \iint \frac{\mathe^{(d-\Delta) [ \omega(y_s) + \omega(x) ]} f(y_s) g(x)}{[ - (x^0-y^0 - \mathi \epsilon)^2 + (\vec{x}-\vec{y})^2 ]^\Delta} \total^d x \total^d y \\
&\quad= \left( \Omega, \hat{\op}(g) \hat{\op}\left( f_{s,\futurecone} \right) \Omega \right) = \left( \Omega, \hat{\op}(g) \sigma^\futurecone_s\left[ \hat{\op}(f) \right] \Omega \right) \eqend{,}
\end{splitequation}
verifying thus the KMS condition~\eqref{eq:intro_kms}. In fact, since the modular Hamiltonian $\ln \Delta_\futurecone = 2 \pi ( D - \tau H )$ agrees with the flat-space one~\eqref{eq:minkowski_modular_light_cone_lndelta}, and the KMS condition holds in flat space for an arbitrary test function~\eqref{eq:minkowski_kms_final}, via the rescaling $\hat{\op}(f) = \op(f_\omega)$~\eqref{eq:conformal_op_def} one obtains directly that the KMS condition holds also in the conformally related spacetime. Nevertheless, it is instructive to see explicitly how the various conformal factors combine in~\eqref{eq:conformal_light_cone_kms_pre} and~\eqref{eq:conformal_light_cone_kms_final} in order for the KMS condition to hold.

\subsection{Modular data for double cones}

As for the future and past light cones, the modular Hamiltonian is the same as in flat space~\eqref{eq:minkowski_modular_double_cone_lndelta}, namely $\ln \Delta_\doublecone = \pi/\ell \left[ ( \ell^2 - \tau^2 ) H + 2 \tau D + K \right]$. However, through the adjoint action~\eqref{eq:conformal_ulambda_op} of the corresponding unitaries the modular automorphism depends on $\omega$, and generalizing the flat-space result~\eqref{eq:minkowski_modular_double_cone_automorphism} and~\eqref{eq:minkowski_modular_double_cone_fs2} we obtain
\begin{equation}
\label{eq:conformal_modular_double_cone_automorphism}
\Delta_\doublecone^{\mathi s} \hat{\op}(f) \Delta_\doublecone^{- \mathi s} = \hat{\op}(f_{s,\doublecone}) \eqend{,} \quad f_{s,\doublecone}(x) = \mathe^{- (d-\Delta) [ \omega(x) - \omega(x_s) ]} \left( \frac{\vec{x}_s^2}{\vec{x}^2} \right)^\frac{d-\Delta}{2} f(x_s)
\end{equation}
with the transformed coordinates~\eqref{eq:minkowski_modular_double_cone_fs_coords2}. The analytic continuation $s \to - \mathi/2$ can be performed straightforwardly and results in the generalization of Eq.~\eqref{eq:minkowski_modular_double_cone_deltahalf}:
\begin{splitequation}
\label{eq:conformal_modular_double_cone_deltahalf}
&\Delta_\doublecone^\frac{1}{2} \op(f) \Delta_\doublecone^{- \frac{1}{2}} = \op\left( \underline{f} \right) \eqend{,} \quad \underline{f}(x) = \mathe^{- (d-\Delta) [ \omega(x) - \omega(\underline{x}) ]} \left( \frac{\ell^2}{\vec{x}^2 - (t-\tau)^2} \right)^{d-\Delta} f(\underline{x}) \\
&\text{with}\quad \underline{t} = \tau - \frac{\ell^2 (t-\tau)}{\vec{x}^2 - (t-\tau)^2} \eqend{,} \quad \underline{\vec{x}} = \frac{\ell^2 \vec{x}}{\vec{x}^2 - (t-\tau)^2} \eqend{.}
\end{splitequation}
As in flat space, this maps a test function $f$ with support in the double cone $\doublecone$ into three regions outside of it, namely into the future light cone $\futurecone$ with tip at $(t,\vec{x}) = (\ell,\vec{0})$ if $x$ lies in the upper quadrant of $\doublecone$ such that $t - \tau - r > 0$, into the past light cone $\pastcone$ with tip at $(t,\vec{x}) = (-\ell,\vec{0})$ if $x$ lies in the lower quadrant of $\doublecone$ such that $t - \tau + r < 0$, and otherwise into a region spacelike separated from $\doublecone$. For $\Delta \in \mathbb{N}$, Huygen's principle holds and we obtain a mapping into the commutant algebra $\mathfrak{A}'$.

For the modular conjugation $J_\doublecone$, we again verify that the same operator
\begin{equation}
\label{eq:conformal_modular_double_cone_j_def}
J_\doublecone = U(\tau) VCT U(-\tau)
\end{equation}
as in flat space~\eqref{eq:minkowski_modular_double_cone_j_def} results in the required mapping
\begin{equation}
\label{eq:conformal_modular_double_cone_j}
J_\doublecone \hat{\op}(f) J_\doublecone^{-1} = \hat{\op}^\dagger(f_J) \eqend{,} \quad f_J(x) = \left( \frac{\ell^2}{\vec{x}^2 - (t-\tau)^2} \right)^{d-\Delta} \mathe^{- (d-\Delta) [ \omega(x) - \omega(\underline{x}) ]} f^*(\underline{x}) \eqend{,}
\end{equation}
with the transformed coordinates the same as in Eq.~\eqref{eq:conformal_modular_double_cone_deltahalf}. For this, in addition to the action of $T$~\eqref{eq:conformal_timereversal} and $C$~\eqref{eq:conformal_chargeconjugation}, we also needed the conformal inversion $V$~\eqref{eq:conformal_inversion}. It thus follows as for the analytically continued automorphism~\eqref{eq:conformal_modular_double_cone_deltahalf} that $J_\doublecone$ maps $\mathfrak{A}$ into the commutant $\mathfrak{A}'$; moreover, from $\underline{\underline{x^\mu}} = x^\mu$ we see that $(f_J)_J = f$ and thus $J_\doublecone^2 = \1$. We can thus assemble the Tomita operator $S_\doublecone = J_\doublecone \Delta_\doublecone^\frac{1}{2}$, which has the correct action
\begin{equation}
S_\doublecone \op(f) S_\doublecone^{-1} = \op^\dagger(f^*) \eqend{.}
\end{equation}

In this case, we do not busy ourselves with the explicit verification of the KMS condition~\eqref{eq:intro_kms}, since it follows straightforwardly from the fact that the modular Hamiltonian $\ln \Delta_\doublecone = \pi/\ell \left[ ( \ell^2 - \tau^2 ) H + 2 \tau D + K \right]$ agrees with the flat-space one~\eqref{eq:minkowski_modular_double_cone_lndelta}, and the KMS condition holds in flat space for an arbitrary test function~\eqref{eq:minkowski_kms_final}, as explained at the end of the last subsection.

\section{de~Sitter spacetime}
\label{sec:desitter}

We now specialize our results to de~Sitter spacetime. We first recall the construction of the de~Sitter spacetime as an embedding surface in a higher-dimensional Minkowski spacetime, various useful coordinate systems covering certain parts of de~Sitter, and the conformal embeddings of these parts. We then give the explicit form of the modular Hamiltonian, the modular flow and the modular conjugation for light cones and diamonds in the expanding Poincaré patch. Finally, we show that for large diamonds the modular Hamiltonian tends to the time translations in the static patch, recovering the known fact that the de~Sitter vacuum state restricted to the static patch is a thermal state with respect to its time translations. We here concentrate on $d \geq 3$ dimensions, since the cases $d = 2$ and $d = 1$ are special (for example, the de~Sitter spacetime is not simply connected for $d = 2$, and for $d = 1$ we have a two-sheeted hyperboloid with each sheet isomorphic to $\mathbb{R}$).

\subsection{Geometry of de~Sitter}
\label{sec:desitter_geometry}

\begin{figure}
\begin{minipage}[t]{0.48\textwidth}
\centering
\includegraphics[]{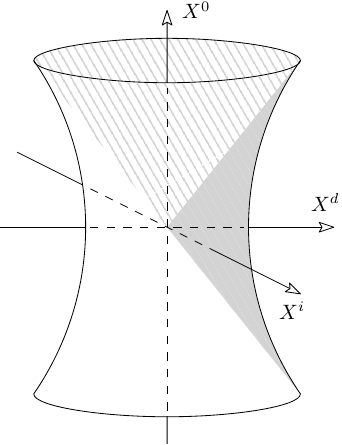}
\caption{De~Sitter spacetime embedded in the higher-dimensional Minkowski spacetime. The expanding Poincaré patch and the static patch are hatched and shaded, respectively; the Poincaré patch includes the static patch.}
\label{fig:desitter}
\end{minipage}
\hfill
\begin{minipage}[t]{0.48\textwidth}
\centering
\includegraphics[]{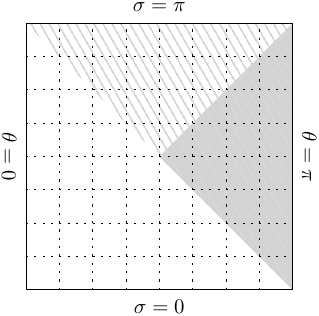}
\caption{Penrose diagram of de~Sitter spacetime, the expanding Poincaré patch and the static patch. Lines of constant $\sigma$ and constant $\theta$ are dotted. Each point in the diagram corresponds to a $(d-2)$-sphere.}
\label{fig:desitter-penrose}
\end{minipage}
\end{figure}

Given a $(d+1)$-dimensional ambient Minkowski spacetime with Cartesian coordinates $X^A$, $d$-dimensional de~Sitter spacetime is defined as the one-sheeted hyperboloid
\begin{equation}
\eta_{AB} X^A X^B = H^{-2} \eqend{,}
\end{equation}
where $H$ is the Hubble constant, the inverse radius of de~Sitter. Global coordinates $t \in \mathbb{R}$ and $n^i \in [-1,1]$ with $\delta_{ij} n^i n^j = 1$ which are defined by
\begin{equation}
X^0 = H^{-1} \sinh(H t) \eqend{,} \quad X^i = H^{-1} \cosh(H t) n^i \quad (i = 1,\ldots,d)
\end{equation}
cover the full hyperboloid, and the induced de~Sitter metric reads
\begin{equation}
\total s^2 = \eta_{AB} \total X^A \total X^B = g_{\mu\nu} \total x^\mu \total x^\nu = - \total t^2 + H^{-2} \cosh^2(H t) \total \Omega_{d-1}^2 \eqend{.}
\end{equation}
In this expression,
\begin{equation}
\label{eq:desitter_metric_omega}
\total \Omega_{d-1}^2 = \delta_{ij} \total n^i \total n^j = \total \theta^2 + \sin^2 \theta \total \Omega_{d-2}^2
\end{equation}
is the line element of the $(d-1)$-sphere with the polar angle $\theta \in [0,\pi)$, and we used that $\delta_{ij} n^i n^j = 1$ implies that $\delta_{ij} n^i \total n^j = 0$. Changing to the new time coordinate $\sigma$ defined as
\begin{equation}
\sigma = 2 \arctan\left[ \tanh\left( \frac{H t}{2} \right) \right] + \frac{\pi}{2} = \mathi \ln\left[ \frac{( 1 - \mathi \mathe^{H t} )^2}{1 + \mathe^{2 H t}} \right] \in (0, \pi)
\end{equation}
with inverse $t = H^{-1} \ln\left( \frac{\sin \sigma}{1 + \cos \sigma} \right)$, the de~Sitter metric reduces to
\begin{equation}
\label{eq:desitter_metric_full_conformal}
\total s^2 = \frac{1}{H^2 \sin^2 \sigma} \left( - \total \sigma^2 + \total \Omega_{d-1}^2 \right) \eqend{.}
\end{equation}
We see that the full de~Sitter spacetime is topologically equal to $\mathbb{R} \times \mathbb{S}^{d-1}$, and conformal to a portion of the Einstein static universe~\cite{hawkingellis}. The full de~Sitter spacetime is depicted in Fig.~\ref{fig:desitter}, and its conformal structure in Fig.~\ref{fig:desitter-penrose}. We note that the Einstein static universe is covered by $\sigma \in \mathbb{R}$, $\theta \in [0,2\pi)$, such that two copies of the Penrose diagram in Fig.~\ref{fig:desitter-penrose} (with one mirrored at $\theta = \pi$) are conformally embedded in the Einstein static universe; compare Candelas and Dowker~\cite{candelasdowker1979}.

On the other hand, the part of de~Sitter that is relevant for cosmology is the expanding Poincaré patch. This patch is best described using the conformally flat coordinates $\eta \in (-\infty,0)$ and $x^i \in \mathbb{R}$ which are defined by
\begin{splitequation}
\label{eq:desitter_embedding_poincare}
&X^0 = \frac{\eta}{2} - \frac{1}{2 H^2 \eta} - \frac{1}{2 \eta} r^2 \eqend{,} \mkern27mu X^i = - \frac{1}{H \eta} x^i \quad (i = 1,\ldots,d-1) \eqend{,} \\
&X^d = - \frac{\eta}{2} - \frac{1}{2 H^2 \eta} + \frac{1}{2 \eta} r^2 \eqend{,} \quad r^2 = \delta_{ij} x^i x^j \eqend{,}
\end{splitequation}
and only cover half of the hyperboloid, namely the part where $X^0 + X^d > 0$. The induced metric reads
\begin{equation}
\label{eq:desitter_metric_poincare}
\total s^2 = \frac{1}{(- H \eta)^2} \left( - \total \eta^2 + \delta_{ij} \total x^i \total x^j \right) \eqend{,}
\end{equation}
and we see that the Poincaré patch is conformal to half of Minkowski spacetime (which itself is conformal to a portion of the Einstein static universe). Changing the sign of $\eta$, we obtain the contracting Poincaré patch, which covers the other half of the hyperboloid where $X^0 + X^d < 0$. To properly display the expanding Poincaré patch in a Penrose diagram, we pass to spherical coordinates and then use the coordinate transformation (slighty modified from~\cite{candelasdowker1979})
\begin{splitequation}
\label{eq:desitter_poincare_conformaltrafo}
\sigma &= \pi + \arctan\left[ H (\eta+r) \right] + \arctan\left[ H (\eta-r) \right] \in (0,\pi) \eqend{,} \\
\theta &= \pi - \arctan\left[ H (\eta+r) \right] + \arctan\left[ H (\eta-r) \right] \in [0,\pi)
\end{splitequation}
to obtain the metric~\eqref{eq:desitter_metric_full_conformal} with~\eqref{eq:desitter_metric_omega}. However, since now the time $\eta$ only takes values in $(-\infty,0)$, we have the constraint $\sigma + \theta > \pi$. For the contracting patch, the same coordinate transformation works, but now we obtain $\sigma \in (\pi,2\pi)$ and the constraint $\sigma - \theta < \pi$.

\begin{figure}
\begin{minipage}[t]{0.48\textwidth}
\centering
\includegraphics[]{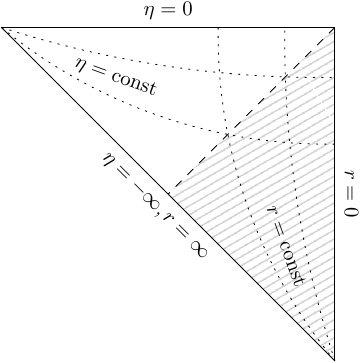}
\caption{Penrose diagram of the expanding Poincaré patch of de~Sitter spacetime. The static patch is hatched, lines of constant $\eta$ and constant $r$ are dotted.}
\label{fig:desitter-poincare}
\end{minipage}
\hfill
\begin{minipage}[t]{0.48\textwidth}
\centering
\includegraphics[]{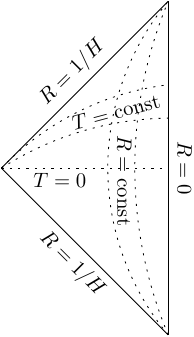}
\caption{Penrose diagram of the static patch of de~Sitter spacetime. Lines of constant $T$ and constant $R$ are dotted.}
\label{fig:desitter-static}
\end{minipage}
\end{figure}

Lastly, a further important part of de~Sitter is the static patch, described by coordinates $T \in \mathbb{R}$, $R \in [0,H^{-1})$ and $n^i \in [-1,1]$ with $\delta_{ij} n^i n^j = 1$ which are defined by
\begin{splitequation}
\label{eq:desitter_embedding_static}
&X^0 = H^{-1} \sqrt{1 - H^2 R^2} \sinh(H T) \eqend{,} \quad X^i = R n^i \quad (i = 1,\ldots,d-1) \eqend{,} \\
&X^d = H^{-1} \sqrt{1 - H^2 R^2} \cosh(H T) \eqend{.}
\end{splitequation}
This patch covers only a quarter of the full de~Sitter spacetime, namely the one with $X^d > 0$ and $X^0 + X^d > 0$; another static patch is obtained by changing the signs of both $X^0$ and $X^d$. The induced de~Sitter metric of the static patch reads
\begin{equation}
\label{eq:desitter_metric_static}
\total s^2 = - ( 1 - H^2 R^2 ) \total T^2 + \frac{1}{1 - H^2 R^2} \total R^2 + R^2 \total \Omega_{d-2}^2 \eqend{,}
\end{equation}
and it is clear that $\partial_T$ is a Killing vector in this patch. The coordinate transformation (slightly modified from~\cite{candelasdowker1979})
\begin{splitequation}
\sigma &= \frac{\pi}{2} + \arctan\left[ \tanh\left( \frac{H T + \arctanh( H R )}{2} \right) \right] + \arctan\left[ \tanh\left( \frac{H T - \arctanh( H R )}{2} \right) \right] \\
&= \mathi \ln\left[ - \frac{\mathi + \sqrt{1 - H^2 R^2} \sinh(H T)}{\sqrt{ 1 + (1 - H^2 R^2) \sinh^2(H T) }} \right] \in (0,\pi) \eqend{,} \\
\theta &= \pi - \arctan\left[ \tanh\left( \frac{H T + \arctanh( H R )}{2} \right) \right] + \arctan\left[ \tanh\left( \frac{H T - \arctanh( H R )}{2} \right) \right] \\
&= \pi - \mathi \ln\left[ \frac{\sqrt{1-H^2 R^2} \cosh(H T) - \mathi H R}{\sqrt{ (1 - H^2 R^2) \cosh^2(H T) + H^2 R^2 }} \right] \in \left( \frac{\pi}{2}, \pi \right]
\end{splitequation}
then results again in the metric~\eqref{eq:desitter_metric_full_conformal}, but now with the constraint $\pi - \theta < \sigma < \theta$ stemming from the condition that $H R < 1$. The Penrose diagrams of the expanding Poincaré patch and the static patch are depicted in Figs.~\ref{fig:desitter-poincare} and~\ref{fig:desitter-static}.

For later use, we also give the coordinate transformation between the expanding Poincaré patch and the static patch, which can be determined from Eqs.~\eqref{eq:desitter_embedding_poincare} and~\eqref{eq:desitter_embedding_static}:
\begin{splitequation}
\label{eq:desitter_coordtrafo}
&T = - \frac{1}{2 H} \ln\left[ H^2 \left( \eta^2 - r^2 \right) \right] \eqend{,} \quad R = - \frac{r}{H \eta} \eqend{,} \\
&\eta = - \frac{\mathe^{- H T}}{H \sqrt{1 - H^2 R^2}} \eqend{,} \quad r = \frac{R \, \mathe^{- H T}}{\sqrt{1 - H^2 R^2}} \eqend{.}
\end{splitequation}
Note that in the static patch, we have $- \eta > r$ as can be seen in Fig.~\ref{fig:desitter-poincare}.

\subsection{Modular data for the light cone}
\label{sec:desitter_modular_light_cone}

The metric of the expanding Poincaré patch~\eqref{eq:desitter_metric_poincare} is already in the conformally flat form \eqref{eq:conformal_metric}, and we read off that
\begin{equation}
\label{eq:desitter_omega_poincare}
\omega = - \ln (- H \eta) \eqend{.}
\end{equation}
We consider the future light cone with tip at $\eta = \tau$, which is the region
\begin{equation}
\label{eq:desitter_lightcone}
\futurecone = \left\{ (\eta,\vec{x})\colon \tau < \eta < 0, \eta-\tau > r = \abs{\vec{x}} \right\} \eqend{.}
\end{equation}

We recall that the smeared operators are defined as~\eqref{eq:conformal_phi_rescaling}
\begin{equation}
\label{eq:desitter_phi_rescaling}
\hat{\op}(f) \equiv \int \hat{\op}(x) f(x) \sqrt{-g} \total^d x = \int \hat{\op}(x) f(x) (- H \eta)^{-d} \total^d x \eqend{.}
\end{equation}
Specialising the general result~\eqref{eq:conformal_modular_light_cone_automorphism} to the expanding Poincaré patch, the candidate modular automorphism reads
\begin{equation}
\label{eq:desitter_lightcone_modularautomorphism}
\Delta_\futurecone^{\mathi s} \, \hat{\op}(f) \Delta_\futurecone^{- \mathi s} = \hat{\op}(f_{s,\futurecone})
\end{equation}
with
\begin{equation}
\label{eq:desitter_lightcone_flow_f}
f_{s,\futurecone}(\eta,\vec{x}) = \left[ 1 + \left( \mathe^{2 \pi s} - 1 \right) \frac{\tau}{\eta} \right]^{\Delta-d} f\left( \tau + \mathe^{-2 \pi s} (\eta - \tau), \mathe^{-2 \pi s} \vec{x} \right) \eqend{.}
\end{equation}
We recall that this result is valid for integer conformal dimension $\Delta$, for example $\Delta = (d-2)/2$ for the free scalar field in even dimensions $d$.

However, the candidate modular automorphism~\eqref{eq:desitter_lightcone_modularautomorphism} with the transformed test function~\eqref{eq:desitter_lightcone_flow_f} has a curious feature. Namely, there exists $s_\text{max}$ (depending on $f$) such that $f_{s,\futurecone} = 0$ for $s > s_\text{max}$, such that the automorphism is not defined for all $s \in \mathbb{R}$. To show this, let us assume that $f(x) = 0$ for $\eta < \eta_0$ inside $\futurecone$ such that $\tau < \eta_0 < 0$. Then we have $f_{s,\futurecone}(x) = 0$ for $\tau + \mathe^{-2 \pi s} (\eta - \tau) < \eta_0$, which translates to $2 \pi s > \ln\left[ (\eta - \tau)/(\eta_0 - \tau) \right]$. Since $\eta < 0$, that means that $2 \pi s_\text{max} = - \ln\left( 1 - \eta_0/\tau \right) > 0$. Strictly speaking, we therefore do not have determined the modular automorphism, since by Tomita--Takesaki theory this should be defined for all $s \in \mathbb{R}$. We only have some sort of modular inclusion since $\Delta_\futurecone^{\mathi s} \, \mathfrak{A} \Delta_\futurecone^{- \mathi s} \subset \mathfrak{A}$ for $s \leq 0$, whis is reminiscent of but different from the so-called half-sided modular inclusions~\cite{wiesbrock1993}.

Nevertheless, the analytic continuation $s \to - \mathi/2$ is well defined, and we have
\begin{equation}
\Delta_\futurecone^\frac{1}{2} \, \hat{\op}(f) \Delta_\futurecone^{- \frac{1}{2}} = \hat{\op}\left( f_{-\frac{\mathi}{2},\futurecone} \right) \quad\text{with}\quad f_{-\frac{\mathi}{2},\futurecone}(\eta,\vec{x}) = \left( 1 - \frac{2 \tau}{\eta} \right)^{\Delta-d} f(2\tau-\eta,-\vec{x}) \eqend{,}
\end{equation}
which always lies inside the expanding Poincaré patch. One would presume that this means that the modular conjugation is also well defined, such that specialising the general result~\eqref{eq:conformal_modular_light_cone_j} to the expanding Poincaré patch, we have
\begin{equation}
\label{eq:desitter_lightcone_j}
J_\futurecone \, \hat{\op}(f) J_\futurecone^{-1} = \hat{\op}^\dagger(f_J) \quad\text{with}\quad f_J(\eta, \vec{x}) = \left( 1 - \frac{2 \tau}{\eta} \right)^{\Delta-d} f^*(2\tau-\eta, -\vec{x}) \eqend{.}
\end{equation}
However, considering the action of the transformations contained in $J_\futurecone$~\eqref{eq:conformal_modular_light_cone_j_def} individually, one sees that the time reversal brings the test function $f$ outside of the expanding Poincaré patch. Since this invalidates the transformation, it is only possible to salvage the action~\eqref{eq:desitter_lightcone_j} of the modular conjugation by considering the full action of $J_\futurecone$ on Fock space, without breaking it up into the individual transformations. This is explicitly determined in appendix~\ref{app:modularfock}, and the result given in Eq.~\eqref{eq:app_modularfock_jfuturecone}.

For the past light cone, we obtain instead the candidate modular automorphism
\begin{equation}
\label{eq:desitter_pastlightcone_modularautomorphism}
\Delta_\pastcone^{\mathi s} \, \hat{\op}(f) \Delta_\pastcone^{- \mathi s} = \hat{\op}(f_{s,\pastcone})
\end{equation}
with the transformed test function
\begin{equation}
\label{eq:desitter_pastlightcone_flow_f}
f_{s,\pastcone}(\eta,\vec{x}) = \left[ 1 + \left( \mathe^{- 2 \pi s} - 1 \right) \frac{\tau}{\eta} \right]^{\Delta-d} f\left( \tau + \mathe^{2 \pi s} (\eta - \tau), \mathe^{2 \pi s} \vec{x} \right) \eqend{,}
\end{equation}
and we recall again that this result is valid for integer conformal dimension $\Delta$. Now there is no issue along the flow for real $s$, since $\eta - \tau$ is negative and thus $\tau + \mathe^{2 \pi s} (\eta - \tau) < \tau$ for any $s \in \mathbb{R}$, which therefore always lies inside the expanding Poincaré patch. However, the analytic continuation $s \to - \mathi/2$ which explicitly reads
\begin{equation}
\Delta_\pastcone^\frac{1}{2} \, \hat{\op}(f) \Delta_\pastcone^{- \frac{1}{2}} = \hat{\op}\left( f_{-\frac{\mathi}{2},\pastcone} \right) \quad\text{with}\quad f_{-\frac{\mathi}{2},\pastcone}(\eta,\vec{x}) = \left( 1 - \frac{2 \tau}{\eta} \right)^{\Delta-d} f(2\tau-\eta,-\vec{x})
\end{equation}
is now problematic, since for $\eta$ sufficiently large and negative we have $2 \tau - \eta > 0$. For the same reason, the modular conjugation
\begin{equation}
\label{eq:desitter_pastlightcone_j}
J_\pastcone \, \hat{\op}(f) J_\pastcone^{-1} = \hat{\op}^\dagger(f_J) \quad\text{with}\quad f_J(\eta, \vec{x}) = \left( 1 - \frac{2 \tau}{\eta} \right)^{\Delta-d} f^*(2\tau-\eta, -\vec{x})
\end{equation}
is problematic, even if we define it directly on Fock space~\eqref{eq:app_modularfock_jpastcone} as for the future light cone.

To resolve these problems, we have to extend the range of conformal time $\eta$ to include also positive values; that is, we adjoin the contracting Poincaré patch of de Sitter spacetime to the future of the expanding patch. While the metric~\eqref{eq:desitter_metric_poincare} itself obviously has a singularity at $\eta = 0$, this singularity resides only in the conformal factor, and is thus transparent to conformal fields. The corresponding Penrose diagram can be obtained by using the coordinate transformation~\eqref{eq:desitter_poincare_conformaltrafo}, which already covers the full range; the union of the contracting and expanding Poincaré patch is then conformal to the region $\{ (\sigma,\theta) \in (0,2\pi) \times [0,\pi) \colon \pi < \sigma + \theta < 2 \pi \} \cup \{ (\sigma,\theta) \in (0,2\pi) \times [0,\pi) \colon 0 < \sigma - \theta < \pi \}$.

\subsection{Modular data for the double cone}
\label{sec:desitter_modular_double_cone}

For the double cone, the situation is similar to the past light cone. Indeed, the double cone of size $\ell$ with center at $\eta = \tau$ is the region
\begin{equation}
\label{eq:desitter_diamond}
\doublecone = \left\{ (\eta,\vec{x})\colon r = \abs{\vec{x}} \in [0, \ell), \eta \in (\tau - \ell + r, \tau + \ell - r) \right\} \eqend{,}
\end{equation}
which for $\ell \leq \abs{\tau}$ lies fully in the expanding Poincaré patch. We thus make this restriction in the following.

\begin{figure}
\begin{minipage}[t]{0.48\textwidth}
\centering
\includegraphics[]{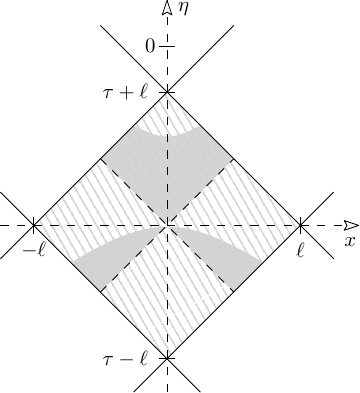}
\caption{The double cone in the expanding Poincaré patch of de Sitter space. In grey are the parts which are mapped to $\eta > 0$ by the action of the analytically continued modular automorphism, the adjoint action of $\Delta_\doublecone^\frac{1}{2}$. Compare Fig.~\ref{fig:diamond2}.}
\label{fig:diamond-ds-map}
\end{minipage}
\hfill
\begin{minipage}[t]{0.48\textwidth}
\centering
\includegraphics[]{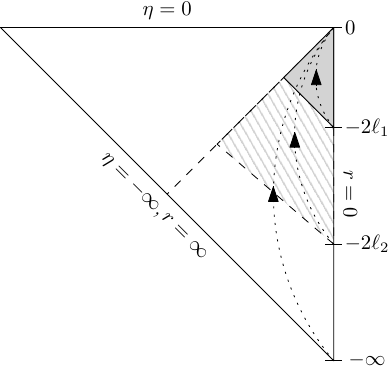}
\caption{A sequence of diamonds inside the static patch of de Sitter space with center at $\eta = - \ell_i$, upper tip at $\eta = 0$ and lower tip at $\eta = - 2 \ell_i$. The arrows indicate the action of the modular automorphism. Compare Fig.~\ref{fig:desitter-poincare}.}
\label{fig:desitter-modular}
\end{minipage}
\end{figure}

Specialising the general result~\eqref{eq:conformal_modular_double_cone_automorphism} with the transformed coordinates~\eqref{eq:minkowski_modular_double_cone_fs_coords2} to the expanding Poincaré patch, the candidate modular automorphism reads
\begin{equation}
\label{eq:desitter_doublecone_modularautomorphism}
\Delta_\doublecone^{\mathi s} \, \hat{\op}(f) \Delta_\doublecone^{- \mathi s} = \hat{\op}(f_{s,\doublecone}) \quad\text{with}\quad f_{s,\doublecone}(\eta, \vec{x}) = \left( \frac{r \, \eta_s}{\eta \, r_s} \right)^{\Delta-d} f\left( \eta_s, \vec{x}_s \right)
\end{equation}
and the transformed coordinates
\begin{splitequation}
\label{eq:desitter_doublecone_flow_coords}
\vec{x}_s &= \frac{2 \ell^2 \vec{x}}{\ell^2 - (\eta-\tau)^2 + r^2 + \left[ \ell^2 + (\eta-\tau)^2 - r^2 \right] \cosh\left( 2 \pi s \right) - 2 \ell (\eta-\tau) \sinh\left( 2 \pi s \right)} \eqend{,} \\
\eta_s &= \tau + \ell \frac{2 \ell (\eta-\tau) \cosh\left( 2 \pi s \right) - \left[ \ell^2 + (\eta-\tau)^2 - r^2 \right] \sinh\left( 2 \pi s \right)}{\ell^2 - (\eta-\tau)^2 + r^2 + \left[ \ell^2 + (\eta-\tau)^2 - r^2 \right] \cosh\left( 2 \pi s \right) - 2 \ell (\eta-\tau) \sinh\left( 2 \pi s \right)} \eqend{.}
\end{splitequation}
We again recall that this result is valid for integer conformal dimension $\Delta$, for example $\Delta = (d-2)/2$ for the free scalar field in even dimensions $d$. As $s \to \pm \infty$, we obtain
\begin{equation}
\vec{x}_s \to 0 \eqend{,} \quad \eta_s \to \tau \mp \ell \eqend{,}
\end{equation}
and the modular flow reaches the (future or past) tip of the double cone. The denominator of the transformed coordinates~\eqref{eq:desitter_doublecone_flow_coords} vanishes for
\begin{equation}
s = \frac{1}{2 \pi} \ln\left[ \frac{\eta - (\tau - \ell \pm r)}{(\tau + \ell \pm r) - \eta} \right] + \mathi \left( k + \frac{1}{2} \right) \eqend{,} \quad k \in \mathbb{Z} \eqend{,}
\end{equation}
where both signs in the logarithm are the same. Since in the diamond~\eqref{eq:desitter_diamond} we have $\eta > \tau - \ell \pm r$ and $\eta < \tau + \ell \pm r$, the argument of the logarithm is always positive, and we can analytically continue to $\Im s = - \mathi/2$ without encountering a singularity. The analytic continuation is thus well defined, and we obtain
\begin{equation}
\Delta_\doublecone^\frac{1}{2} \, \hat{\op}(f) \Delta_\doublecone^{- \frac{1}{2}} = \hat{\op}\left( f_{- \frac{\mathi}{2},\doublecone} \right)
\end{equation}
with
\begin{equation}
\label{eq:desitter_doublecone_flow_fcont}
f_{- \frac{\mathi}{2},\doublecone}(\eta, \vec{x}) = \left[ \frac{\tau [ \vec{x}^2 - (\eta-\tau)^2 ] - \ell^2 (\eta-\tau)}{\ell^2 \eta} \right]^{\Delta-d} f\left[ \tau - \frac{\ell^2 (\eta-\tau)}{\vec{x}^2 - (\eta-\tau)^2}, \frac{\ell^2 \vec{x}}{\vec{x}^2 - (\eta-\tau)^2} \right] \eqend{.}
\end{equation}
As in flat space, the support of $f_{- \frac{\mathi}{2},\doublecone}$ splits in three disconnected regions: the past light cone with tip at $(\eta,\vec{x}) = (\tau-\ell,\vec{0})$, the future light cone with tip at $(\eta,\vec{x}) = (\tau+\ell,\vec{0})$, and a region which is spacelike separated from the double cone. While the past light cone lies fully within the expanding Poincaré patch, parts of both the future light cone and the spacelike separated region extend to $\eta > 0$. In particular, the region $\{ (\eta,\vec{x}) \in \doublecone \colon 0 > \vec{x}^2 - (\eta-\tau)^2 > (\eta/\tau-1) \ell^2 \}$ of the double cone is mapped into the region $\{ (\eta,\vec{x}) \in \futurecone \colon \eta > 0 \}$ of the future light cone with tip at $(\eta,\vec{x}) = (\tau+\ell,\vec{0})$, and the region $\{ (\eta,\vec{x}) \in \doublecone \colon 0 < \vec{x}^2 - (\eta - \tau)^2 < (\eta/\tau-1) \ell^2 \}$ of the double cone is mapped into the region $\{ (\eta,\vec{x})\colon \abs{x} > \ell, (\abs{x}-\ell)^2 > (\eta-\tau)^2, \eta > 0 \}$. Those regions are depicted in Fig.~\ref{fig:diamond-ds-map}.

Therefore, as for the light cone we have to adjoin the contracting Poincaré patch of de Sitter spacetime to the future of the expanding patch to obtain a fully well-defined modular action. The modular conjugation is then also well defined, and specialising the general result~\eqref{eq:conformal_modular_double_cone_j} to the expanding Poincaré patch we obtain
\begin{equation}
J_\doublecone \, \hat{\op}(f) J_\doublecone^{-1} = \hat{\op}^\dagger(f_J) \eqend{,} \quad f_J(x) = \left[ f_{- \frac{\mathi}{2},\doublecone}(x) \right]^* \eqend{.}
\end{equation}

Lastly, let us consider the limit $\ell \to \infty$, where the diamond becomes the full static patch. We thus take a sequence of diamonds with $\tau = - \ell$ whose future tip coincides with the far future $r = \eta = 0$ of an observer at rest, and which are depicted in Fig.~\ref{fig:desitter-modular}. For such a sequence, the transformed coordinates~\eqref{eq:desitter_doublecone_flow_coords} of the modular automorphism~\eqref{eq:desitter_doublecone_modularautomorphism} simplify to
\begin{splitequation}
\label{eq:desitter_doublecone_flow_coords_future}
\vec{x}_s &= \frac{\ell^2 \vec{x}}{\ell^2 \, \mathe^{- 2 \pi s} + \ell \eta (\mathe^{- 2 \pi s} - 1) + (\eta^2-r^2) \sinh^2(\pi s)} \to \mathe^{2 \pi s} \vec{x} \quad (\ell \to \infty) \eqend{,} \\
\eta_s &= \frac{\ell^2 \eta - \ell (\eta^2-r^2) \mathe^{\pi s} \sinh(\pi s)}{\ell^2 \, \mathe^{- 2 \pi s} + \ell \eta (\mathe^{- 2 \pi s} - 1) + (\eta^2-r^2) \sinh^2(\pi s)} \to \mathe^{2 \pi s} \eta \quad (\ell \to \infty) \eqend{.}
\end{splitequation}
In the limit $\ell \to \infty$, we thus obtain a simple rescaling of the coordinates of the expanding Poincaré patch. The modular automorphism~\eqref{eq:desitter_doublecone_modularautomorphism} thus becomes in this limit
\begin{equation}
\Delta_\doublecone^{\mathi s} \, \hat{\op}(f) \Delta_\doublecone^{- \mathi s} = \hat{\op}(f_{s,\doublecone}) \quad\text{with}\quad f_{s,\doublecone}(\eta, \vec{x}) = f\left( \mathe^{2 \pi s} \eta, \mathe^{2 \pi s} \vec{x} \right) \eqend{,}
\end{equation}
and the analytic continuation $s \to - \mathi/2$ now results in
\begin{equation}
\Delta_\doublecone^\frac{1}{2} \, \hat{\op}(f) \Delta_\doublecone^{- \frac{1}{2}} = \hat{\op}\left( f_{- \frac{\mathi}{2},\doublecone} \right) \quad\text{with}\quad f_{- \frac{\mathi}{2},\doublecone}(\eta, \vec{x}) = f(-\eta, -\vec{x}) \eqend{.}
\end{equation}
(The same result is obtained by setting $\tau = -\ell$ in the result~\eqref{eq:desitter_doublecone_flow_fcont} and taking the limit $\ell \to \infty$ there.) We see that in this limit all of the diamond is mapped into the contracting Poincaré patch, and in fact into the corresponding diamond there, which is another static patch. We thus recover the known results for the full static patch.

To better interpret these results, we express them using the static coordinates~\eqref{eq:desitter_metric_static}, whose relation to the expanding Poincaré coordinates is given in~\eqref{eq:desitter_coordtrafo}. To keep the expressions manageable, we restrict to the above sequence of diamonds with $\tau = - \ell$. The diamond $\doublecone$ itself~\eqref{eq:desitter_diamond} in these coordinates is the region
\begin{equation}
\label{eq:desitter_diamond_static}
\doublecone = \left\{ (T,\vec{X})\colon H R = H \abs{\vec{X}} < 1, H T > - \ln\left( 2 H \ell \sqrt{\frac{1 - H R}{1 + H R}} \right) \right\} \eqend{,}
\end{equation}
which has a somewhat more complicated expression than the one in conformally flat coordinates~\eqref{eq:desitter_diamond}. The transformed test function of the modular automorphism~\eqref{eq:desitter_doublecone_modularautomorphism} reads
\begin{equation}
\label{eq:desitter_doublecone_modularautomorphism_static_func}
f_{s,\doublecone}(T,\vec{X}) = \left( 1 + \frac{\mathe^{- H T} \sqrt{1 - H^2 R^2} \, \left( \mathe^{2 \pi s} - 1 \right)}{2 H \ell} \right)^{\Delta-d} f\left( T_s, \vec{X}_s \right)
\end{equation}
with the transformed coordinates
\begin{splitequation}
\label{eq:desitter_doublecone_flow_coords_static}
\vec{X}_s &= \frac{2 H \ell \vec{X}}{2 H \ell + \mathe^{- H T} \sqrt{1 - H^2 R^2} \, \left( \mathe^{2 \pi s} - 1 \right)} \eqend{,} \\
T_s &= \frac{1}{2 H} \ln\left[ \mathe^{2 (H T - 2 \pi s)} + \frac{\left( 1 - \mathe^{- 2 \pi s} \right)}{H \ell \sqrt{1-H^2 R^2}} \mathe^{H T - 2 \pi s} + \frac{\left( 1 - \mathe^{- 2 \pi s} \right)^2}{4 H^2 \ell^2} \right] \eqend{.}
\end{splitequation}
As $s \to \infty$, we obtain
\begin{equation}
\vec{X}_s \to \vec{0} \eqend{,} \quad T_s \to - \frac{1}{H} \ln\left( 2 H \ell \right) \eqend{,}
\end{equation}
which is the lower tip of the diamond~\eqref{eq:desitter_diamond_static}, and as $s \to - \infty$ we have
\begin{equation}
\vec{X}_s \to \frac{2 H \ell \vec{X}}{2 H \ell - \mathe^{- H T} \sqrt{1 - H^2 R^2}} \eqend{,} \quad T_s \to \infty \eqend{,}
\end{equation}
which is the far future. That is, in static coordinates the modular flow of the spatial coordinates saturates at a finite value, which is reasonable since for the sequence of diamonds~\eqref{eq:desitter_diamond_static} that we consider, the tip of the diamond includes the whole future of an observer at rest. In conformally flat coordinates however, all points with finite $R$ tend to the same point as $T \to \infty$, namely the origin $r = 0$ as can be seen from the relation between the conformally flat and static coordinates~\eqref{eq:desitter_coordtrafo}.

We then again consider the limit $\ell \to \infty$, where the diamond becomes the full static patch. From the results~\eqref{eq:desitter_doublecone_modularautomorphism_static_func} and~\eqref{eq:desitter_doublecone_flow_coords_static}, we easily obtain the transformed test function
\begin{equation}
\label{eq:desitter_modular_double_cone_flow_staticpatch}
f_{s,\doublecone}(T,\vec{X}) = f\left( T - 2 \pi H^{-1} s, \vec{X} \right) \eqend{,}
\end{equation}
and we see that in this limit the modular automorphism becomes a time translation (in static time). Moreover, since the parameter $s$ is normalized such that the KMS condition~\eqref{eq:intro_kms} is fulfilled with an inverse temperature $\beta = 1$, the result~\eqref{eq:desitter_modular_double_cone_flow_staticpatch} shows that the inverse temperature of the de~Sitter vacuum state restricted to the static patch is equal to $\beta_\text{dS} = 2 \pi H^{-1}$ when we normalize with respect to time translations, such that we recover the known results for the full static patch.

Finally, let us determine explicitly the modular Hamiltonian $\ln \Delta_\doublecone$ in static coordinates. For the family of diamonds~\eqref{eq:desitter_diamond_static} with $\tau = - \ell$ that we consider, the modular Hamiltonian~\eqref{eq:minkowski_modular_double_cone_lndelta} reduces to $\ln \Delta_\doublecone = - 2 \pi D + \pi/\ell \, K$. From the transformation laws~\eqref{eq:conformal_hdk_phi_commutator}, specialized to the expanding Poincaré patch of de~Sitter space with $\omega = - \ln(- H \eta)$~\eqref{eq:desitter_omega_poincare}, we obtain $\mathi \left[ \ln \Delta_\doublecone, \hat{\op}(f) \right] = \hat{\op}(\hat{f})$ with
\begin{equation}
\hat{f}(x) = 2 \pi x^\mu \partial_\mu f(x) + \frac{\pi}{\ell} \left[ 2 \eta \vec{x}^i \partial_i + ( \vec{x}^2 + \eta^2 ) \partial_\eta - ( \vec{x}^2 - \eta^2 ) \frac{d-\Delta}{\eta} \right] f(x) \eqend{,}
\end{equation}
and we recall that $\Delta$ is the conformal dimension of $\hat{\op}$, equal to $(d-2)/2$ for a free scalar field. Taking the limit where $f(x) \to (- H \eta)^d \delta(\eta-\eta_0) \delta^{d-1}(\vec{x}-\vec{x}_0)$ for some point $x_0 = (\eta_0,\vec{x}_0)$ such that $\hat{\op}(f) \to \hat{\op}(x_0)$~\eqref{eq:desitter_phi_rescaling}, a short computation shows that
\begin{equation}
\mathi \left[ \ln \Delta_\doublecone, \hat{\op}(x_0) \right] = - 2 \pi x_0^\mu \partial_\mu \hat{\op}(x_0) - \frac{\pi}{\ell} \left( 2 \eta_0 x_0^\mu \partial_\mu + x_0^2 \partial_\eta - \Delta \frac{x_0^2}{\eta_0} \right) \hat{\op}(x_0) \eqend{.}
\end{equation}
We can now transform this expression to static coordinates using the coordinate transformation~\eqref{eq:desitter_coordtrafo}, and (renaming $x_0 \to x$) it follows that
\begin{equation}
\label{eq:desitter_modular_double_cone_hamiltonian_static}
\mathi \left[ \ln \Delta_\doublecone, \hat{\op}(x) \right] = \frac{2 \pi}{H} \partial_T \hat{\op}(x) - \frac{\pi}{H^2 \ell} \frac{\mathe^{- H T}}{\sqrt{1 - H^2 R^2}} \left[ \partial_T - H \left( 1 - H^2 R^2 \right) \left( R \partial_R + \Delta \right) \right] \hat{\op}(x) \eqend{.}
\end{equation}
Clearly, in the limit $\ell \to \infty$ as the diamonds become big, the second term vanishes and the modular Hamiltonian is the generator of time translations in the static patch, up to a rescaling which fixes the inverse temperature of the de~Sitter vacuum state restricted to the static patch to $\beta_\text{dS} = 2 \pi H^{-1}$. For diamonds of finite size, we have instead a correction term which also depends on the position of the operator.

\section{Conclusions}
\label{sec:conclusions}

We have determined the full Tomita--Takesaki modular data --- modular Hamiltonian, modular flow and modular conjugation --- for a conformal scalar field in conformally flat spacetimes in the conformal vacuum state, restricted to light cones and double cones. We have then specialized our results to de Sitter spacetime, and in particular derived the modular Hamiltonian $\ln \Delta_\doublecone$ for diamonds of size $\ell$ inside the static patch, Eq.~\eqref{eq:desitter_modular_double_cone_hamiltonian_static}. In the limit $\ell \to \infty$, where the diamonds fill the full static patch, the modular Hamiltonian becomes (up to a rescaling) the generator of time translations (as was known before), but for finite size it is more complicated.

An important application of the modular Hamiltonian is in the computation of the relative entropy between two ``excited'' states relative to a ``vacuum'' state, which can be done using the modular Hamiltonian for the ``vacuum'' state via the Araki formula~\eqref{eq:intro_araki2}. With the explicit expression for the modular Hamiltonian, we can now compute this entropy exactly, at least for conformal fields~\cite{dangeloetal2023}. The results can then shed further light on the subject of the entropy of de Sitter space, and in particular on the proposal that empty de Sitter space has maximum entropy. Namely, it has been argued that the addition of any matter with its own entropy $\mathcal{S}_\text{M}$ has the effect of decreasing the horizon area $A_\text{dS}$, in such a way that the generalized entropy $A_\text{dS}/(4 G_\text{N}) + \mathcal{S}_\text{M}$ decreases, see for example Refs.~\cite{giddingsmarolf2007,banihashemijacobsonsveskovisser2023,balasubramaniannomuraugajin2023}. In the context of the de~Sitter holographic correspondence~\cite{strominger2001,alishahihakarchsilversteintong2004,alishahihakarchsilverstein2005}, the horizon entropy can be understood as the entanglement entropy between the right and left dual conformal field theories (CFTs) that appear in the correspondence, or between the past and future dual CFTs~\cite{nguyen2017,narayan2018,dongsilversteintorroba2018,narayan2019,genggrieningerkarch2019,ariasdiazsundell2020,geng2020,geng2021}, and then the variation of the generalized entropy can be studied by considering the variation of the entanglement entropy of the CFTs when matter is added to de~Sitter space. However, it should in principle be possible to understand this issue from a purely field-theoretic point of view without assuming holography, but taking into account the backreaction of the matter on the geometry. This is of course a difficult problem; for a concrete model of de~Sitter space including the gravitational dynamics of an observer see~\cite{chandrasekaranlongopeningtonwitten2023,balasubramaniannomuraugajin2023}.

Further applications arise in the study of thermodynamics of causal diamonds~\cite{deboerhaehlhellermyers2016}, and in particular in the definition of local temperatures~\cite{buchholzschlemmer2007,solveen2010,solveen2012,sanders2017,ariasblancocasinihuerta2017,mistrypinzul2023}. Indeed, Tomita--Takesaki theory shows that the state $\Omega$ is a KMS state with respect to the time evolution generated by the modular Hamiltonian, and by relating this time evolution to the one generated by the true Hamiltonian of the theory one can define a local temperature~\cite{connesrovelli1994}. In flat space, this connection has been explored for wedges and diamonds~\cite{longomartinettirehren2010,martinettirovelli2003}, while in de~Sitter space only results for wedges are available~\cite{froebmuchpapadopoulos2023}. With our results, this can now be generalized to diamonds in de~Sitter space.

It would be interesting to see whether the method presented here also works in more general cases, where the modular flow is non-local. Examples where the flow is known explicitly include chiral scalars and fermions in two dimensions, restricted to regions which are not simply connected or in spacetimes with non-trivial topology~\cite{casinihuerta2009a,longomartinettirehren2010,rehrentedesco2013,ariascasinihuertapontello2018,hollands2021,blancopereznadal2019,friesreyes2019a,friesreyes2019b,erdmengerfriesreyessimon2020,mintchevtonni2021a,mintchevtonni2021b,gravakelstonni2021,eislertonnipeschel2022,abateblancokoifmanpereznadal2023}. Since these are conformal fields, in principle it should be possible to generalize our method also to these cases, even though the details of the computation are probably intricate. We thus leave such a computation to future work.

\begin{acknowledgments}
This work has been funded by the Deutsche Forschungsgemeinschaft (DFG, German Research Foundation) --- project no. 396692871 within the Emmy Noether grant CA1850/1-1 and project no. 406116891 within the Research Training Group RTG 2522/1. It is a pleasure to thank Christoph Minz, Albert Much, Alessio Ranallo, Leonardo Sangaletti and Alexander Stottmeister for discussions about modular theory and the KMS condition, Stefano Galanda for pointing out a sign mistake, and Rainer Verch for useful comments and references.
\end{acknowledgments}

\appendix

\section{Lorentzian inversions}
\label{app:inversion}

In this appendix, we want to show that the operator $V$ acting on the Fock space $\mathcal{F}$ of the free massless scalar field and defined by~\eqref{eq:minkowski_modular_double_cone_inversion}
\begin{equation}
V a(f) V^{-1} = a(f_V) \eqend{,} \quad V a^\dagger(f) V^{-1} = a^\dagger(f_V) \eqend{,} \quad V \Omega = \Omega
\end{equation}
with~\eqref{eq:minkowski_modular_double_cone_inversion_fv}
\begin{equation}
\label{eq:app_inversion_tilde_fv}
\widetilde{f_V}(\vec{p}) = \int \sqrt{ \frac{q^0}{p^0} } \, \tilde{f}(\vec{q}) \int \left( \frac{\ell^2}{\vec{x}^2} \right)^\frac{d-2}{2} \mathe^{\mathi \vec{q} \vec{x} - \mathi \ell^2 (\vec{p} \vec{x})/\vec{x}^2} \total^{d-1} \vec{x} \frac{\total^{d-1} \vec{q}}{(2\pi)^{d-1}} \eqend{.}
\end{equation}
indeed results in a Lorentzian inversion for the field $\phi$, and moreover is unitary. This was first derived by Swieca and V{\"o}lkel~\cite{swiecavoelkel1973}; with hindsight we can streamline the computations somewhat. Throughout this section, we assume that the scaling dimension $\Delta = (d-2)/2$ of the scalar field $\phi$ is an integer, i.e., that $d$ is even.

We first note that from the definition~\eqref{eq:minkowski_phi_def} of $\phi$ we have
\begin{equation}
V \phi(f) V^{-1} = V a(F_-) V^{-1} + V a^\dagger(F_+) V^{-1} = a(F^V_-) + a^\dagger(F^V_+)
\end{equation}
with
\begin{splitequation}
\widetilde{F^V_+}(\vec{p}) &= \frac{1}{\sqrt{2 p^0}} \int \left. \widetilde{f}(q) \right\rvert_{q^0 = \abs{\vec{q}}} \int \left( \frac{\ell^2}{\vec{x}^2} \right)^\frac{d-2}{2} \mathe^{\mathi \vec{q} \vec{x} - \mathi \ell^2 (\vec{p} \vec{x})/\vec{x}^2} \total^{d-1} \vec{x} \frac{\total^{d-1} \vec{q}}{(2\pi)^{d-1}} \\
&= \frac{1}{\sqrt{2 p^0}} \int f(y) \int \mathe^{\mathi \abs{\vec{q}} y^0 - \mathi \vec{q} \vec{y}} \int \left( \frac{\ell^2}{\vec{x}^2} \right)^\frac{d-2}{2} \mathe^{\mathi \vec{q} \vec{x} - \mathi \ell^2 (\vec{p} \vec{x})/\vec{x}^2} \total^{d-1} \vec{x} \frac{\total^{d-1} \vec{q}}{(2\pi)^{d-1}} \total^d y
\end{splitequation}
and
\begin{equation}
\widetilde{F^V_-}(\vec{p}) = \frac{1}{\sqrt{2 p^0}} \int f^*(y) \int \mathe^{\mathi \abs{\vec{q}} y^0 - \mathi \vec{q} \vec{y}} \int \left( \frac{\ell^2}{\vec{x}^2} \right)^\frac{d-2}{2} \mathe^{\mathi \vec{q} \vec{x} - \mathi \ell^2 (\vec{p} \vec{x})/\vec{x}^2} \total^{d-1} \vec{x} \frac{\total^{d-1} \vec{q}}{(2\pi)^{d-1}} \total^d y \eqend{.}
\end{equation}

We then perform the integral over $\vec{q}$, the Fourier transform of $\mathe^{\mathi \abs{\vec{q}} y^0}$:
\begin{splitequation}
\label{eq:app_inversion_fourier_pre}
\int \mathe^{\mathi \abs{\vec{q}} a - \mathi \vec{q} \vec{b}} \frac{\total^{d-1} \vec{q}}{(2\pi)^{d-1}} &= \frac{\Omega_{d-3}}{(2\pi)^{d-1}} \int_0^\infty \int_0^\pi \mathe^{\mathi q a - \mathi q \abs{\vec{b}} \cos \theta} \sin^{d-3} \theta \, q^{d-2} \total \theta \total q \\
&= (2\pi)^\frac{1-d}{2} \abs{\vec{b}}^\frac{3-d}{2} \lim_{\epsilon \to 0^+} \int_0^\infty \mathe^{- \epsilon q + \mathi a q} q^\frac{d-1}{2} \bessel{J}{\frac{d-3}{2}}{\abs{\vec{b}} q} \total q \eqend{,}
\end{splitequation}
where $\Omega_N = 2 \pi^\frac{N+1}{2}/\Gamma\left( \frac{N+1}{2} \right)$ is the area of the unit $N$-sphere, and $\bessel{J}{\nu}{z}$ is the Bessel function of order $\nu$ obtained from the integral~\cite[Eq.~(10.9.4)]{dlmf}. Here we have introduced a regulator $\mathe^{- \epsilon q}$ to ensure convergence of the integral at infinity, which means that we compute this Fourier transform as a tempered distribution. The integral over $q$ can be performed using~\cite[Eq.~(10.22.49)]{dlmf}, and we obtain
\begin{equation}
\label{eq:app_inversion_fourier}
\int \mathe^{\mathi \abs{\vec{q}} a - \mathi \vec{q} \vec{b}} \frac{\total^{d-1} \vec{q}}{(2\pi)^{d-1}} = - \mathi a \pi^{- \frac{d}{2}} \Gamma\left( \frac{d}{2} \right) \lim_{\epsilon \to 0^+} \left[ - (a + \mathi \epsilon)^2 + \vec{b}^2 \right]^{-\frac{d}{2}} \eqend{.}
\end{equation}
It follows that
\begin{splitequation}
\widetilde{F^V_+}(\vec{p}) = \frac{- \mathi}{\sqrt{2 p^0}} \pi^{- \frac{d}{2}} \Gamma\left( \frac{d}{2} \right) \lim_{\epsilon \to 0^+} &\int f(y) \int y^0 \left[ - (y^0 + \mathi \epsilon)^2 + (\vec{y}-\vec{x})^2 \right]^{-\frac{d}{2}} \\
&\quad\times \left( \frac{\ell^2}{\vec{x}^2} \right)^\frac{d-2}{2} \mathe^{- \mathi \ell^2 (\vec{p} \vec{x})/\vec{x}^2} \total^{d-1} \vec{x} \total^d y
\end{splitequation}
and the analogous expression for $\widetilde{F^V_-}(\vec{p})$ with $f$ replaced by $f^*$. Performing now the change of variables $\vec{x} \to \vec{z} = \ell^2 \vec{x}/\vec{x}^2$, an inversion whose Jacobian reads
\begin{equation}
\label{eq:app_inversion_inv_jacobian}
\abs{ \det\left( \frac{\partial \vec{x}^i}{\partial \vec{z}^j} \right)_{i,j=1}^{d-1} } = \left( \frac{\ell^2}{\vec{z}^2} \right)^{d-1} = \left( \frac{\vec{x}^2}{\ell^2} \right)^{d-1} \eqend{,}
\end{equation}
and the Lorentzian inversion
\begin{equation}
\label{eq:app_inversion_lorentzian_inv}
y^0 \to w^0 = \frac{\ell^2 y^0}{\vec{y}^2 - (y^0)^2} \eqend{,} \quad \vec{y} \to \vec{w} = \frac{\ell^2 \vec{y}}{\vec{y}^2 - (y^0)^2} \eqend{,}
\end{equation}
whose Jacobian reads (recall that $d$ is even)
\begin{equation}
\abs{ \det\left( \frac{\partial y^\mu}{\partial w^\nu} \right)_{i,j=1}^d } = \left( \frac{\ell^2}{\vec{w}^2 - (w^0)^2} \right)^d = \left( \frac{\vec{y}^2 - (y^0)^2}{\ell^2} \right)^d \eqend{,}
\end{equation}
this reduces to
\begin{splitequation}
\widetilde{F^V_+}(\vec{p}) &= \frac{- \mathi}{\sqrt{2 p^0}} \pi^{- \frac{d}{2}} \Gamma\left( \frac{d}{2} \right) \lim_{\epsilon \to 0^+} \int f\left( \frac{\ell^2 w^0}{\vec{w}^2 - (w^0)^2}, \frac{\ell^2 \vec{w}}{\vec{w}^2 - (w^0)^2} \right) \\
&\quad\times \int \frac{\ell^2 w^0}{\vec{w}^2 - (w^0)^2} \left[ - \left( \frac{\ell^2 w^0}{\vec{w}^2 - (w^0)^2} + \mathi \epsilon \right)^2 + \left( \ell^2 \frac{\vec{w}}{\vec{w}^2 - (w^0)^2} - \ell^2 \frac{\vec{z}}{\vec{z}^2} \right)^2 \right]^{-\frac{d}{2}} \\
&\qquad\times \left( \frac{\ell^2}{\vec{w}^2 - (w^0)^2} \right)^d \left( \frac{\ell^2}{\vec{z}^2} \right)^\frac{d}{2} \mathe^{- \mathi \vec{p} \vec{z}} \total^{d-1} \vec{z} \total^d w \eqend{,} \raisetag{2.2em}
\end{splitequation}
and the analogous expression for $\widetilde{F^V_-}(\vec{p})$ with $f$ replaced by $f^*$. Expanding the term in brackets and using that in the limit $\epsilon \to 0^+$ only the sign (but not the magnitude) of $\epsilon$ is relevant, we can further simplify this and obtain
\begin{splitequation}
\widetilde{F^V_+}(\vec{p}) &= \frac{- \mathi}{\sqrt{2 p^0}} \pi^{- \frac{d}{2}} \Gamma\left( \frac{d}{2} \right) \lim_{\epsilon \to 0^+} \int f\left( \frac{\ell^2 w^0}{\vec{w}^2 - (w^0)^2}, \frac{\ell^2 \vec{w}}{\vec{w}^2 - (w^0)^2} \right) \\
&\quad\times \int w_0 \left[ (\vec{w} - \vec{z})^2 - (w^0 + \mathi \epsilon)^2 \right]^{-\frac{d}{2}} \left( \frac{\ell^2}{\vec{w}^2 - (w^0)^2} \right)^\frac{d+2}{2} \mathe^{- \mathi \vec{p} \vec{z}} \total^{d-1} \vec{z} \total^d w \eqend{,}
\end{splitequation}
and the analogous expression for $\widetilde{F^V_-}(\vec{p})$ with $f$ replaced by $f^*$.

Inverting the Fourier transform~\eqref{eq:app_inversion_fourier}, we have
\begin{equation}
\lim_{\epsilon \to 0^+} \int \left[ - (a + \mathi \epsilon)^2 + \vec{b}^2 \right]^{-\frac{d}{2}} \mathe^{\mathi \vec{p} \vec{b}} \total^{d-1} \vec{b} = \frac{\mathi \pi^\frac{d}{2}}{a \Gamma\left( \frac{d}{2} \right)} \mathe^{\mathi \abs{\vec{p}} a} \eqend{,}
\end{equation}
and thus
\begin{equation}
\widetilde{F^V_+}(\vec{p}) = \frac{1}{\sqrt{2 p^0}} \int f\left( \frac{\ell^2 w^0}{\vec{w}^2 - (w^0)^2}, \frac{\ell^2 \vec{w}}{\vec{w}^2 - (w^0)^2} \right) \left( \frac{\ell^2}{\vec{w}^2 - (w^0)^2} \right)^\frac{d+2}{2} \mathe^{\mathi \abs{\vec{p}} w^0 - \mathi \vec{p} \vec{w}} \total^d w \eqend{,}
\end{equation}
which is of the form~\eqref{eq:minkowski_phi_def} with
\begin{equation}
\label{eq:app_inversion_finv}
f(x) \to f_V(x) = f\left( \frac{\ell^2 x^0}{\vec{x}^2 - (x^0)^2}, \frac{\ell^2 \vec{x}}{\vec{x}^2 - (x^0)^2} \right) \left( \frac{\ell^2}{\vec{x}^2 - (x^0)^2} \right)^{d-\Delta} \eqend{.}
\end{equation}
The analogous result is obtained for $\widetilde{F^V_-}(\vec{p})$, such that indeed $V \phi(f) V^{-1} = \phi(f_V)$. Since the inversion~\eqref{eq:app_inversion_finv} squares to the identity, we also have $V^{-1} = V$.

To show that $V$ is a unitary operator on Fock space, it is enough to show that it does preserve the scalar product between states of the form $a^\dagger(f_1) \cdots a^\dagger(f_k) \Omega \in \mathcal{F}$ with finite $k$ and Schwartz functions $f_j \in \mathcal{S}(\mathbb{R}^{d-1})$ for all $j$, since they generate the dense subset $\mathcal{D} \subset \mathcal{F}$, and $V$ can then be extended to the full Fock space by continuity. We compute
\begin{splitequation}
&\left( a^\dagger(f_1) \cdots a^\dagger(f_k) \Omega, V^\dagger V a^\dagger(g_1) \cdots a^\dagger(g_j) \Omega \right) = \left( V a^\dagger(f_1) \cdots a^\dagger(f_k) \Omega, V a^\dagger(g_1) \cdots a^\dagger(g_j) \Omega \right) \\
&= \left( a^\dagger(f_1^V) \cdots a^\dagger(f_k^V) \Omega, a^\dagger(g_1^V) \cdots a^\dagger(g_j^V) \Omega \right) = \left( \Omega, a(f_k^V) \cdots a(f_1^V) a^\dagger(g_1^V) \cdots a^\dagger(g_j^V) \Omega \right) \eqend{,}
\end{splitequation}
where we used that $V \Omega = \Omega$. Using the commutation relations~\eqref{eq:minkowski_aadagger}, this gives
\begin{equation}
\left( a^\dagger(f_1) \cdots a^\dagger(f_k) \Omega, V^\dagger V a^\dagger(g_1) \cdots a^\dagger(g_j) \Omega \right) = \delta_{jk} \sum_\pi \prod_{i=1}^k \left( f_i^V, g_{\pi(i)}^V \right) \eqend{,}
\end{equation}
where the sum runs over all permutations $\pi$ of $1,\ldots,k$. To show that $V^\dagger V = \1$ and thus $V^\dagger = V^{-1} = V$ on $\mathcal{D}$ and by continuity on the full Fock space $\mathcal{F}$, we have to prove that this is equal to
\begin{equation}
\left( a^\dagger(f_1) \cdots a^\dagger(f_k) \Omega, a^\dagger(g_1) \cdots a^\dagger(g_j) \Omega \right) = \delta_{jk} \sum_\pi \prod_{i=1}^k \left( f_i, g_{\pi(i)} \right) \eqend{,}
\end{equation}
which follows if $(f,g) = (f_V,g_V)$ for arbitrary Schwartz functions $f$ and $g$. Using the definition~\eqref{eq:app_inversion_tilde_fv} of $f_V$, we obtain
\begin{splitequation}
(f_V,g_V) &= \int f_V^*(\vec{x}) g_V(\vec{x}) \total^{d-1} \vec{x} = \int \left[ \widetilde{f_V}(\vec{p}) \right]^* \widetilde{g_V}(\vec{p}) \frac{\total^{d-1} \vec{p}}{(2\pi)^{d-1}} \\
&= \int \iint \sqrt{ \frac{\abs{\vec{q}}}{\abs{\vec{p}}} } \, \left[ \tilde{f}(\vec{q}) \right]^* \left( \frac{\ell^2}{\vec{x}^2} \right)^\frac{d-2}{2} \mathe^{- \mathi \vec{q} \vec{x} + \mathi \ell^2 (\vec{p} \vec{x})/\vec{x}^2} \total^{d-1} \vec{x} \frac{\total^{d-1} \vec{q}}{(2\pi)^{d-1}} \\
&\quad\times \iint \sqrt{ \frac{\abs{\vec{k}}}{\abs{\vec{p}}} } \, \tilde{g}(\vec{k}) \left( \frac{\ell^2}{\vec{y}^2} \right)^\frac{d-2}{2} \mathe^{\mathi \vec{k} \vec{y} - \mathi \ell^2 (\vec{p} \vec{y})/\vec{y}^2} \total^{d-1} \vec{y} \frac{\total^{d-1} \vec{k}}{(2\pi)^{d-1}} \frac{\total^{d-1} \vec{p}}{(2\pi)^{d-1}} \eqend{.}
\end{splitequation}
We first perform the integral over $\vec{p}$, for which we compute analogously to Eqs.~\eqref{eq:app_inversion_fourier_pre} and~\eqref{eq:app_inversion_fourier} that
\begin{equation}
\label{eq:app_inversion_fourier_2}
\int \abs{\vec{q}}^{-1} \mathe^{- \mathi \vec{q} \vec{b}} \frac{\total^{d-1} \vec{q}}{(2\pi)^{d-1}} = \frac{1}{2} \pi^{-\frac{d}{2}} \Gamma\left( \frac{d-2}{2} \right) \left( \vec{b}^2 \right)^{1-\frac{d}{2}} \eqend{,}
\end{equation}
such that
\begin{splitequation}
(f_V,g_V) &= \frac{1}{2} \pi^{-\frac{d}{2}} \Gamma\left( \frac{d-2}{2} \right) \iiiint \sqrt{ \abs{\vec{q}} \abs{\vec{k}} } \, \left[ \tilde{f}(\vec{q}) \right]^* \mathe^{- \mathi \vec{q} \vec{x}} \tilde{g}(\vec{k}) \, \mathe^{\mathi \vec{k} \vec{y}} \left[ \left( \frac{\ell^2 \vec{x}}{\vec{x}^2} - \frac{\ell^2 \vec{y}}{\vec{y}^2} \right)^2 \right]^{1-\frac{d}{2}} \\
&\hspace{8em}\times \left( \frac{\ell^2}{\vec{x}^2} \right)^\frac{d-2}{2} \left( \frac{\ell^2}{\vec{y}^2} \right)^\frac{d-2}{2} \total^{d-1} \vec{x} \total^{d-1} \vec{y} \frac{\total^{d-1} \vec{q}}{(2\pi)^{d-1}} \frac{\total^{d-1} \vec{k}}{(2\pi)^{d-1}} \\
&= \frac{1}{2} \pi^{-\frac{d}{2}} \Gamma\left( \frac{d-2}{2} \right) \iiiint \sqrt{ \abs{\vec{q}} \abs{\vec{k}} } \, \left[ \tilde{f}(\vec{q}) \right]^* \mathe^{- \mathi \vec{q} \vec{x}} \tilde{g}(\vec{k}) \, \mathe^{\mathi \vec{k} \vec{y}} \\
&\hspace{8em}\times \left[ \left( \vec{x} - \vec{y} \right)^2 \right]^{-\frac{d-2}{2}} \total^{d-1} \vec{x} \total^{d-1} \vec{y} \frac{\total^{d-1} \vec{q}}{(2\pi)^{d-1}} \frac{\total^{d-1} \vec{k}}{(2\pi)^{d-1}} \eqend{.} \raisetag{6em}
\end{splitequation}
We can now use the Fourier transform~\eqref{eq:app_inversion_fourier_2} again and perform the integrals over $\vec{x}$ and $\vec{y}$ to obtain
\begin{splitequation}
(f_V,g_V) &= \iiiint \sqrt{ \abs{\vec{q}} \abs{\vec{k}} } \, \left[ \tilde{f}(\vec{q}) \right]^* \mathe^{- \mathi \vec{q} \vec{x}} \tilde{g}(\vec{k}) \, \mathe^{\mathi \vec{k} \vec{y}} \\
&\qquad\times \int \abs{\vec{p}}^{-1} \mathe^{- \mathi \vec{p} (\vec{x}-\vec{y})} \frac{\total^{d-1} \vec{p}}{(2\pi)^{d-1}} \total^{d-1} \vec{x} \total^{d-1} \vec{y} \frac{\total^{d-1} \vec{q}}{(2\pi)^{d-1}} \frac{\total^{d-1} \vec{k}}{(2\pi)^{d-1}} \\
&= \int \left[ \tilde{f}(\vec{p}) \right]^* \tilde{g}(\vec{p}) \frac{\total^{d-1} \vec{p}}{(2\pi)^{d-1}} = (f,g)
\end{splitequation}
as required.

\section{Modular data in Fock space}
\label{app:modularfock}

In this appendix, we want to derive the modular automorphisms $\sigma_s$ as well as the action of the modular conjugation $J$ directly in Fock space $\mathcal{F}$. For this, we have to first determine the adjoint action~\eqref{eq:minkowski_ulambda_phi} for creation and annihilation operators $a$ and $a^\dagger$ instead of the field $\phi$. Using the definition~\eqref{eq:minkowski_q_def} of $U(\lambda)$ and the commutators~\eqref{eq:minkowski_hdk_commutator}, from the adjoint action
\begin{equation}
U(\lambda) a^\dagger(f) U(-\lambda) = a^\dagger(f_\lambda)
\end{equation}
we obtain the first-order differential equation
\begin{equation}
\label{eq:app_modularfock_ode_f}
\partial_\lambda f_\lambda(\vec{x}) = \mathi a (f_H)_\lambda(\vec{x}) + 2 \mathi b (f_D)_\lambda(\vec{x}) + \mathi c (f_K)_\lambda(\vec{x}) \eqend{,}
\end{equation}
which can now be solved for $f_\lambda$. For this, we make the ansatz
\begin{equation}
\label{eq:app_modularfock_ansatz_f}
f_\lambda(\vec{x}) = \int K_\lambda(\vec{x},\vec{y}) f(\vec{y}) \total^{d-1} \vec{y} = \int \tilde{K}_\lambda(\vec{x},\vec{p}) \tilde{f}(\vec{p}) \frac{\total^{d-1} \vec{p}}{(2\pi)^{d-1}}
\end{equation}
with a kernel $K_\lambda$ and the partial Fourier transform
\begin{equation}
\tilde{K}_\lambda(\vec{x},\vec{p}) \equiv \int K_\lambda(\vec{x},\vec{y}) \, \mathe^{\mathi \vec{p} \vec{y}} \total^{d-1} \vec{y} \eqend{.}
\end{equation}
The Fourier transforms of $f_H$, $f_D$ and $f_K$ are given in Eqs.~\eqref{eq:minkowski_hdk_commutator}, and the differential equation~\eqref{eq:app_modularfock_ode_f} turns into one for the kernel $K_\lambda$ that reads
\begin{splitequation}
\label{eq:app_modularfock_ode_k}
\partial_\lambda \tilde{K}_\lambda(\vec{x},\vec{p}) &= \mathi a p^0 \tilde{K}_\lambda(\vec{x},\vec{p}) - 2 b \left( \vec{p}^i \partial_{\vec{p}^i} + \frac{d-1}{2} \right) \tilde{K}_\lambda(\vec{x},\vec{p}) \\
&\quad+ \mathi c \left( p^0 \laplace_{\vec{p}} + \frac{\vec{p}^i}{p^0} \partial_{\vec{p}^i} + \frac{2d-5}{4 p^0} \right) \tilde{K}_\lambda(\vec{x},\vec{p}) \eqend{.}
\end{splitequation}
The initial condition for this equation is obtained by requiring that $f_0(\vec{x}) = f(\vec{x})$, which results in $\tilde{K}_0(\vec{x},\vec{p}) = \mathe^{\mathi \vec{p} \vec{x}}$.

To solve the differential equation~\eqref{eq:app_modularfock_ode_k}, we see that we can make the ansatz that $\tilde{K}_\lambda(\vec{x},\vec{p})$ is a function of $\vec{p} \vec{x}$ and $p^0$ only, $\tilde{K}_\lambda(\vec{x},\vec{p}) = \kappa_\lambda(\vec{p} \vec{x}, p^0)$. We then obtain
\begin{splitequation}
\partial_\lambda \kappa_\lambda(p,q) &= \mathi a q \kappa_\lambda(p,q) - 2 b \left( p \partial_p + q \partial_q + \frac{d-1}{2} \right) \kappa_\lambda(p,q) \\
&\quad+ \mathi c \left[ q \vec{x}^2 \partial_p^2 + 2 p \partial_q \partial_p + q \partial_q^2 + (d-1) \partial_q + \frac{p}{q} \partial_p + \frac{2d-5}{4 q} \right] \kappa_\lambda(p,q) \eqend{.}
\end{splitequation}
This equation is easily solved if $c = 0$, where we obtain
\begin{equation}
\label{eq:app_modularfock_kappalambda}
\kappa_\lambda(p,q) = \exp\left[ - b (d-1) \lambda + \mathi \mathe^{- 2 b \lambda} p - \frac{\mathi a}{2 b} \left( \mathe^{- 2 b \lambda} - 1 \right) q \right]
\end{equation}
and thus, employing the inverse Fourier transform~\eqref{eq:app_inversion_fourier},
\begin{splitequation}
\label{eq:app_modularfock_klambda_c0}
&K_\lambda(\vec{x},\vec{y}) = \int \tilde{K}_\lambda(\vec{x},\vec{p}) \, \mathe^{- \mathi \vec{p} \vec{y}} \frac{\total^{d-1} \vec{p}}{(2\pi)^{d-1}} \\
&\quad= \int \exp\left[ - b (d-1) \lambda + \mathi \mathe^{- 2 b \lambda} \vec{p} \vec{x} - \frac{\mathi a}{2 b} \left( \mathe^{- 2 b \lambda} - 1 \right) p^0 \right] \mathe^{- \mathi \vec{p} \vec{y}} \frac{\total^{d-1} \vec{p}}{(2\pi)^{d-1}} \\
&\quad= - \mathi \frac{a}{b} \sinh(b \lambda) \pi^{- \frac{d}{2}} \Gamma\left( \frac{d}{2} \right) \mathe^{- b d \lambda} \lim_{\epsilon \to 0^+} \left[ - \left( \frac{a}{2 b} \left( \mathe^{- 2 b \lambda} - 1 \right) - \mathi \epsilon \right)^2 + \left( \mathe^{- 2 b \lambda} \vec{x} - \vec{y} \right)^2 \right]^{-\frac{d}{2}} \eqend{.} \raisetag{8.4em}
\end{splitequation}
However, for non-vanishing $c$ it does not seem that we can obtain an explicit closed-form solution. Fortunately, this is enough to determine at least the action of the modular conjugation $J$. Namely, in the limit $b \to 0$ the result~\eqref{eq:app_modularfock_klambda_c0} reduces to
\begin{splitequation}
\label{eq:app_modularfock_klambda_bc0}
K_\lambda(\vec{x},\vec{y}) &= \int \exp\left[ \mathi \vec{p} (\vec{x}-\vec{y}) + \mathi a \lambda p^0 \right] \frac{\total^{d-1} \vec{p}}{(2\pi)^{d-1}} \\
&= - \mathi a \lambda \pi^{- \frac{d}{2}} \Gamma\left( \frac{d}{2} \right) \lim_{\epsilon \to 0^+} \left[ - \left( a \lambda + \mathi \epsilon \right)^2 + \left( \vec{x} - \vec{y} \right)^2 \right]^{-\frac{d}{2}} \\
&= - \frac{\mathi}{2 a} \pi^{- \frac{d}{2}} \Gamma\left( \frac{d-2}{2} \right) \lim_{\epsilon \to 0^+} \partial_\lambda \left[ - \left( a \lambda + \mathi \epsilon \right)^2 + \left( \vec{x} - \vec{y} \right)^2 \right]^{1-\frac{d}{2}} \eqend{,}
\end{splitequation}
and thus we have
\begin{equation}
U(\tau) a^\dagger(f) U(-\tau) = a^\dagger(f_\tau) \eqend{,} \quad U(\tau) a(f) U(-\tau) = a(f_\tau)
\end{equation}
for $U(\tau) = \mathe^{\mathi \tau H}$ and with
\begin{equation}
\label{eq:app_modularfock_ftau}
f_\tau(\vec{x}) = - \frac{\mathi}{2} \pi^{- \frac{d}{2}} \Gamma\left( \frac{d-2}{2} \right) \lim_{\epsilon \to 0^+} \partial_\tau \int \left[ - \left( \tau + \mathi \epsilon \right)^2 + \left( \vec{x} - \vec{y} \right)^2 \right]^{1-\frac{d}{2}} f(\vec{y}) \total^{d-1} \vec{y} \eqend{.}
\end{equation}
In Fourier space, this reads
\begin{equation}
\label{eq:app_modularfock_ftau_fourier}
\widetilde{f_\tau}(\vec{p}) = \mathe^{\mathi \tau \abs{\vec{p}}} \tilde{f}(\vec{p}) \eqend{,}
\end{equation}
where we used the form~\eqref{eq:app_modularfock_klambda_bc0} of the kernel $K_\lambda$ in Fourier space, taking $a = 1$ and $\lambda = \tau$.

With this, we can compute the action of the modular conjugation for light cones~\eqref{eq:minkowski_modular_light_cone_j_def}, \eqref{eq:minkowski_modular_light_cone_past_lndelta_j} and double cones~\eqref{eq:minkowski_modular_double_cone_j_def} in Minkowski spacetime directly in Fock space. Using the action of the discrete transformations $C$~\eqref{eq:minkowski_chargeconjugation}, $P$~\eqref{eq:minkowski_parity_fock} and $T$~\eqref{eq:minkowski_timereversal_fock} on Fock space, it follows that
\begin{equation}
\label{eq:app_modularfock_jfuturecone}
J_\futurecone \, a^\dagger(f) J_\futurecone^{-1} = a^\dagger(f_{J,\futurecone}) \eqend{,} \quad J_\futurecone \, a(f) J_\futurecone^{-1} = a(f_{J,\futurecone})
\end{equation}
for the future light cone with
\begin{splitequation}
\label{eq:app_modularfock_fjfuturecone}
f_{J,\futurecone}(\vec{x}) &= - \frac{1}{4} \pi^{-d} \Gamma^2\left( \frac{d-2}{2} \right) (-1)^{\Delta-d} \lim_{\epsilon,\delta \to 0^+} \partial_\tau \partial_{\tau'} \iint \left[ - \left( \tau + \mathi \epsilon \right)^2 + \left( \vec{x} - \vec{y} \right)^2 \right]^{1-\frac{d}{2}} \\
&\hspace{8em}\times \left[ - \left( \tau' + \mathi \delta \right)^2 + \left( \vec{y} + \vec{z} \right)^2 \right]^{1-\frac{d}{2}} f^*(\vec{z}) \total^{d-1} \vec{z} \total^{d-1} \vec{y} \Big\rvert_{\tau' = \tau} \\
&= (-1)^{\Delta-d} \iint \mathe^{2 \mathi \tau \abs{\vec{q}} + \mathi \vec{q} (\vec{x}+\vec{z})} f^*(\vec{z}) \frac{\total^{d-1} \vec{q}}{(2\pi)^{d-1}} \total^{d-1} \vec{z} \\
&= - \frac{\mathi (-1)^{\Delta-d}}{4 \pi^\frac{d}{2}} \Gamma\left( \frac{d-2}{2} \right) \lim_{\epsilon \to 0^+} \partial_\tau \int \left[ - (2 \tau + \mathi \epsilon)^2 + (\vec{x}+\vec{z})^2 \right]^{1-\frac{d}{2}} f^*(\vec{z}) \total^{d-1} \vec{z} \eqend{,}
\end{splitequation}
where we used the Fourier space relation~\eqref{eq:app_modularfock_ftau_fourier} for the second equality and the result~\eqref{eq:app_inversion_fourier} for the integral over $\vec{q}$. Analogously, we compute
\begin{equation}
\label{eq:app_modularfock_jpastcone}
J_\pastcone \, a^\dagger(f) J_\pastcone^{-1} = a^\dagger(f_{J,\pastcone}) \eqend{,} \quad J_\pastcone \, a(f) J_\pastcone^{-1} = a(f_{J,\pastcone}) \quad\text{with}\quad f_{J,\pastcone}(\vec{x}) = f_{J,\futurecone}(\vec{x})
\end{equation}
for the past light cone. The limit $\tau \to 0$ is taken most easily using the second equality in Eq.~\eqref{eq:app_modularfock_fjfuturecone}, and results in
\begin{equation}
\label{eq:app_modularfock_jfuturecone_tau0}
\lim_{\tau \to 0} f_{J,\futurecone}(\vec{x}) = (-1)^{\Delta-d} f^*(\vec{x}) \eqend{.}
\end{equation}

With the action of $V$~\eqref{eq:minkowski_modular_double_cone_inversion} on Fock space, we also obtain
\begin{equation}
\label{eq:app_modularfock_jdoublecone}
J_\doublecone \, a^\dagger(f) J_\doublecone^{-1} = a^\dagger(f_{J,\doublecone}) \eqend{,} \quad J_\doublecone \, a(f) J_\doublecone^{-1} = a(f_{J,\doublecone})
\end{equation}
with
\begin{equation}
f_{J,\doublecone}(\vec{x}) = \iint \sqrt{ \frac{\abs{\vec{q}}}{\abs{\vec{p}}} } \, \mathe^{\mathi \tau (\abs{\vec{p}}+\abs{\vec{q}})} [ \tilde{f}(- \vec{q}) ]^* \int \left( \frac{\ell^2}{\vec{y}^2} \right)^\frac{d-2}{2} \mathe^{\mathi \vec{q} \vec{y} - \mathi \ell^2 (\vec{p} \vec{y})/\vec{y}^2} \total^{d-1} \vec{y} \frac{\total^{d-1} \vec{q}}{(2\pi)^{d-1}} \mathe^{\mathi \vec{p} \vec{x}} \frac{\total^{d-1} \vec{p}}{(2\pi)^{d-1}}
\end{equation}
for the double cone. To simplify this, we need the generalization of the result~\eqref{eq:app_inversion_fourier}, which can be obtained from~\eqref{eq:app_inversion_fourier_pre} and the integral~\cite[Eq.~(10.22.49)]{dlmf}:
\begin{equations}
\begin{split}
\int \mathe^{\mathi \abs{\vec{q}} a + \mathi \vec{q} \vec{b}} \abs{\vec{q}}^\frac{1}{2} \frac{\total^{d-1} \vec{q}}{(2\pi)^{d-1}} &= \frac{2^{2-d} \pi^\frac{1-d}{2} \Gamma\left( d - \frac{1}{2} \right)}{\Gamma\left( \frac{d-1}{2} \right)} \\
&\quad\times \lim_{\epsilon \to 0^+} \left[ \left( \epsilon - \mathi a \right)^\frac{1-2d}{2} \hypergeom{2}{1}\left( \frac{2d-1}{4}, \frac{2d+1}{4}; \frac{d-1}{2}; \frac{\vec{b}^2}{(a + \mathi \epsilon)^2} \right) \right] \eqend{,} \raisetag{7.8em}
\end{split} \\
\begin{split}
\int \mathe^{\mathi \abs{\vec{q}} a + \mathi \vec{q} \vec{b}} \abs{\vec{q}}^{-\frac{1}{2}} \frac{\total^{d-1} \vec{q}}{(2\pi)^{d-1}} &= \frac{2^{2-d} \pi^\frac{1-d}{2} \Gamma\left( d - \frac{3}{2} \right)}{\Gamma\left( \frac{d-1}{2} \right)} \\
&\quad\times \lim_{\epsilon \to 0^+} \left[ \left( \epsilon - \mathi a \right)^\frac{3-2d}{2} \hypergeom{2}{1}\left( \frac{2d-3}{4}, \frac{2d-1}{4}; \frac{d-1}{2}; \frac{\vec{b}^2}{(a + \mathi \epsilon)^2} \right) \right] \eqend{,} \raisetag{5.6em}
\end{split}
\end{equations}
where $\hypergeom{2}{1}$ is the Gau{\ss} hypergeometric function. It then follows that
\begin{splitequation}
\label{eq:app_modularfock_jdoublecone_f_coord}
f_{J,\doublecone}(\vec{x}) &= \frac{4^{2-d} \pi^{1-d} \Gamma\left( d - \frac{1}{2} \right) \Gamma\left( d - \frac{3}{2} \right)}{\Gamma\left( \frac{d-1}{2} \right) \Gamma\left( \frac{d-1}{2} \right)} \\
&\quad\times \lim_{\delta,\epsilon \to 0^+} \iint \left( \delta - \mathi \tau \right)^\frac{3-2d}{2} \hypergeom{2}{1}\left( \frac{2d-3}{4}, \frac{2d-1}{4}; \frac{d-1}{2}; \frac{\left( \vec{x} - \ell^2 \vec{y}/\vec{y}^2 \right)^2}{(\tau + \mathi \delta)^2} \right) \\
&\hspace{5em}\times \left( \epsilon - \mathi \tau \right)^\frac{1-2d}{2} \hypergeom{2}{1}\left( \frac{2d-1}{4}, \frac{2d+1}{4}; \frac{d-1}{2}; \frac{(\vec{y}-\vec{z})^2}{(\tau + \mathi \epsilon)^2} \right) \\
&\hspace{5em}\times \left( \frac{\ell^2}{\vec{y}^2} \right)^\frac{d-2}{2} f^*(\vec{z}) \total^{d-1} \vec{y} \total^{d-1} \vec{z} \eqend{,}
\end{splitequation}
which unfortunately is not very nice. The limit $\tau \to 0$ can be taken after using a hypergeometric identity transforming the variable $z \to z^{-1}$~\cite[Eq.~(15.8.2)]{dlmf}, and results in the simpler expression
\begin{splitequation}
\lim_{\tau \to 0} f_{J,\doublecone}(\vec{x}) &= - 2^{-d} \pi^\frac{1-2d}{2} \Gamma\left( d - \frac{3}{2} \right) \iint \left[ \left( \vec{x} - \frac{\ell^2 \vec{y}}{\vec{y}^2} \right)^2 \right]^\frac{3-2d}{4} \\
&\hspace{5em}\times \left[ (\vec{y}-\vec{z})^2 \right]^\frac{1-2d}{4} \left( \frac{\ell^2}{\vec{y}^2} \right)^\frac{d-2}{2} f^*(\vec{z}) \total^{d-1} \vec{y} \total^{d-1} \vec{z} \eqend{.}
\end{splitequation}

\section{Alternative form of the modular automorphism}
\label{app:alt}

In this appendix, we want to derive a different expression for the modular automorphisms, which connects more closely to the results of Buchholz~\cite{buchholz1977} and Hislop and Longo~\cite{hisloplongo1982,hislop1988} for light cones with tip and double cones with center at the origin. We recall that for the future light cone with tip at $(t,\vec{x}) = (\tau,\vec{0})$ we have $\ln \Delta_\futurecone = 2 \pi ( D - \tau H )$~\eqref{eq:minkowski_modular_light_cone_lndelta} and hence for the modular automorphism~\eqref{eq:minkowski_modular_light_cone_automorphism}
\begin{equation}
\label{eq:app_alt_futurecone_1}
\Delta_\futurecone^{\mathi s} \op(f) \Delta_\futurecone^{- \mathi s} = \mathe^{2 \pi \mathi s ( D - \tau H )} \op(f) \, \mathe^{- 2 \pi \mathi s ( D - \tau H )} \eqend{.}
\end{equation}
On the other hand, we expect that the same automorphism is obtained by first translating the cone to the origin, then applying the modular automorphism with $\tau = 0$, and translating back. That is, we should have
\begin{equation}
\label{eq:app_alt_futurecone_2}
\Delta_\futurecone^{\mathi s} \op(f) \Delta_\futurecone^{- \mathi s} = U(-\tau) \, \mathe^{2 \pi \mathi s D} U(\tau) \op(f) U(-\tau) \, \mathe^{- 2 \pi \mathi s D} U(\tau)
\end{equation}
with the unitary $U(\tau) = \mathe^{\mathi \tau H}$~\eqref{eq:minkowski_q_def}, analogous to the translation in the modular conjugation $J_\futurecone$~\eqref{eq:minkowski_modular_light_cone_j_def}. To show that this indeed holds, we first note that it is certainly true for $s = 0$. If we can then show that the differential equations
\begin{splitequation}
\label{eq:app_alt_futurecone_ode}
\left[ \partial_s - 2 \pi \mathi (D - \tau H) \right] \mathe^{2 \pi \mathi s ( D - \tau H )} &= \left[ \partial_s - 2 \pi \mathi (D - \tau H) \right] \left[ \mathe^{- \mathi \tau H} \mathe^{2 \pi \mathi s D} \mathe^{\mathi \tau H} \right] \eqend{,} \\
\left[ \partial_s + 2 \pi \mathi (D - \tau H) \right] \mathe^{- 2 \pi \mathi s ( D - \tau H )} &= \left[ \partial_s + 2 \pi \mathi (D - \tau H) \right] \left[ \mathe^{- \mathi \tau H} \mathe^{- 2 \pi \mathi s D} \mathe^{\mathi \tau H} \right]
\end{splitequation}
hold, it follows that the right-hand sides of~\eqref{eq:app_alt_futurecone_1} and~\eqref{eq:app_alt_futurecone_2} fulfill the same first-order ODE with the same boundary condition, and are thus equal. Here and in the following, we assume implicitly that the operators in~\eqref{eq:app_alt_futurecone_ode} are acting on vectors in the dense subset $\mathcal{D}$ of the full Fock space $\mathcal{F}$, such that we can use the commutators~\eqref{eq:minkowski_hdk_algebra}. Because the adjoint action of the modular automorphism~\eqref{eq:app_alt_futurecone_1} is defined on all of Fock space by continuity, this is sufficient to conclude that~\eqref{eq:app_alt_futurecone_1} and~\eqref{eq:app_alt_futurecone_2} agree on the full Fock space.

Since the first equation of~\eqref{eq:app_alt_futurecone_ode} is obtained from the second one by changing $s \to -s$, it is enough to show the second one. Obviously the left-hand side vanishes, and for the right-hand side to vanish we have to show that
\begin{equation}
\label{eq:app_alt_futurecone_identity}
\mathe^{- \mathi \tau H} D \, \mathe^{- 2 \pi \mathi s D} \mathe^{\mathi \tau H} = (D - \tau H) \, \mathe^{- \mathi \tau H} \mathe^{- 2 \pi \mathi s D} \mathe^{\mathi \tau H} \quad\Leftrightarrow\quad D - \tau H = \mathe^{- \mathi \tau H} D \, \mathe^{\mathi \tau H} \eqend{.}
\end{equation}
Again we see that the last equality is true for $\tau = 0$, and taking a derivative with respect to $\tau$ we obtain that it holds in general if
\begin{equation}
H = - \mathi \mathe^{\mathi \tau H} [ H, D ] \, \mathe^{- \mathi \tau H} = \mathe^{\mathi \tau H} H \, \mathe^{- \mathi \tau H} = H \eqend{,}
\end{equation}
where we used the commutator~\eqref{eq:minkowski_hdk_algebra}. So we have indeed
\begin{equation}
\Delta_\futurecone^{\mathi s} = U(-\tau) \left[ \Delta_\futurecone^{\mathi s} \right]_{\tau = 0} U(\tau) \eqend{,} \quad \Delta_\futurecone^{-\mathi s} = U(-\tau) \left[ \Delta_\futurecone^{-\mathi s} \right]_{\tau = 0} U(\tau) \eqend{;}
\end{equation}
in other words, the modular automorphism for the translated future light cone with tip at $(t,\vec{x}) = (\tau,\vec{0})$ is obtained by conjugating the modular automorphism of the centered light cone with tip at $x = 0$ by the adjoint action of the unitaries that implement time translations.

Analogously, for the double cone with center at $(t,\vec{x}) = (\tau,\vec{0})$ we recall that $\ln \Delta_\doublecone = \frac{\pi}{\ell} \left[ ( \ell^2 - \tau^2 ) H + 2 \tau D + K \right]$~\eqref{eq:minkowski_modular_double_cone_lndelta} and hence for the modular automorphism~\eqref{eq:minkowski_modular_double_cone_automorphism2}
\begin{equation}
\label{eq:app_alt_doublecone_1}
\Delta_\doublecone^{\mathi s} \op(f) \Delta_\doublecone^{- \mathi s} = \mathe^{\mathi s \frac{\pi}{\ell} \left[ ( \ell^2 - \tau^2 ) H + 2 \tau D + K \right]} \op(f) \, \mathe^{- \mathi s \frac{\pi}{\ell} \left[ ( \ell^2 - \tau^2 ) H + 2 \tau D + K \right]} \eqend{.}
\end{equation}
Again, we expect that the same automorphism is obtained by first translating the double cone to the origin, applying the modular automorphism with $\tau = 0$, and translating back. That is, we should have
\begin{equation}
\label{eq:app_alt_doublecone_2}
\Delta_\doublecone^{\mathi s} \op(f) \Delta_\doublecone^{- \mathi s} = U(-\tau) \, \mathe^{\mathi \pi s \left( \ell H + \frac{1}{\ell} K \right)} U(\tau) \op(f) U(-\tau) \, \mathe^{- \mathi \pi s \left( \ell H + \frac{1}{\ell} K \right)} U(\tau)
\end{equation}
with the same unitary $U(\tau) = \mathe^{\mathi \tau H}$~\eqref{eq:minkowski_q_def} as before. We proceed in the same way: this holds certainly for $s = 0$, and then we have to show that the differential equations
\begin{splitequation}
\label{eq:app_alt_doublecone_ode}
\begin{split}
&\left[ \partial_s - \mathi \frac{\pi}{\ell} \left[ ( \ell^2 - \tau^2 ) H + 2 \tau D + K \right] \right] \mathe^{\mathi s \frac{\pi}{\ell} \left[ ( \ell^2 - \tau^2 ) H + 2 \tau D + K \right]} \\
&\quad= \left[ \partial_s - \mathi \frac{\pi}{\ell} \left[ ( \ell^2 - \tau^2 ) H + 2 \tau D + K \right] \right] \left[ \mathe^{- \mathi \tau H} \mathe^{\mathi \pi s \left( \ell H + \frac{1}{\ell} K \right)} \mathe^{\mathi \tau H} \right] \eqend{,}
\end{split} \\
\begin{split}
&\left[ \partial_s + \mathi \frac{\pi}{\ell} \left[ ( \ell^2 - \tau^2 ) H + 2 \tau D + K \right] \right] \mathe^{- \mathi s \frac{\pi}{\ell} \left[ ( \ell^2 - \tau^2 ) H + 2 \tau D + K \right]} \\
&\quad= \left[ \partial_s + \mathi \frac{\pi}{\ell} \left[ ( \ell^2 - \tau^2 ) H + 2 \tau D + K \right] \right] \left[ \mathe^{- \mathi \tau H} \mathe^{- \mathi \pi s \left( \ell H + \frac{1}{\ell} K \right)} \mathe^{\mathi \tau H} \right]
\end{split}
\end{splitequation}
hold such that the right-hand sides of~\eqref{eq:app_alt_doublecone_1} and~\eqref{eq:app_alt_doublecone_2} fulfill the same first-order ODE with the same boundary condition, and thus must be equal. To use the commutators~\eqref{eq:minkowski_hdk_algebra}, we also again assume implicitly that the operators in~\eqref{eq:app_alt_doublecone_ode} acts on vectors in the dense subset $\mathcal{D}$ of Fock space, and then conclude equality of~\eqref{eq:app_alt_doublecone_1} and~\eqref{eq:app_alt_doublecone_2} on the full Fock space by continuity of the adjoint action of the modular automorphism.

The left-hand sides of~\eqref{eq:app_alt_doublecone_ode} clearly vanish, and the second equation is obtained from the first by changing $s \to -s$, such that it is enough to show that the right-hand side of the first equation vanishes. For that, we have to show that
\begin{splitequation}
&\mathe^{- \mathi \tau H} \left( \ell^2 H + K \right) \mathe^{\mathi \pi s \left( \ell H + \frac{1}{\ell} K \right)} \mathe^{\mathi \tau H} = \left[ ( \ell^2 - \tau^2 ) H + 2 \tau D + K \right] \mathe^{- \mathi \tau H} \mathe^{\mathi \pi s \left( \ell H + \frac{1}{\ell} K \right)} \mathe^{\mathi \tau H} \\
&\quad\Leftrightarrow ( \ell^2 - \tau^2 ) H + 2 \tau D + K = \mathe^{- \mathi \tau H} \left( \ell^2 H + K \right) \, \mathe^{\mathi \tau H} \eqend{.} \raisetag{1.8em}
\end{splitequation}
We see that the last equality holds for $\tau = 0$, and taking a derivative with respect to $\tau$ we obtain that it holds in general if
\begin{equation}
- \tau H + D = - \frac{\mathi}{2} \mathe^{- \mathi \tau H} \left[ H, \ell^2 H + K \right] \, \mathe^{\mathi \tau H} = \mathe^{- \mathi \tau H} D \, \mathe^{\mathi \tau H} \eqend{,}
\end{equation}
where we used the commutator~\eqref{eq:minkowski_hdk_algebra}. Since we have already shown~\eqref{eq:app_alt_futurecone_identity} that this last equality holds, we have indeed
\begin{equation}
\Delta_\doublecone^{\mathi s} = U(-\tau) \left[ \Delta_\doublecone^{\mathi s} \right]_{\tau = 0} U(\tau) \eqend{,} \quad \Delta_\doublecone^{-\mathi s} = U(-\tau) \left[ \Delta_\doublecone^{-\mathi s} \right]_{\tau = 0} U(\tau) \eqend{,}
\end{equation}
and the modular automorphism for the translated double cone with center at $(t,\vec{x}) = (\tau,\vec{0})$ is obtained by conjugating the modular automorphism of the double cone with center at $x = 0$ by the adjoint action of the unitaries that implement time translations.

\bibliographystyle{JHEP}
\bibliography{literature}

\end{document}